\documentclass[11pt,a4paper]{article}
\usepackage{graphicx}
\usepackage{float}
\usepackage{afterpage}
\usepackage{epsfig,cite}
\usepackage{amssymb}
\usepackage{amsmath}
\usepackage{dsfont}
\usepackage{multirow}
\usepackage{url,hyperref}
\usepackage{booktabs}
\usepackage{tabularx}
\usepackage{xcolor}
\usepackage{bm}
\usepackage{textcomp}
\usepackage{caption}
\usepackage{slashed}
\usepackage{changes}
\usepackage{url}
\usepackage{hyperref}

\usepackage[normalem]{ulem}

\usepackage{url}
\usepackage{hyperref}

\newcounter{comment}

\newcommand{\HMcolor}{red}

\newcommand{\CMcolor}{orange}

\definecolor{DarkGreen}{RGB}{0,105,62}
\newcommand{\VBcolor}{DarkGreen}

{\refstepcounter{comment}%
\begin{quote}
\color{\VBcolor}\small$\blacksquare$ \textbf{\underline{Comment} $\sharp$\thecomment.~VB:}}%
{\end{quote}}

{\refstepcounter{comment}%
\begin{quote}
\color{\HMcolor}\small$\blacksquare$ \textbf{\underline{Comment} $\sharp$\thecomment.~HM:}}%
{\end{quote}}

{\refstepcounter{comment}%
\begin{quote}
\color{\CMcolor}\small$\blacksquare$ \textbf{\underline{Comment} $\sharp$\thecomment.~CM:}}%
{\end{quote}}

\definechangesauthor[color=\CMcolor]{CM}

\begin{document}

\vspace{-2.0cm}

\begin{center} 

  {\Large \bf Revisiting evolution equations\\ for generalised parton distributions}
  \vspace{.7cm}

Valerio~Bertone$^1$, Herv\'e~Dutrieux$^1$, C\'edric~Mezrag$^1$, Jos\'e
M. Morgado$^2$, Herv\'e~Moutarde$^1$

\vspace{.3cm}
{\it ~$^1$ IRFU, CEA, Universit\'e Paris-Saclay, F-91191
  Gif-sur-Yvette, France.}\\
{\it ~$^2$Department of Integrated Sciences and Center for Advanced Studies in Physics,\\ Mathematics and Computation, University of Huelva, E-21071 Huelva, Spain.\\}

\end{center}   

\vspace{0.1cm}

\begin{center}
  {\bf \large Abstract}\\
\end{center}

We revisit the evolution of generalised parton distributions (GPDs) in
momentum space.  We formulate the evolution kernels at one-loop in
perturbative Quantum Chromodynamics (pQCD) in a form suitable for
numerical implementation and that allows for an accurate study of
their properties. This leads to the first open-source implementation
of GPD evolution equations able to cover the entire kinematic region
and allowing for heavy-quark-threshold crossings. The numerical
implementation of the GPD evolution equations is publicly accessible
through the {\tt APFEL++} evolution library and is available within
the {\tt PARTONS} framework.

Our formulation makes use of the operator definition of GPDs in
light-cone gauge renormalised in the $\overline{\mbox{MS}}$ scheme.
For the sake of clarity, we recompute the evolution kernels at
one-loop in pQCD, confirming previous calculations. We obtain general
conditions on the evolution kernels deriving from the GPD sum rules
and show that our formulation obeys these conditions. We analytically
show that our calculation reproduces the DGLAP and the ERBL equations
in the appropriate limits and that it guarantees the continuity of
GPDs. We numerically check that the evolved GPDs fulfil DGLAP and ERBL
limits, continuity, and polynomiality.  We benchmark our numerical
implementation against analytical evolution in conformal
space. Finally, we perform a numerical comparison to an existing
implementation of GPD evolution finding a general good agreement on
the kinematic region accessible to the latter.

This work provides a pedagogical description of GPD evolution
equations which benefits from a renewed interest as future colliders,
such as the electron-ion colliders in the US and in China, are being
designed. It also paves the way to the extension of GPD evolution
codes to higher accuracies in pQCD desirable for precision
phenomenology at these facilities.

\clearpage

\tableofcontents

\section{Introduction}
\label{sec:introduction}

Generalised parton distributions (GPDs) have been introduced in the
1990s~\cite{Mueller:1998fv,Ji:1996nm,
  Ji:1996ek,Radyushkin:1996ru,Radyushkin:1997ki} and have been
thoroughly studied ever since (see \emph{e.g.} the review papers in
Refs.~\cite{Diehl:2003ny,Belitsky:2005qn,Kumericki:2016ehc}). There
are many reasons for their interest. GPDs can be interpreted in terms
of partonic probability densities in longitudinal momentum and
transverse position~\cite{Burkardt:2000za,Diehl:2002he}. Therefore, their knowledge
would allow us to obtain a spacial picture of hadrons (hadron
tomography) that is not achievable otherwise. Moreover, GPDs are
closely related to the form factors of the energy-momentum tensor,
allowing for a gauge-invariant spin decomposition of the
hadron~\cite{Ji:1996ek} and for a formal analogy with pressure and
shear forces distributions~\cite{Polyakov:2002yz}. GPDs emerge from
the factorisation of exclusive hard processes such as deeply-virtual
Compton scattering~\cite{Ji:1996nm,Collins:1998be}. This ultimately gives us the
possibility to achieve an experimentally-driven tomography of
hadrons. In fact, this has been one of the main motivations to invest
in current experimental programmes, such as the Jefferson Laboratory
upgrade to 12 GeV, and in future facilities, like the electron-ion
colliders in the US (EIC)~\cite{Accardi:2012qut,AbdulKhalek:2021gbh}
and in China (EicC)~\cite{Anderle:2021wcy}.

Already in the early days of GPDs, and guided by both the work done on
parton distribution functions (PDFs) and distribution amplitudes
(DAs), several groups have derived evolution equations for GPDs,
generalising both the Dokshitzer-Gribov-Lipatov-Altarelli-Parisi
(DGLAP) and Efremov-Radyushkin-Brodsky-Lepage (ERBL) evolution
equations. Leading-order (LO) results were readily
obtained~\cite{Mueller:1998fv,Ji:1996nm,Radyushkin:1997ki,Balitsky:1997mj,Radyushkin:1998es,Blumlein:1997pi,Blumlein:1999sc}
followed shortly after by the calculation of the next-to-leading order
(NLO)
corrections~\cite{Belitsky:1998vj,Belitsky:1998gc,Belitsky:1999gu,Belitsky:1999fu,Belitsky:1999hf}
that were recently confirmed by an independent
study~\cite{Braun:2019qtp} and even extended to three loops (NNLO) in
the non-singlet case~\cite{Braun:2017cih}.

On the phenomenological side, early efforts were devoted to developing
GPD evolution codes.  A. Vinnikov~\cite{Vinnikov:2006xw} developed the
first open-source code in momentum space able to evolve GPDs at LO
accuracy. However, the code webpage does not exist anymore and, as far
as we can tell, the only public version of this code is its
implementation in the {\tt PARTONS}
framework~\cite{Berthou:2015oaw}. A few years before, A. Freund and
M. McDermott~\cite{Freund:2001bf} developed a code able to evolve GPDs
at NLO tailored to the computation of deeply-virtual Compton
scattering. However, to the best of our knowledge, this code was never
made fully open-source and as of today it is difficult to find a clean
copy. In parallel, a strong effort was put into obtaining an evolution
procedure at NLO in conformal space (see \emph{e.g.}
Ref.~\cite{Mueller:2005ed}), yielding the only public NLO evolution
code available today~\cite{Kumericki:2007sa, gepard}. We point out
that all the codes mentioned above are rigidly associated to specific
GPD models or families of parameterisations, and can hardly be used
out of the box to evolve different input GPDs. Moreover, to the best
of our knowledge, none of them allows for the treatment of heavy
flavours while a significant amount of current experimental data lies
above the charm threshold. In the last decade, these codes have not
taken the front stage mainly because the latest and most precise
experimental data related to GPDs were obtained in relatively small
ranges and at relatively small values of the hard scale $Q^2$. The
necessity of using evolution equations for a consistent theoretical
analysis of experimental data was jeopardised by the poor accuracy of
LO perturbative QCD at low scales. This has made evolution of GPDs
less critical for phenomenological purposes. However, with the
forthcoming EIC and EicC the situation is expected to change
drastically as exclusive processes will be measured in a larger
kinematic range making the need for evolution pressing.

In this paper, we revisit the LO evolution equations of GPDs in
momentum space computing the one-loop unpolarised anomalous dimensions
renormalised in the $\overline{\mbox{MS}}$ scheme in the light-cone
gauge. In order to make the paper self-contained, we provide a
pedagogical description of the computation targeting newcomers
unfamiliar with the most technical aspects of the field, a community
which is expected to grow in view of the timeline of the EIC and EicC
projects. We formalise our results in a way that allows us to study
their properties and that facilitates the numerical
implementation. The solution of the evolution equations is implemented
in the open-source code {\tt APFEL++}~\cite{Bertone:2013vaa,
  Bertone:2017gds} that is interfaced to the {\tt PARTONS} framework.

In section~\ref{sec:evolutionequations}, we derive the GPD evolution
equations and present our calculation of the kernels. These equations
are presented in a form that resembles the DGLAP equations thus
allowing to exploit the capabilities of existing evolution codes such
as {\tt APFEL++} for their solution. In section~\ref{sec:properties},
we present a thorough study of the analytic properties of the ensuing
evolution kernels. In section~\ref{sec:numerics}, we discuss the
numerical implementation and provide quantitative evidence that the
evolution fulfils fundamental requirements such as correct DGLAP and
ERBL limits, continuity, polynomiality, and equivalence with the
conformal-space approach. To the best of our knowledge, these
numerical consistency checks have not been discussed in a detailed
manner in the existing literature concerning GPD evolution codes.
Finally, in section~\ref{sec:conclusions} we summarise and give some
concluding remarks. Appendices are devoted to some technical
aspects. Appendix~\ref{app:gpdnormlisation} discusses the general
method used to compute the evolution kernels by the introduction of
the parton-in-parton GPDs, appendix~\ref{app:splittingfunctions} gives
some details concerning the explicit calculation of the one-loop
evolution kernel $\mathcal{P}_{q/q}^{[0]}$, and
appendix~\ref{app:conformalmoments} presents the explicit calculation
of its conformal moments.

\section{Operator definition of GPDs and evolution equations}
\label{sec:evolutionequations}

GPDs enjoy an operator definition that results from the collinear
factorisation of processes like deeply-virtual Compton scattering and
deeply-virtual meson
production~\cite{Collins:1996fb,Collins:1998be}. This operator
definition is affected by UV divergences related to the integration
over the transverse momenta $k_T$ of the constituent partons and that
need to be renormalised. As customary, the renormalisation procedure
introduces an unphysical scale, $\mu$, that roughly speaking
corresponds to a cutoff on the integral in $k_T$.  The fact that
unrenormalised (bare) GPDs do not depend on $\mu$ allows one to derive
a set of renormalisation-group equations (RGEs) that governs the
dependence of the renormalised GPDs on $\mu$: the evolution
equations. The anomalous dimensions (sometimes referred to as
evolution kernels or splitting functions) of these evolution equations
can be computed in perturbation theory by isolating the coefficient of
the UV divergences of the bare GPDs order by order in the expansion in
powers of the strong coupling $\alpha_s$. GPDs cannot be computed in
perturbation theory but to the purpose of extracting the UV poles, one
can replace the hadronic states that enter their operator definition
with partonic states thus enabling an explicit computation. This
descends from the fact that GPDs emerge from factorisation theorems
that apply to \textit{any} target. As a consequence, the extraction of
the anomalous dimensions related to UV poles (as well as of the
partonic cross sections) is conveniently done using partonic on-shell
targets~\cite{Collins:2011}. Although GPDs are not physical
observables, they are gauge-invariant quantities.  In a covariant
formulation, gauge invariance is guaranteed by the presence of the
so-called Wilson line that connects the bi-local GPD operator along
the light-cone direction.

When using the operator definition of GPDs in perturbative
calculations, the presence of the Wilson line introduces substantial
complications~\cite{Collins:2011zzd}. This is due to the fact that a
Wilson line can be pictured as the radiation of an arbitrary number of
collinear gluons with scalar polarisation (\textit{i.e.} with
polarisation proportional to the gluon momentum and thus to the
collinear direction) that massively increase the number of diagrams to
be considered at any given perturbative order. This problem can be
overcome by adopting an axial gauge, $n\cdot A = 0$, in which the
gauge vector $n$ is on the light cone, $n^2=0$: this is usually called
light-cone gauge. By definition, scalar gluons are absent in the
light-cone gauge thus enormously reducing the number of diagrams to be
considered. In addition, in light-cone gauge there are no
ghosts~\cite{Lepage:1980fj} which further reduces the complication of
the calculation. These simplifications however come at the price of a
complication of the gluon propagator that in the light-cone gauge
takes the form:
\begin{equation}
\mathcal{D}_{\mu\nu}(k) =
\frac{1}{k^2+i\varepsilon}\left(-g_{\mu\nu}+\frac{k_\mu
    n_\nu+k_\nu n_\mu}{k\cdot n}\right)\,.
\label{eq:lightconepropagator}
\end{equation}
A particularly unpleasant feature of this propagator is that it
develops a spurious pole at $k\cdot n=0$. However, it has been argued
that poles deriving from the gluon propagator in the light-cone gauge
must cancel in gauge invariant
quantities~\cite{Curci:1980uw}. Therefore, when computing GPD
anomalous dimensions, it is enough to regularise these poles on a
diagram-by-diagram basis using a suitable prescription bearing in mind
that they eventually cancel when summing up all
diagrams.\footnote{Different regularisation prescriptions exist. The
  authors of Ref.~\cite{Curci:1980uw} originally introduced a simple
  principal-value regularisation. Later, also other prescriptions were
  introduced~\cite{Kovchegov:1997pc, Marlen-Heinrich:1998acz}, see
  also Ref.~\cite{Chirilli:2015fza}. We also mention that in
  Refs.~\cite{Gaunt:2014xga, Gaunt:2014cfa} it was observed that, up
  to two-loop accuracy, most of the poles at $k\cdot n=0$ can be
  regularised by means of dimensional regularisation. In addition,
  those that cannot be regularised in this way (only one specific
  virtual three-point integral, see Appendix~C of
  Refs.~\cite{Gaunt:2014cfa}) give a result that is largely
  independent of the regularisation procedure. We thank the referee
  for drawing our attention to these calculations.}

Finally, the operator defining the bare quark and gluon unpolarised
GPDs of a generic hadron species $H$ in the light-cone gauge with
gauge vector $n$ reads:
\begin{equation}
\begin{array}{l}
  \displaystyle \hat{F}_{q/H}(x,\xi, \Delta^2)
  =\displaystyle \int\frac{dy}{2\pi}e^{-ix(n\cdot P)y}\left\langle
  P-\Delta\left|\overline{\psi}_q\left(\frac{yn}2\right)\frac{\slashed{n}}2
  \psi_q\left(-\frac{yn}2\right)\right|P+\Delta\right\rangle\\
  \\
  \displaystyle \hat{F}_{g/H}(x,\xi, \Delta^2) =\displaystyle \frac{n_\mu n_\nu }{x(n\cdot P)}\int\frac{dy}{2\pi}e^{-ix(n\cdot P)y}\left\langle
  P-\Delta\left|F_{a}^ {\mu j}\left(\frac{yn}2\right)
  F_{a}^{\nu j}\left(-\frac{yn}2\right)\right|P+\Delta\right\rangle\,,\\
\end{array}
\label{eq:GPDdefinition}
\end{equation}
where $\psi_q$ is the quark field for the flavour $q $ in the
fundamental colour representation and $F_{a}^ {\mu \nu}$ is the gluon
field strength for the colour configuration $a$ in the adjoint
representation. The integrals are understood to run between $-\infty$
and $+\infty$.  The variable $x$ is the longitudinal fraction of the
average momentum $P$ carried by the parton while $\xi$, often referred
to as skewness, is the longitudinal fraction of the momentum transfer
$\Delta$. In addition, an average over the initial-state spin/helicity
physical states is understood. The index $j$ in the gluon distribution
runs over the longitudinal components ($j=1,2$) and is summed over as
well as the colour index $a$. Notice the absence of the Wilson line as
a consequence of the light-cone gauge. A further simplification
induced by this gauge is that the contraction of the gauge vector with
the gluon field strength reduces to
$n_\mu F_a^{\mu j}(x)=\left(n\cdot\partial\right)A_a^j(x)$. The
tensorial decomposition of the correlators in
Eq.~(\ref{eq:GPDdefinition}) leads to the actual definition of the
bare GPDs $\hat{H}_{i/H}$ and $\hat{E}_{i/H}$~\cite{Diehl:2003ny}:
\begin{equation}
\begin{array}{rcl}
  \displaystyle \hat{F}_{i/H}(x,\xi, \Delta^2)
  &=&\displaystyle \frac{1}{n\cdot P}\bigg[\hat{H}_{i/H}(x,\xi,
      \Delta^2)\overline{u}(P-\Delta)\frac{\slashed{n}}{2}u(P+\Delta)\\
  \\
  &+&\displaystyle \hat{E}_{i/H}(x,\xi,
      \Delta^2)\overline{u}(P-\Delta)\frac{i\sigma^{\mu\nu}n_\mu\Delta_\nu}{4M}u(P+\Delta)\bigg]\,,
\end{array}
\end{equation}
with $i=q,g$ and where $u$ is spinor of the external state $H$ and $M$
is its mass. Notice that the definitions in
Eq.~(\ref{eq:GPDdefinition}) are such that GPDs in the forward limit
$\Delta\rightarrow 0$ \textit{exactly} reproduce the standard
collinear parton distribution functions (PDFs):
\begin{equation}
\begin{array}{l}
  \displaystyle \lim_{\Delta\rightarrow 0}\hat{F}_{q/H}(x,\xi, \Delta^2) =
  \hat{f}_{q/H}(x)\,,\\
  \\
  \displaystyle \lim_{\Delta\rightarrow 0}\hat{F}_{g/H}(x,\xi, \Delta^2) = \hat{f}_{g/H}(x)\,.
\end{array}
\label{eq:GPDlimit}
\end{equation}
In order to fulfil Eq.~(\ref{eq:GPDlimit}) for the gluon, we adopt the
off-forward generalisation of the definition of gluon PDF given in
Ref.~\cite{Collins:1981uw}. This differs by a factor $2/(n\cdot P)$
w.r.t. Ref.~\cite{Ji:1996nm} and by factor $1/x$
w.r.t. Ref.~\cite{Diehl:2003ny}. From now on, we will drop the
dependence on the total momentum transfer $\Delta^2$ because it does
not participate in the evolution of GPDs.

\begin{figure}[t]
  \centering
    \includegraphics[width=0.9\textwidth]{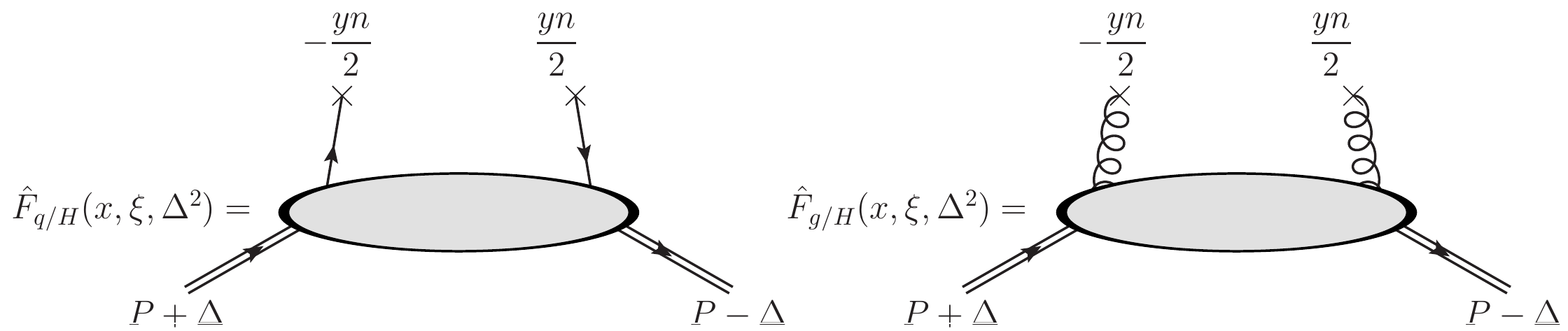}
    \vspace{15pt}
    \caption{Graphical representation of the parton-in-hadron GPDs
      defined in Eq.~(\ref{eq:GPDdefinition}).\label{fig:HadronGPDs}}
\end{figure}

A graphical representation of the GPDs defined in
Eq.~(\ref{eq:GPDdefinition}) is displayed in
Fig.~\ref{fig:HadronGPDs}.
In these graphs, the crosses represent the operator insertion and the
integration over $y$, that is:
\begin{figure}[h!]
  \centering
    \includegraphics[width=0.5\textwidth]{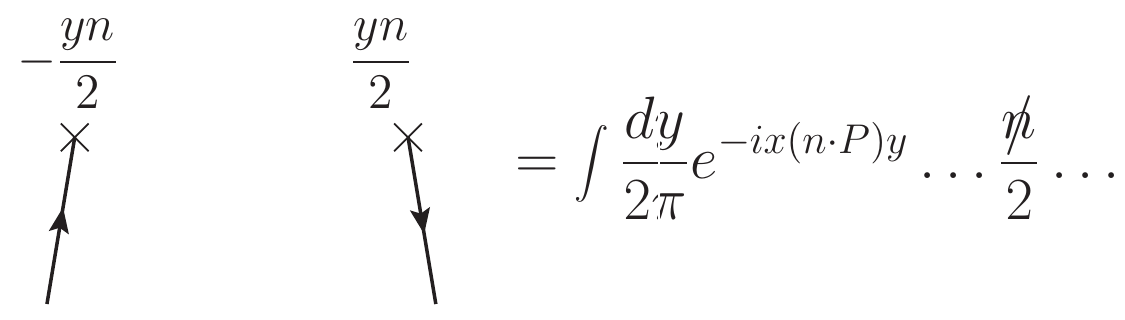}
\end{figure}

\noindent for quarks and:
\begin{figure}[h!]
  \centering
  \includegraphics[width=0.55\textwidth]{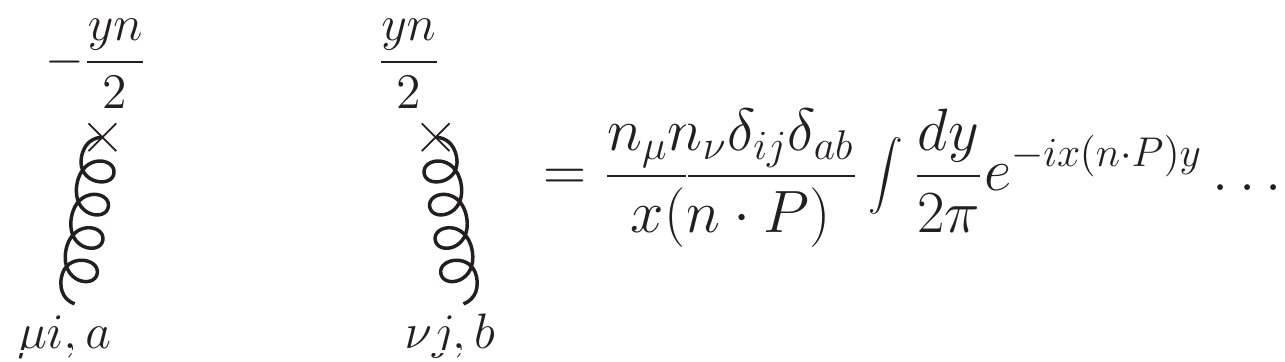}
\end{figure}

\noindent for gluons.

Assuming dimensional regularisation in $4-2\varepsilon$ dimension,
with $\varepsilon>0$, the bare GPD correlator in
Eq.~(\ref{eq:GPDdefinition}) can be renormalised in the
$\overline{\mbox{MS}}$ scheme as follows:
\begin{equation}
  F_{i/H}(x,\xi,\mu) =
  \sum_{j=q,g}\int_{-1}^1\frac{dy}{|y|}Z_{ij}\left(\frac{x}{y},\frac{\xi}{x},\alpha_s(\mu),\varepsilon\right)\hat{F}_{j/H}(y,\xi ,\varepsilon)\,,\quad
  i = q,g\,,
\label{eq:renormalisation}
\end{equation}
where the sum runs over all active quark flavours at the scale
$\mu$. Due to longitudinal boost invariance, the
$\overline{\mbox{MS}}$ renormalisation constants $Z_{ij}$ can only be
functions of ratios of momentum fractions, of the coupling $\alpha_s$,
and of the regulator $\varepsilon$, and can be expanded
as~\cite{Collins:2011}:\footnote{In the \textit{modified}
  minimal-subtraction ($\overline{\mbox{MS}}$) scheme the poles are
  actually embedded in powers of $S_{\varepsilon}/\varepsilon$ with:
\begin{equation}
  S_\varepsilon=\frac{(4\pi)^\varepsilon}{\Gamma(1-\varepsilon)} = 1 +
  \varepsilon\left(\ln 4\pi-\gamma_{\rm
      E}\right)+\mathcal{O}(\varepsilon^2)\,,
\label{eq:Sepsilon}
\end{equation}
where $\gamma_{\rm E}$ is the Euler constant. To simplify the
notation, in the following we omit the factor $S_{\varepsilon}$.}
\begin{equation}
  Z_{ij}\left(\frac{x}{y},\frac{\xi}{x},\alpha_s,\varepsilon\right)
  = \sum_{n=0}^{\infty}a_s^n
  Z_{ij}^{[n]}\left(\frac{x}{y},\frac{\xi}{x},\varepsilon\right)=\delta_{ij}\delta\left(1-\frac{x}{y}\right)+\sum_{n=1}^{\infty}a_s^n
  \sum_{p=1}^{n}\frac{1}{\varepsilon^p}Z_{ij}^{[n,p]}\left(\frac{x}{y},\frac{\xi}{x}\right)\,,
\label{eq:Zexpansion}
\end{equation}
where we have defined $a_s=g^2/16\pi^2=\alpha_s/4\pi$. Exploiting the
fact that $\hat{F}$ does not depend on the renormalisation scale
$\mu$, the logarithmic derivative w.r.t. $\mu$ of
Eq.~(\ref{eq:renormalisation}) gives:
\begin{equation}
  \frac{dF_{i/H}(x,\xi,\mu)}{d\ln\mu^2} =
  \sum_{k=q,g}\int_{-1}^1\frac{dz}{|z|}\mathcal{P}_{i/k}\left(\frac{x}{z},\frac{\xi}{x},\alpha_s(\mu)\right)F_{k/H}(z,\xi,\mu)\,,
\end{equation}
with:
\begin{equation}
\mathcal{P}_{i/k}\left(\frac{x}{z},\frac{\xi}{x},\alpha_s\right)=\lim_{\varepsilon\rightarrow
0}\sum_{j=q,g}\int_{-1}^1\frac{dy}{|y|}\frac{dZ_{ij}\left(\frac{x}{y},\frac{\xi}{x},\alpha_s(\mu),\varepsilon\right)
}{d\ln\mu^2}Z_{jk}^{-1}\left(\frac{y}{z},\frac{\xi}{y},\alpha_s,\varepsilon\right)\,,
\label{eq:kernels}
\end{equation}
where $Z_{kj}^{-1}$ is defined by means of the following equality:
\begin{equation}
\sum_{j=q,g}\int_{-1}^1\frac{dx}{|x|}Z_{kj}^{-1}\left(\frac{z}{x},\frac{\xi}{z},\alpha_s,\varepsilon\right)Z_{ji}\left(\frac{x}{y},\frac{\xi}{x},\alpha_s,\varepsilon\right)=\delta_{ki}\delta\left(1-\frac{z}{y}\right)\,.
\end{equation}
Notice that the definition of $\mathcal{P}_{i/k}$ allows one to take
the limit $\varepsilon\rightarrow 0$ because these quantities are
finite order by order in perturbation theory and therefore admit the
perturbative expansion:
\begin{equation}
\mathcal{P}_{i/k}\left(\frac{x}{z},\frac{\xi}{x},\alpha_s\right)=
\sum_{n=0}^{\infty}a_s^{n+1}
\mathcal{P}_{i/k}^{[n]}\left(\frac{x}{z},\frac{\xi}{x}\right)\,.
\label{eq:splittingfunctionsexp}
\end{equation}
From Eq. \eqref{eq:kernels}, $\mathcal{P}_{i/k}$ can be seen as an
$x$-dependent generalisation of the anomalous dimension introduced in
the renormalisation of local operators.  Exploiting the fact that
$Z_{ij}$ depend on the scale $\mu$ only through the strong coupling
$\alpha_s$, one can further manipulate the derivative in
Eq.~(\ref{eq:kernels}) as follows:
\begin{equation}
\begin{array}{rcl}
\displaystyle \frac{dZ_{ij}\left(\frac{x}{y},\frac{\xi}{x},\alpha_s(\mu),\varepsilon\right)
}{d\ln\mu^2} &=& \displaystyle \left(-\varepsilon a_s+\beta(a_s)\right) \frac{dZ_{ij}\left(\frac{x}{y},\frac{\xi}{x},\alpha_s,\varepsilon\right)
}{da_s}\\
\\
&=&\displaystyle \left(-\varepsilon
  a_s+\beta(a_s)\right)\sum_{n=1}^{\infty}na_s^{n-1}\sum_{p=0}^{n}
  \frac1{\varepsilon^p}Z_{ij}^{[n,p]}\left(\frac{x}{y},\frac{\xi}{x}\right)
\end{array}
\end{equation}
where we have used the $(4-2\varepsilon)$-dimensional RGE for the
strong coupling:
\begin{equation}
  \frac{da_s}{d\ln\mu^2}=-\varepsilon a_s+\beta(a_s)\,.
\end{equation}
Since in this paper we are mainly concerned with the leading-order
contribution to perturbative expansion of the anomalous dimensions in
Eq.~(\ref{eq:splittingfunctionsexp}), considering that
$\beta(a_s)=\mathcal{O}(\alpha_s^2)$, we find:
\begin{equation}
\mathcal{P}_{i/k}^{[0]}\left(\frac{x}{z},\frac{\xi}{x}\right)=-Z_{ik}^{[1,1]}\left(\frac{x}{z},\frac{\xi}{x}\right)\,.
\label{eq:splittingfromZ}
\end{equation}
Therefore, the calculation of the one-loop anomalous dimension of the
GPD evolution boils down to computing the coefficient of the
divergence of the one-loop renormalisation constant of the bare GPDs
themselves. However, the procedure is totally general and can be
extended to any fixed order in perturbation theory.

The calculation of the renormalisation constants can be accomplished
by using the parton-in-parton GPDs defined in
Appendix~\ref{app:gpdnormlisation}. As mentioned above, owing to the
universality of the UV structure of the partonic correlator, one can
replace the hadronic states in Eq.~(\ref{eq:GPDdefinition}) with
partonic states thus enabling a perturbative calculation. Therefore,
both the bare and renormalised parton-in-parton GPDs enjoy the
perturbative expansions:
\begin{equation}
\begin{array}{l}
\hat{F}_{i/j}(x,\xi ,\varepsilon) =\displaystyle \sum_{n=0}^{\infty}a_s^n \hat{F}_{i/j}^{[n]}(x,\xi ,\varepsilon)\,,\\
\\
F_{i/j}(x,\xi, \mu) =\displaystyle \sum_{n=0}^{\infty}a_s^n F_{i/j}^{[n]}(x,\xi,\mu)\,,
\end{array}
\end{equation}
that plugged into Eq.~(\ref{eq:renormalisation}), along with the
expansion in Eq.~(\ref{eq:Zexpansion}), allow us to relate bare and
renormalised parton-in-parton GPDs order by order in $\alpha_s$:
\begin{equation}
F_{i/k}^{[n]}(x,\xi,\mu) =
\sum_{j=q,g}\sum_{p=0}^{n}\int_{-1}^1\frac{dy}{|y|}
  Z_{ij}^{[p]}\left(\frac{x}{y},\frac{\xi}{x},\varepsilon\right) \hat{F}_{j/k}^{[{n-p}]}(y,\xi ,\varepsilon)\,.
\end{equation}
The first two orders explicitly read:
\begin{equation}\label{eq:normpartoninpartonGPDs}
\begin{array}{rcl}
  F_{i/k}^{[0]}(x,\xi,\mu) &=&\displaystyle
                               \hat{F}_{i/k}^{[0]}(x,\xi
                               ,\varepsilon)\equiv D_i(\xi)\delta_{ik}\delta(1-x)\,,\\
  \\
  F_{i/k}^{[1]}(x,\xi,\mu) &=&\displaystyle \hat{F}_{i/k}^{[1]}(x,\xi ,\varepsilon) 
                               +Z_{ik}^{[1]}\left(x,\frac{\xi}{x},\varepsilon\right) D_k(\xi)\,,
\end{array}
\end{equation}
where the first equality is the result of a tree-level computation
using the definitions in Eq.~(\ref{eq:GPDdefinition}) (see
Appendix~\ref{app:gpdnormlisation} where the factors $D_i$ are also
derived). The second equality instead allows us to extract
$Z_{ik}^{[1,1]}$ in Eq.~(\ref{eq:splittingfromZ}) by requiring that
$F_{i/k}^{[1]}(x,\xi,\mu)$ be finite in the $\varepsilon\rightarrow 0$
limit, finally obtaining:
\begin{equation}
  \mathcal{P}_{i/k}^{[0]}\left(x,\frac{\xi}{x}\right) =  {\rm
    P.P.}\left[\hat{F}_{i/k}^{[1]}(x,\xi ,\varepsilon)\right]D_k^{-1}(\xi)\,,
\label{eq:splittingfromGPD}
\end{equation}
where P.P. stands for ``$\overline{\mbox{MS}}$ UV pole part''.
$\hat{F}_{i/k}^{[1]}$ can be obtained through the calculation of the
appropriate one-loop diagrams. Using the definitions given in
Appendix~\ref{app:gpdnormlisation}, we have computed the one-loop
corrections to all (non-vanishing) parton-in-parton GPDs and extracted
the pole part. Finally, using Eq.~(\ref{eq:splittingfromGPD}), we
found that the one-loop anomalous dimensions have the following
structure:\footnote{The integral appearing in the second line of
  Eq.~(\ref{eq:polepartGPDFs}) is clearly divergent. However, this
  expression is to be intended in the sense of a distribution that
  acquires a meaning only upon integration. In this respect, the
  diverging integral has the scope of subtracting an opposite
  divergence generated by the first line of
  Eq.~(\ref{eq:polepartGPDFs}).}
\begin{equation}
\begin{array}{rcl}
  \displaystyle \mathcal{P}_{i/k}^{[0]}\left(x,\frac{\xi}{x}\right)
  &=&\displaystyle  
      \theta(1-x) \left[\theta\left(x+\xi\right) p_{ik}\left(x,\frac{\xi}{x}\right)+\theta\left(x-\xi\right) p_{ik}\left(x,-\frac{\xi}{x}\right)\right]\\
  \\
  &+&\displaystyle \delta_{ik}\delta(1-x) 2C_i\left[K_i-2\int_0^1 \frac{dz}{1-z}-\ln\left(\left|1-\frac{\xi^2}{x^2}\right|\right)\right] \,,
\end{array}
\label{eq:polepartGPDFs}
\end{equation}
where:
\begin{equation}
\begin{array}{rcl}
  p_{qq}(y,\kappa) &=& \displaystyle C_F \frac{(1+\kappa)
                       (1-y+2\kappa y)}{\kappa (1+\kappa y)(1-y)}\,,\\
  \\
  p_{qg}(y,\kappa) &=& \displaystyle T_R\frac{(1+\kappa)
                       (1-2y+\kappa y)}{\kappa (1+\kappa y) (1-\kappa^2 y^2)} \,,\\
  \\
  p_{gq}(y,\kappa) &=& \displaystyle C_F\frac{(1+\kappa )
                       (2-y+\kappa y)}{\kappa y (1+\kappa y)}\,,\\
  \\
  p_{gg}(y,\kappa) &=&\displaystyle -C_A\frac{1-\kappa^2}{\kappa (1-\kappa^2 y^2)}\left[1-\frac{2\kappa y}{1-y}-\frac{2(1+y^2)}{y (1-\kappa ) (1+\kappa y)}\right]\,,
\end{array}
\label{eq:explicitpik}
\end{equation}
and:
\begin{equation}
K_q = \frac32\,,\quad K_g = \frac{11C_A-4n_fT_R}{6C_A}\,,
\end{equation}
with $C_g=C_A=N_c=3$, $C_q=C_F=(N_c^2-1)/2N_c=4/3$, $T_R=1/2$, and
$n_f$ the number of active quark flavours. For the sake of
illustration, the explicit calculation of $\hat{F}_{q/q}^{[1]}$, that
allowed us to extract $\mathcal{P}_{q/q}^{[0]}$, is presented in
Appendix~\ref{app:splittingfunctions}. The remaining one-loop
parton-in-parton GPDs and the corresponding anomalous dimensions can
be computed in a similar fashion.

\begin{figure}[t]
  \centering
  \includegraphics[width=0.7\textwidth]{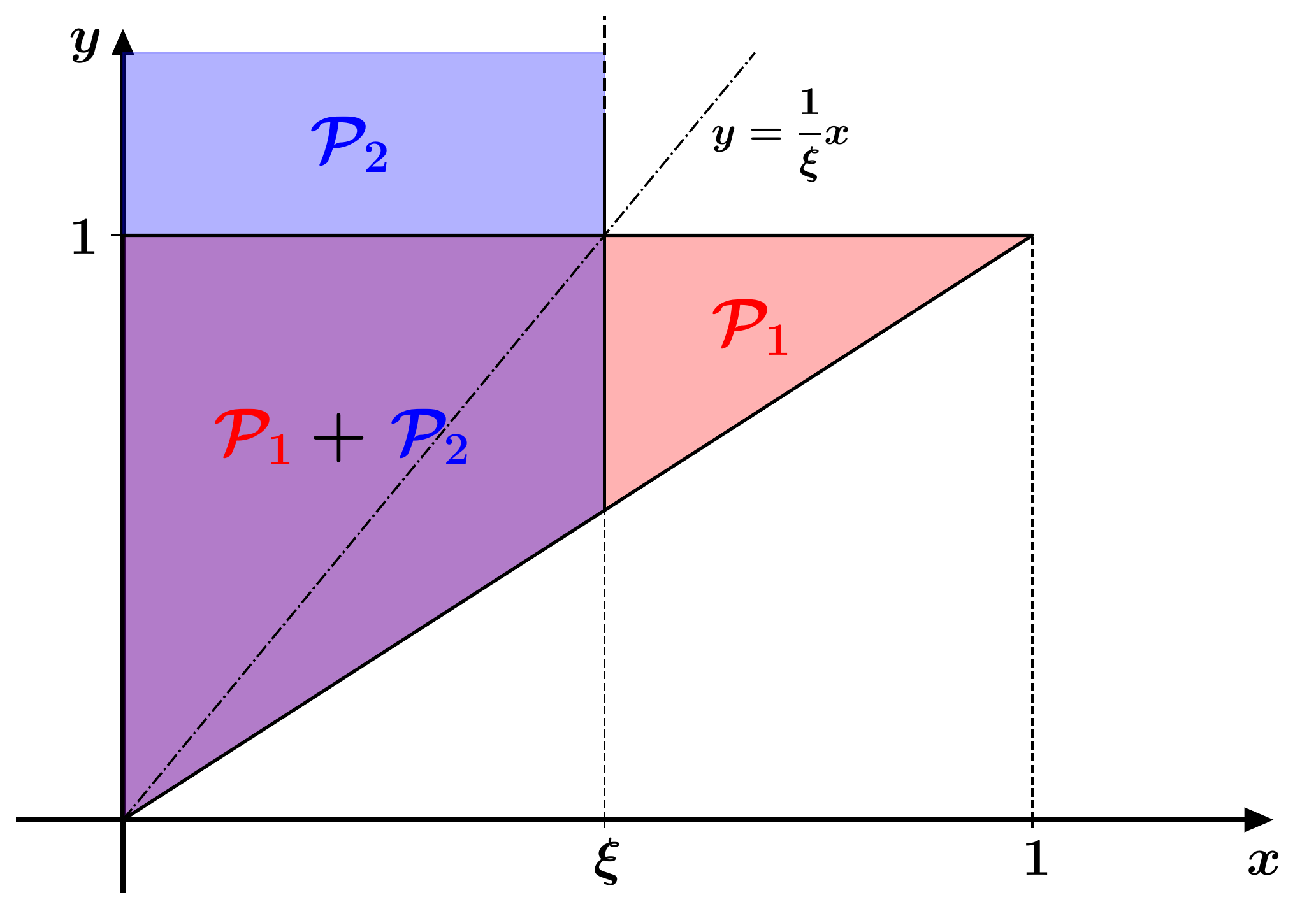}
  \vspace{15pt}
  \caption{Integration domain covered by the convolution integral in
    the r.h.s. of the evolution equations in
    Eq.~(\ref{eq:regdglap}). The coverage of the single functions
    $\mathcal{P}_1$ and $\mathcal{P}_2$ according to the decomposition
    in Eq.~(\ref{eq:KernelDecomposition}) is shown in red and blue,
    respectively. The dot-dashed line corresponding to $y = x/\xi$ is
    relevant in that along this line both $\mathcal{P}_1$ and
    $\mathcal{P}_2$ separately diverge (see
    Sect.~\ref{subsec:continuity}). \label{fig:GPDIntDomain}}
\end{figure}

In the following, we will formulate the GPD evolution equations in a
form that resembles the DGLAP equations for PDFs. On the one hand,
this facilitates the implementation in existing computer codes able to
compute the DGLAP evolution. Indeed, relying on solid and
well-established numerical techniques, several DGLAP evolution codes
have nowadays reached a numerical accuracy well below the per-mil
level~\cite{Salam:2008qg,Botje:2010ay,Bertone:2013vaa,Bertone:2017gds,Diehl:2021gvs}.
On the other hand, this formulation allows us to highlight some
interesting properties of the anomalous dimensions. In order to do so,
we restrict the longitudinal momentum fraction $x$ to be
non-negative. This can be done first by observing that, using the
definition in Eq.~(\ref{eq:GPDdefinition}), the gluon GPD is an odd
function of $x$, so that
${F}_{g/H}(-x,\xi, \Delta^2)=-{F}_{g/H}(x,\xi, \Delta^2)$, and second
by defining the anti-quark GPDs as
${F}_{\overline{q}/H}(x,\xi, \Delta^2)=-{F}_{q/H}(-x,\xi,
\Delta^2)$. In addition, from Eq.~(\ref{eq:polepartGPDFs}) it is
apparent that evolution kernels and thus GPDs are symmetric under the
transformation $\xi\rightarrow -\xi$. Therefore, without loss of
generality we can restrict to considering non-negative values of
$\xi$.  We can then write leading-order evolution equations for quark,
antiquark, and gluon GPDs separately as:
\begin{equation}
\begin{array}{rcl}
\displaystyle   \frac{dF_{q/H}( x,\xi,\mu)}{d\ln\mu^2} &=&\displaystyle 
  \frac{\alpha_s(\mu)}{4\pi}\Bigg\{\int_{0}^1\frac{dz}{z}\mathcal{P}^{[0]}_{q/q}\left(\frac{x}{z},\frac{\xi}{x}\right)F_{q/H}(z,\xi,\mu)\\
\\
&-&\displaystyle \int_{0}^1\frac{dz}{z}\mathcal{P}^{[0]}_{q/q}\left(-\frac{x}{z},\frac{\xi}{x}\right)F_{\overline{q}/H}(z,\xi,\mu)\\
\\
&+&\displaystyle 
  \int_{0}^1\frac{dz}{z}\left[\mathcal{P}^{[0]}_{q/g}\left(\frac{
    x}{z},\frac{\xi}{x}\right)-\mathcal{P}^{[0]}_{q/g}\left(-\frac{
    x}{z},\frac{\xi}{x}\right)\right]F_{g/H}(z,\xi,\mu)\Bigg\}\,,
\end{array}
\end{equation}
\begin{equation}
\begin{array}{rcl}
\displaystyle   \frac{dF_{\overline{q}/H}(x,\xi,\mu)}{d\ln\mu^2} &=&\displaystyle \frac{\alpha_s(\mu)}{4\pi}\Bigg\{
  -\int_{0}^1\frac{dz}{z}\mathcal{P}^{[0]}_{q/q}\left(-\frac{x}{z},\frac{\xi}{x}\right)F_{q/H}(z,\xi,\mu)\\
\\
&+&\displaystyle \int_{0}^1\frac{dz}{z}\mathcal{P}^{[0]}_{q/q}\left(\frac{x}{z},\frac{\xi}{x}\right)F_{\overline{q}/H}(z,\xi,\mu)\\
\\
&+&\displaystyle 
  \int_{0}^1\frac{dz}{z}\left[\mathcal{P}^{[0]}_{q/g}\left(\frac{
    x}{z},\frac{\xi}{x}\right)-\mathcal{P}^{[0]}_{q/g}\left(-\frac{
    x}{z},\frac{\xi}{x}\right)\right]F_{g/H}(z,\xi,\mu) \Bigg\}\,,
\end{array}
\end{equation}
\begin{equation}
\begin{array}{rcl}
\displaystyle   \frac{dF_{g/H}( x,\xi,\mu)}{d\ln\mu^2} &=&\displaystyle 
 \frac{\alpha_s(\mu)}{4\pi}\Bigg\{\int_{0}^1\frac{dz}{z}\mathcal{P}^{[0]}_{g/q}\left(\frac{x}{z},\frac{\xi}{x}\right)F_{q/H}(z,\xi,\mu)\\
\\
&-&\displaystyle \int_{0}^1\frac{dz}{z}\mathcal{P}^{[0]}_{g/q}\left(-\frac{x}{z},\frac{\xi}{x}\right)F_{\overline{q}/H}(z,\xi,\mu)\\
\\
&+&\displaystyle 
  \int_{0}^1\frac{dz}{z}\left[\mathcal{P}^{[0]}_{g/g}\left(\frac{x}{z},\frac{\xi}{x}\right)-\mathcal{P}^{[0]}_{g/g}\left(-\frac{x}{z},\frac{\xi}{x}\right)\right]F_{g/H}(z,\xi,\mu) \Bigg\}\,,
\end{array}
\end{equation}
where we have used the following equality:
\begin{equation}
  \mathcal{P}_{i/k}^{[0]}\left(x,-\frac{\xi}{x}\right)=  \mathcal{P}_{i/k}^{[0]}\left(x,\frac{\xi}{x}\right)\,,
\end{equation}
that is a consequence of a general symmetry of GPDs and follows
immediately from Eq.~(\ref{eq:polepartGPDFs}). It is now possible to
define parton-in-hadron GPD combinations that maximally diagonalise
the matrix of one-loop anomalous dimensions
$\mathcal{P}_{i/k}^{[0]}$. More precisely, one defines the
total-valence non-singlet GPD as:
\begin{equation}\label{eq:totalvalence}
  F^- = \sum_{q=1}^{n_f}F_{q/H}-F_{\overline{q}/H}\,,
\end{equation}
and a bidimensional vector of GPDs made of the total-singlet and the
gluon GPDs, often collectively referred to as singlet:
\begin{equation}
  F^+=
\begin{pmatrix}
  \displaystyle \sum_{q=1}^{n_f}F_{q/H}+F_{\overline{q}/H}\\
  F_{g/H}
\end{pmatrix}\,.
\end{equation}
These combinations obey the following evolution equations:\footnote{We
  point out that having only one non-singlet evolution equation is the
  consequence of working at one-loop accuracy. As mentioned in
  Appendix~\ref{app:gpdnormlisation}, in massless QCD with more than
  one quark flavour, there are in general seven independent evolution
  kernels that can be arranged in a way that four of them are
  responsible for the evolution of the singlet and the remaining three
  for the evolution of three independent sets of non-singlet
  combinations. At one loop, all non-singlet combinations evolve
  through the same kernel $\mathcal{P}^{-,[0]}$ which allows us to
  consider only the total-valence distribution $F^-$ in
  Eq.~(\ref{eq:totalvalence}).}
\begin{equation}
\frac{dF^{\pm}(x,\xi,\mu)}{d\ln\mu^2} =\frac{\alpha_s(\mu)}{4\pi}
\int_x^\infty\frac{dy}{y}\mathcal{P}^{\pm,[0]}\left(y,\kappa\right)F^{\pm}\left(\frac{x}{y},\xi,\mu\right)\,,
\label{eq:regdglap}
\end{equation}
with $\kappa=\xi/x$. The evolution kernel of the non-singlet GPD is
given by:
\begin{equation}
  \mathcal{P}^{-,[0]}\left(y,\kappa\right)
  =\mathcal{P}_{q/q}^{[0]}\left(y,\kappa\right) + \mathcal{P}_{q/q}^{[0]}\left(-y,\kappa\right)\,,
\end{equation}
while that of the singlet is given by:
\begin{equation}
  \mathcal{P}^{+,[0]}\left(y,\kappa\right) =
\begin{pmatrix}
  \mathcal{P}_{q/q}^{[0]}\left(y,\kappa\right)-\mathcal{P}_{q/q}^{[0]}\left(-y,\kappa\right)
& 2n_f\left(\mathcal{P}_{q/g}^{[0]}\left(y,\kappa\right)-\mathcal{P}_{q/g}^{[0]}\left(-y,\kappa\right)\right)\\
\mathcal{P}_{g/q}^{[0]}\left(y,\kappa\right)-\mathcal{P}_{g/q}^{[0]}\left(-y,\kappa\right)
& \mathcal{P}_{g/g}^{[0]}\left(y,\kappa\right)-\mathcal{P}_{g/g}^{[0]}\left(-y,\kappa\right)
\end{pmatrix}\,.
\end{equation}
Using Eq.~(\ref{eq:polepartGPDFs}), it is easy to see that for $x>0$
and $\xi\geq 0$:
\begin{equation}
\mathcal{P}_{i/k}^{[0]}\left(-y,\kappa\right)=\theta\left(\kappa-1\right)p_{ik}\left(-y,-\kappa\right)\,,
\end{equation}
so that the splitting kernels can be recasted as:
\begin{equation}
\mathcal{P}^{\pm,[0]}\left(y,\kappa\right) = \theta(1-y)
\mathcal{P}_1^{\pm,[0]}\left(y,\kappa\right)+\theta(\kappa-1)
\mathcal{P}_2^{\pm,[0]}\left(y,\kappa\right)\,.
\label{eq:KernelDecomposition}
\end{equation}
For the non-singlet evolution kernel one finds:\footnote{Notice that,
  for the sake of compactness, in the definition of $\mathcal{P}_1$ in
  both Eqs.~(\ref{eq:P1andP2NS}) and~(\ref{eq:P1andP2SG}) we have
  factored out $\theta(1-y)$ also from the term proportional to
  $\delta(1-y)$ essentially assuming
  $\delta(1-y)=\theta(1-y)\delta(1-y)$. Of course, this is not
  strictly true and is simply meant to simplify the notation.}
\begin{equation}
\begin{array}{rcl}
  \displaystyle\mathcal{P}_1^{-,[0]}\left(y,\kappa\right)&=&\displaystyle
                                                                    p_{qq}\left(y,\kappa\right)+p_{qq}\left(y,-\kappa\right)\\
\\
&+&\displaystyle \delta(1-y) 2C_q\left[K_q-2\int_0^1 \frac{dz}{1-z}-\ln\left(\left|1-\kappa^2\right|\right)\right]\,,\\
\\
\mathcal{P}_2^{-,[0]}\left(y,\kappa\right)&=&\displaystyle
                                                     -p_{qq}\left(y,-\kappa\right)
                                                     +p_{qq}\left(-y,-\kappa\right)\,,
\label{eq:P1andP2NS}
\end{array}
\end{equation}
while for the single components of the matrix associated to the
singlet evolution:
\begin{equation}
\begin{array}{rcl}
\displaystyle\mathcal{P}_{1,ik}^{+,[0]}\left(y,\kappa\right)&=&\displaystyle p_{ik}\left(y,\kappa\right)+p_{ik}\left(y,-\kappa\right) \\
\\
&+&\displaystyle \delta_{ik}\delta(1-y) 2C_i\left[K_i-2\int_0^1 \frac{dz}{1-z}-\ln\left(\left|1-\kappa\right|\right)\right]\,,\\
\\
\displaystyle\mathcal{P}_{2,ik}^{+,[0]}\left(y,\kappa\right)&=&\displaystyle
                                                     -p_{ik}\left(y,-\kappa\right)
                                                     -p_{ik}\left(-y,-\kappa\right)\,.
\label{eq:P1andP2SG}
\end{array}
\end{equation}
The decomposition in Eq.~(\ref{eq:KernelDecomposition}) is
particularly convenient. The $\mathcal{P}_1$ terms, being proportional
to $\theta(1-y)$, reduce Eq.~(\ref{eq:regdglap}) to the exact same
form of a DGLAP evolution equation. As a matter of fact, we will show
below that in the limit $\xi\rightarrow 0$ the one-loop
$\mathcal{P}_1$ kernels exactly reduce to the one-loop DGLAP splitting
functions. The $\mathcal{P}_2$ terms instead come into play for
$\kappa>1$ ($x<\xi$) and thus represent the contribution to the
evolution due to the ERBL region. Of course, for $\xi\rightarrow 0$
these terms do not contribute leaving only the DGLAP kernels. A
graphical representation of the integration domain covered by
$\mathcal{P}_1$ and $\mathcal{P}_2$ is displayed in
Fig.~\ref{fig:GPDIntDomain}.

Using the $p_{ik}$ functions given in Eq.~(\ref{eq:explicitpik}), we
can obtain the explicit expressions for the $\mathcal{P}_{1,2}$
kernels. For the non-singlet sector they read:
\begin{equation}
\left\{\begin{array}{rcl}
\mathcal{P}_{1}^{-,[0]}(y,\kappa) &=& \displaystyle 2C_F\left\{\left(\frac{2}{1-y}\right)_+-\frac{1
  +y}{1-\kappa^2y^2}+\delta(1-y)\left[\frac32-
  \ln\left(|1-\kappa^2|\right)\right]\right\}\,,\\
\\
\mathcal{P}_{2}^{-,[0]}(y,\kappa) &=& \displaystyle 2C_F\left[\frac{1+(1+\kappa)y+(1+\kappa
      -\kappa^2)y^2}{(1+y)(1-\kappa^2y^2)}-\left(\frac{1}{1-y}\right)_{++}\right]\,,
\end{array}\right.
\label{eq:PnsGPD}
\end{equation}
while for the singlet sector we find:
\begin{equation}
\left\{\begin{array}{rcl}
\mathcal{P}_{1, qq}^{+,[0]}(y,\kappa) &=& \mathcal{P}_{1}^{-,[0]}(y,\kappa)\,,\\
\\
\mathcal{P}_{2, qq}^{+,[0]}(y,\kappa) &=& \displaystyle
                                           2C_F\left[\frac{1+y+\kappa y+\kappa^3y^2}{\kappa(1+y)(1-\kappa^2y^2)}-\left(\frac{1}{1-y}\right)_{++}\right]\,,
\end{array}\right.
\label{eq:PqqGPD}
\end{equation}

\begin{equation}
\left\{\begin{array}{rcl}
\mathcal{P}_{1, qg}^{+,[0]}(y,\kappa) &=& \displaystyle 4n_f T_R\left[\frac{y^2+(1-y)^2-\kappa^2y^2}{(1-\kappa^2y^2)^2}\right]\,,\\
\\
\mathcal{P}_{2, qg}^{+,[0]}(y,\kappa) &=& \displaystyle 4n_f T_R(1-\kappa)\left[\frac{1-\kappa(\kappa+2)y^2}{\kappa(1-\kappa^2y^2)^2}\right]\,,
\end{array}\right.
\label{eq:PqgGPD}
\end{equation}

\begin{equation}
\left\{\begin{array}{rcl}
  \mathcal{P}_{1, gq}^{+,[0]}(y,\kappa) &=& \displaystyle 2C_F\left[\frac{1+(1-y)^2-\kappa^2y^2}{y(1-\kappa^2y^2)}\right]\,,\\
  \\
  \mathcal{P}_{2, gq}^{+,[0]}(y,\kappa) &=& \displaystyle-2C_F\frac{(1-\kappa)^2}{\kappa(1-\kappa^2 y^2)}\,,
\end{array}\right.
\label{eq:PgqGPD}
\end{equation}

\begin{equation}
  \left\{\begin{array}{rcl}
           \mathcal{P}_{1, gg}^{+,[0]}(y,\kappa) &=& \displaystyle
                                                     4C_A\left[\left(\frac{1}{1-y}\right)_+-\frac{1+\kappa^2y}{1-\kappa^2y^2}+\frac{1}{(1-\kappa^2y^2)^2}\left(\frac{1-y}{y}+y(1-y)\right)\right]\\
           \\
                                                 &+&\displaystyle\delta(1-y)\left[\left(\frac{11C_A-4 n_f
                                                     T_R}{3}\right) -2C_A\ln(|1-\kappa^2|)\right]\,,\\
           \\
           \mathcal{P}_{2, gg}^{+,[0]}(y,\kappa) &=& \displaystyle
                                                     2C_A\left[\frac{2(1-\kappa)(1+y^2)}{(1-\kappa^2y^2)^2}+\frac{\kappa^2(1+y)}{1-\kappa^2y^2}+\frac{1-\kappa^2}{1-\kappa^2y^2}\left(2-\frac{1}{\kappa}-\frac{1}{1+y}\right)
                                                     -\left(\frac{1}{1-y}\right)_{++}\right]
                                                     \,.
\end{array}\right.
\label{eq:PggGPD}
\end{equation}
In the expressions above two kinds of distributions are present. The
first is the familiar $+$-distribution (with round brackets) that only
appears in the $\mathcal{P}_1$ terms (and thus in the DGLAP region)
and is defined upon integration with a test non-singular function $f$
as:
\begin{equation}
\int_x^1dy\left(\frac{1}{1-y}\right)_+f(y) =
\int_x^1dy\frac{f(y)-f(1)}{1-y} + f(1)\ln(1-x)\,.
\label{eq:plusprescr1}
\end{equation}
The $+$-prescription is a consequence of the cancellation of soft
divergences between real and virtual diagrams and emerges thanks to
the divergent integral in
Eq.~(\ref{eq:polepartGPDFs})~\cite{Curci:1980uw}. The second
distribution is the $++$-distribution that only appears in the
$\mathcal{P}_2$ terms. This distribution is meant to provide a
numerically amenable implementation of the Cauchy principal-value
distribution for integrals of the following kind:
\begin{equation}
I=\int_x^\infty dy\,\frac{f(y)}{1-y}\,.
\end{equation}
If one subtracts and adds back the divergence at $y=1$, \textit{i.e.}:
\begin{equation}
f(1)\int_0^1\frac{dy}{1-y}\,,
\end{equation}
one can rearrange the integral $I$ as follows:
\begin{equation}
I=\int_x^\infty\frac{dy}{1-y}\left[f(y)-f(1)\left(1+\theta(y-1)\frac{1-y}{y}\right)\right]+f(1)\ln(1-x)\equiv
\int_x^\infty dy\left(\frac{1}{1-y}\right)_{++}f(y)\,,
\label{eq:plusplusdist}
\end{equation}
which effectively defines the $++$-distribution. The advantage of this
rearrangement is that the integrand is free of the divergence at $y=1$
making the numerical computation easier. Interestingly, the
$++$-distribution reduces to the standard $+$-distribution when the
upper integration bound is one rather than infinity. In this sense,
the $++$-distribution generalises the $+$-distribution to integrals in
the ERBL region.

\section{Properties of the evolution kernels}\label{sec:properties}

In the previous section we have given a thorough derivation of the
leading-order evolution equations for unpolarised GPDs, providing
explicit expressions for the evolution kernels. In this section, we
analyse these kernels in detail highlighting some prominent
properties.

\subsection{The DGLAP limit}\label{sec:DGLAPlimit}

One of the most important requirements for the GPD evolution equations
is that they reduce to the DGLAP evolution
equations~\cite{Altarelli:1977zs,Dokshitzer:1977sg,Gribov:1972ri} in
the forward limit $\xi\rightarrow 0$. As already mentioned, the
decomposition in Eq.~(\ref{eq:KernelDecomposition}) nicely isolates
the DGLAP contribution to the evolution kernels into the
$\mathcal{P}_1$ functions making the ERBL contribution embedded in
$\mathcal{P}_2$ automatically drop out for $\xi\rightarrow
0$. Therefore, to ensure that our GPD evolution tends to the DGLAP, it
is enough to show that the forward limit of the $\mathcal{P}_1$
functions coincides with the one-loop DGLAP splitting functions. This
is easily done by taking the limit for $\kappa\rightarrow0$ of the
expressions given in Eqs.~(\ref{eq:PnsGPD})-(\ref{eq:PggGPD}):
\begin{equation}
\begin{array}{rcl}
\displaystyle \lim_{\kappa\rightarrow 0}\mathcal{P}_{1}^{-,[0]}(y,\kappa) &=& \displaystyle 2C_F\left[\frac{y^2+1}{(1-y)_+}+\frac32\delta(1-y)\right]\,,\\
\\
\displaystyle \lim_{\kappa\rightarrow 0}\mathcal{P}_{1, qq}^{+,[0]}(y,\kappa) &=& \displaystyle \lim_{\kappa\rightarrow 0}\mathcal{P}_{1}^{-,[0]}(y,\kappa)\,, \\
\\
\displaystyle \lim_{\kappa\rightarrow 0}\mathcal{P}_{1, qg}^{+,[0]}(y,\kappa) &=& \displaystyle 4n_f T_R\left[y^2+(1-y)^2\right]\,, \\
\\
\displaystyle \lim_{\kappa\rightarrow 0}  \mathcal{P}_{1, gq}^{+,[0]}(y,\kappa) &=& \displaystyle 2C_F\left[\frac{1+(1-y)^2}{y}\right]\,, \\
\\
\displaystyle \lim_{\kappa\rightarrow 0}  \mathcal{P}_{1, gg}^{+,[0]}(y,\kappa) &=& \displaystyle
                                                      4C_A\left[\frac{y}{(1-y)_+}+\frac{1-y}{y}+y(1-y)\right]\\
\\
&+&\displaystyle \delta(1-y)\left[\left(\frac{11C_A-4 n_f
                                                      T_R}{3}\right)\right]\,,
\label{eq:DGLAPlimit}
\end{array}
\end{equation}
that indeed are equal to the one-loop DGLAP splitting functions (see
\textit{e.g.} Ref.~\cite{Ellis:1996mzs}).

\subsection{The ERBL limit}

Sound GPD evolution equations also need to reproduce the ERBL
evolution equations~\cite{Lepage:1980fj,Efremov:1978rn} that govern
the evolution of DAs in the $\xi\rightarrow1$ limit. To prove that
this is the case, it is useful to rearrange Eq.~(\ref{eq:regdglap}) as
follows:
\begin{equation}
\frac{d}{d\ln\mu^2} F^{\pm}(x,\xi,\mu)=
\frac{\alpha_s(\mu)}{4\pi}\int_{-1}^1\frac{dy}{|\xi|} \mathbb{V}^{\pm,[0]}\left(\frac{x}{\xi},\frac{y}{\xi}\right) F^{\pm}(y,\xi,\mu)\,,
\label{eq:ERBLlikeeveq}
\end{equation}
with:
\begin{equation}
\begin{array}{rcl}
\displaystyle \frac{1}{|\xi|}\mathbb{V}_{ik}^{+,[0]}\left(\frac{x}{\xi}, \frac{y}{\xi}\right) &=& \displaystyle
\frac{1}{y}\bigg\{\left[\theta(x-\xi) \theta(y-x)-\theta(-x-\xi)
                                              \theta(x-y)\right] \left[p_{ik}\left(\frac{x}{y}, \frac{\xi}{x}\right)+p_{ik}\left(\frac{x}{y}, -\frac{\xi}{x}\right)\right]\\
\\
&+&\displaystyle \theta(\xi-x)\theta(x+\xi) \left[\theta(y-x) p_{ik}\left(\frac{x}{y}, \frac{\xi}{x}\right)-\theta(x-y) p_{ik}\left(\frac{x}{y}, -\frac{\xi}{x}\right)\right]\\
\\
&+&\displaystyle \delta\left(1-\frac{x}{y}\right)
    \delta_{ik}2C_i\left[K_i+\int_\xi^{x}\frac{dz}{z-x}+\int_{-\xi}^{x}\frac{dz}{z-x}\right]\bigg\}\,,\\
\\
\displaystyle \frac{1}{|\xi|}\mathbb{V}^{-,[0]}\left(\frac{x}{\xi}, \frac{y}{\xi}\right)&=&\displaystyle \frac{1}{|\xi|} \mathbb{V}_{qq}^{+,[0]}\left(\frac{x}{\xi}, \frac{y}{\xi}\right)\,.
\end{array}
\label{eq:evkernelERBLgenxi}
\end{equation}
For the moment we are allowing again $x$ to be negative. However, the
combinations $F^{\pm}$ have a definite behaviour upon sign change of
$x$, that is $F^{\pm}(-x,\xi,\mu)=\mp F^{\pm}(x,\xi,\mu)$ and
therefore the negative branch in $x$ is determined in terms of the
positive one. Before taking the limit, it is convenient to introduce
the variables $t$ and $u$ defined as:
\begin{equation}
t = \frac12(x+1)\,,\quad\mbox{and}\quad u=\frac12(y+1)\,,
\end{equation}
spanning the range $[0,1]$ and to write the ERBL evolution equation in
a more conventional form as:
\begin{equation}
  \frac{d}{d\ln\mu^2} \Phi^{\pm}(t,\mu)=
  \frac{\alpha_s(\mu)}{4\pi}\int_{0}^1du V^{\pm,[0]}(t,u) \Phi^{\pm}(u,\mu)\,,
\end{equation}
such that:
\begin{equation}
\Phi^{\pm}(t,\mu) = \lim_{\xi\rightarrow 1}F^{\pm}(2t-1,\xi,\mu)\,,
\end{equation}
and:
\begin{equation}
V^{\pm,[0]}(t,u) = \lim_{\xi\rightarrow1}\frac{1}{|\xi|}\mathbb{V}^{\pm,[0]}\left(\frac{2t-1}{\xi}, \frac{2u-1}{\xi}\right)\,.
\end{equation}
For the non-singlet anomalous dimension, we find:
\begin{equation}
  V^{-,[0]}(t,u)=
  C_F\left\{\left[\frac{\theta(u-t)}{u-t}\right]_++\theta(u-t)\frac{t-1}{u}-\left[\frac{\theta(t-u)}{u-t}\right]_+-\theta(t-u)\frac{t}{1-u} + \frac{3}{2}\delta\left(u-t\right) \right\}\,,
\end{equation}
which reproduces the results of Refs.~\cite{Lepage:1980fj,
  Blumlein:1999sc} where the $+$-prescription (with square brackets)
here has to be interpreted as:
\begin{equation}
  [f(t,u)]_+=f(t,u) - \delta(u-t)\int_0^1dt\, f(t,u)\,,
  \label{eq:plusprescr2}
\end{equation}
which generalises the definition in Eq.~(\ref{eq:plusprescr1}) to a
two-variable function with support $t,u\in[0,1]$ with a single pole at
$t=u$. One can also check that the integral of $V^{-,[0]}$ over $t$
vanishes:
\begin{equation}
  \int_0^1dt\,V^{-,[0]}(t,u) = 0\,,
\label{eq:nonzeroplusERBL}
\end{equation}
that allows us to write it in a fully $+$-prescribed form as:
\begin{equation}
  V^{-,[0]}(t,u) =
  C_F\left[\theta(u-t)\left(\frac{t-1}{u}+\frac{1}{u-t}\right)-\theta(t-u)\left(\frac{t}{1-u}+\frac{1}{u-t}\right)\right]_+\,.
\end{equation}
This property was also explicitly derived in
Ref.~\cite{Mikhailov:1984ii} and it was argued that it must hold for
symmetry reasons. For the singlet sector instead we find:
\begin{equation}
\begin{array}{rcl}
  V_{qq}^{+,[0]}(t,u)&=&V^{-,[0]}(t,u)\,,\\
\\
  V_{qg}^{+,[0]}(t,u)&=&\displaystyle T_R\frac{2u-1}{2u(u-1)}\left[\theta(u-t)\frac{t(u-2t+1)}{u}+\theta(t-u)\frac{(t-1)(u-2t)}{u-1}\right]\,,\\
\\

  V_{gq}^{+,[0]}(t,u)&=&\displaystyle \frac{2C_F}{2t-1}\left[\theta(u-t)\frac{t(2u-t)}{u}+\theta(t-u)\frac{(t-1)(2t-t-1)}{u-1}\right]\,,\\
\\
  V_{gg}^{+,[0]}(t,u)&=&\displaystyle
                         C_A\frac{t(t-1)(2u-1)}{u(u-1)(2t-1)}\Bigg[\frac{1}{u-t}-\theta(u-t)\left(1-\frac{2t^2+2u^2-2t-2u+1}{2u(t-1)(2t-1)^2}\right)\\
\\
&+&\displaystyle \theta(t-u)\left(1-\frac{2t^2+2u^2-2t-2u+1}{2t(u-1)(2t-1)^2}\right)\Bigg]\,.
\end{array}
\end{equation}
We could not find any reference reporting the explicit ERBL singlet
kernels to compare our results to.

\subsection{Spurious divergences and continuity of GPDs at $x=\xi$}\label{subsec:continuity}

All the expressions for the GPD evolution kernels given in
Eqs.~(\ref{eq:PnsGPD})-(\ref{eq:PggGPD}) are affected by a
non-integrable singularity at $y=\kappa^{-1}$ denoted by the
dot-dashed line in Fig.~\ref{fig:GPDIntDomain}. These singularities
may potentially spoil the convergence of the integral in the r.h.s. of
Eq.~(\ref{eq:regdglap}) but fortunately they cancel between the
$\mathcal{P}_1$ and $\mathcal{P}_2$ contributions to the evolution
kernels. As a matter of fact, they appear in the region $\kappa>1$ in
which both $\mathcal{P}_1$ and $\mathcal{P}_2$ contribute. In
addition, for each single kernel the coefficient of the divergence of
$\mathcal{P}_1$ is equal in absolute value but opposite in sign
w.r.t. that of $\mathcal{P}_2$ so that they finally cancel out
yielding a convergent integral. The value of the coefficient of the
divergences can be explicitly computed by taking the appropriate
limits.  For the non-singlet kernels, $\mathcal{P}_1^{-,[0]}$ and
$\mathcal{P}_2^{-,[0]}$, we find:
\begin{equation}
\lim_{y\rightarrow \kappa^{-1}} (1-\kappa^2y^2)
\mathcal{P}_1^{-,[0]}(y,\kappa) = - \lim_{y\rightarrow \kappa^{-1}} (1-\kappa^2y^2)
\mathcal{P}_2^{-,[0]}(y,\kappa) = -2 C_F \frac{1+\kappa}{\kappa}\,,
\end{equation}
while for the singlet kernels we find:
\begin{equation}
\begin{array}{l}
\displaystyle \lim_{y\rightarrow \kappa^{-1}} (1-\kappa^2y^2)
\mathcal{P}_{1, qq}^{+,[0]}(y,\kappa) = - \lim_{y\rightarrow \kappa^{-1}} (1-\kappa^2y^2)
\mathcal{P}_{2, qq}^{+,[0]}(y,\kappa) =
  -2C_F\frac{1+\kappa}{\kappa}\,,\\
\\
\displaystyle \lim_{y\rightarrow \kappa^{-1}} (1-\kappa^2y^2)^2
\mathcal{P}_{1, qg}^{+,[0]}(y,\kappa) = - \lim_{y\rightarrow \kappa^{-1}} (1-\kappa^2y^2)^2
\mathcal{P}_{2, qg}^{+,[0]}(y,\kappa) = \frac{8 n_f
  T_R(1-\kappa)}{\kappa}\,,\\
\\
\displaystyle \lim_{y\rightarrow \kappa^{-1}} (1-\kappa^2y^2)
\mathcal{P}_{1, gq}^{+,[0]}(y,\kappa) = - \lim_{y\rightarrow \kappa^{-1}} (1-\kappa^2y^2)
\mathcal{P}_{2, gq}^{+,[0]}(y,\kappa) =
  2C_F\frac{(1-\kappa)^2}{\kappa}\,,\\
\\
\displaystyle \lim_{y\rightarrow \kappa^{-1}} (1-\kappa^2y^2)^2
\mathcal{P}_{1, gg}^{+,[0]}(y,\kappa) = - \lim_{y\rightarrow \kappa^{-1}} (1-\kappa^2y^2)^2
\mathcal{P}_{2, gg}^{+,[0]}(y,\kappa) =
  4C_A\frac{(\kappa-1)(\kappa^2+1)}{\kappa^2}\,.
\end{array}
\end{equation}
Importantly, all the coefficients above are finite at $\kappa=1$,
\textit{i.e.} at the cross-over point $x=\xi$ between DGLAP and ERBL
regions. This is a requisite to ensure that GPDs remain finite at the
cross-over point upon evolution. The continuity of GPDs at the
crossing point is finally ensured by the following additional
property:\footnote{Despite not directly visible from
  Eqs.~(\ref{eq:PnsGPD}), (\ref{eq:PqqGPD}), and (\ref{eq:PggGPD}),
  $\mathcal{P}_{2}^{-,[0]}$, $\mathcal{P}_{2,qq}^{+,[0]}$, and
  $\mathcal{P}_{2,gg}^{+,[0]}$ also enjoy the property of
  Eq.~(\ref{eq:continuity}).}
\begin{equation}
\mathcal{P}_2^{\pm,[0]}(y,k)\propto (1-\kappa)\,.
\label{eq:continuity}
\end{equation}
Since the $\mathcal{P}_2$ functions multiply $\theta(\kappa-1)$,
Eq.~(\ref{eq:continuity}) guarantees a continuous transition from the
DGLAP region ($\kappa<1$) into the ERBL region ($\kappa>1$). However,
this property does not guarantee that GPDs remain smooth at the
cross-over point upon evolution. In fact, the very presence of the
term proportional to $\theta(\kappa-1)$ in
Eq.~(\ref{eq:KernelDecomposition}) makes the derivative w.r.t. to $x$
of the kernels discontinuous at $x = \xi$. Therefore, the evolution is
expected to generate a cusp at $x=\xi$.

\subsection{Sum rules}\label{subsec:sumrules}

A very important aspect of GPDs is that their first two Mellin moments
can be connected to physical quantities. In order to exemplify the
discussion, let us first consider the forward limit of GPDs,
\textit{i.e.} PDFs. It is well-known that PDFs must obey the so-called
valence (or counting) and momentum sum rules. The valence sum rule
ensures the conservation of the flavour quantum numbers and reads:
\begin{equation}
  \int_0^1dx \left[f_{q/H}(x,\mu)-f_{\overline{q}/H}(x,\mu)\right] =
  \int_0^1dx f^-(x,\mu) = c_q\,,
\label{eq:valencesumrule}
\end{equation}
where the $c_q$'s are constants depending on the valence structure of
the hadron $H$ (for example for the proton $c_u=2$, $c_d=1$, and
$c_q=0$ for all other flavours). The momentum sum rule guarantees that
the total momentum carried by all partons equals the momentum of the
parent hadron and reads:
\begin{equation}
  \int_0^1dx\,x\left[\sum_{q}(f_{q/H}(x,\mu)+f_{\overline{q}/H}(x,\mu))+f_{g/H}(x,\mu)\right]
  =   \int_0^1dx\,x\left[f^{+}(x,\mu)+f_{g/H}(x,\mu)\right] = 1\,.
\label{eq:momentumsumrule}
\end{equation}
The fact that both Eqs.~(\ref{eq:valencesumrule})
and~(\ref{eq:momentumsumrule}) are independent of the factorisation
scale $\mu$ implies a set of constraints on the DGLAP splitting
functions. Specifically, denoting with $P^{\pm,[n]}$ the $(n+1)$-loop
contribution to the singlet and non-singlet splitting functions, the
valence sum rule implies:
\begin{equation}
\int_0^1dx P^{-,[n]}(x) = 0\,,
\label{eq:valsumruleP}
\end{equation}
while the momentum sum rule implies:
\begin{equation}
\begin{array}{l}
\displaystyle \int_0^1dx \,x\left[P_{qq}^{+,[n]}(x)+P_{gq}^{+,[n]}(x)\right] =
  0\,,\\
\\
\displaystyle \int_0^1dx \,x\left[P_{qg}^{+,[n]}(x)+P_{gg}^{+,[n]}(x)\right] =
  0\,,
\end{array}
\label{eq:momsumruleP}
\end{equation}
that must hold for any $n$.

It turns out that the GPD evolution kernels must also fulfil similar
relations that generalise those for the DGLAP splitting functions. The
generalisation of the valence sum rule descends from the fact that the
integral of a non-singlet GPD is:
\begin{equation}
  \int_0^1dx F^-(x,\xi,\mu) = G\,,
\label{eq:valencesumruleGPD}
\end{equation}
where $G$ is an observable (Dirac or Pauli) elastic form factor that
cannot depend on $\mu$.\footnote{In fact, due to polynomiality, $G$
  does not depend on $\xi$ either but it can depend on $\Delta^2$.}
Therefore, one can follow the same reasoning applied to the DGLAP
splitting function to obtain the following order-by-order constraint
on the non-singlet GPD evolution kernels:
\begin{equation}
  \int_0^1dz\,\mathcal{P}_1^{-,[n]}\left(z,\frac{\xi}{yz}\right)+
  \int_0^{\xi/y} dz\,
  \mathcal{P}_2^{-,[n]}\left(z,\frac{\xi}{yz}\right)=0\,.
\label{eq:valsumruleGPDkernel}
\end{equation}
Notice that for $\xi\rightarrow 0$ the equality above reduces to
Eq.~(\ref{eq:valsumruleP}). It is interesting to verify
Eq.~(\ref{eq:valsumruleGPDkernel}) plugging in the explicit one-loop
expressions for $\mathcal{P}_1^{-,[0]}$ and $\mathcal{P}_2^{-,[0]}$
given in Eq.~(\ref{eq:PnsGPD}). One finds that:
\begin{equation}
  \int_0^1dz\,\mathcal{P}_1^{-,[0]}\left(z,\frac{\xi}{yz}\right)=-2C_F\left[\frac{3}{2}\frac{\xi^2}{\xi^2-y^2}+\ln\left(\left|1-\frac{\xi^2}{y^2}\right|\right)\right]\,,
\end{equation}
that correctly tends to zero as $\xi\rightarrow 0$, and:
\begin{equation}
  \int_0^{\xi/y} dz\, \mathcal{P}_2^{-,[0]}\left(z,\frac{\xi}{yz}\right)=2C_F\left[\frac{3}{2}\frac{\xi^2}{\xi^2-y^2}+\ln\left(\left|1-\frac{\xi^2}{y^2}\right|\right)\right]\,,
\end{equation}
such that Eq.~(\ref{eq:valsumruleGPDkernel}) is indeed fulfilled.

Now we move to considering the generalisation of the momentum sum
rule. To do so, we use the property of polynomiality of GPDs, given in
Eq.~(\ref{Eq:polynomiality}) below, to write:\footnote{Also in this
  case, the coefficients $A_{q(g)}^{F}$ and $D_{q(g)}^{F}$ generally
  depend on $\Delta^2$.}
\begin{equation}
  \int_0^1dx\,xF_{q(g)}^+(x,\xi,\mu) = A_{q(g)}^{F}(\mu)+\xi^2 D_{q(g)}^{F} (\mu)\,.
\end{equation}
However, it is well-known that unpolarised helicity-conserving ($H$)
and helicity-flip ($E$) GPDs have the same $D$-term but with opposite
sign~\cite{Diehl:2003ny}, \textit{i.e.}
$D_{q(g)}^{H} (\mu)= -D_{q(g)}^{E}$.  Therefore, if we assume for the
moment $F=H+E$, the $\xi$-dependent term cancels out. In addition,
Ji's sum rule~\cite{Ji:1996ek} ensures that the sum
$A_{q}^{F}+A_{g}^{F}$ has to be independent of the factorisation scale
because related to the physically-observable total angular momentum of
the hadron. Therefore, one finally has:
\begin{equation}
  \int_0^1dx\,x\left[F_{q}^+(x,\xi,\mu)+F_{g}^+(x,\xi,\mu)\right] = \mbox{constant}\,.
\end{equation}
Since $H$ and $E$ obey the same evolution equations so does their
sum. This allows us to take the derivative with respect to $\ln\mu^2$
of both sides of the equation above and use the evolution equations to
obtain the following order-by-order constraints on the GPD evolution
kernels:
\begin{equation}
\begin{array}{rcl}
  \displaystyle 
  \int_0^1dz\,z\left[\mathcal{P}_{1, qq}^{+,[n]}\left(z,\frac{\xi}{yz}\right)+\mathcal{P}_{1, gq}^{+,[n]}\left(z,\frac{\xi}{yz}\right)\right]+
  \int_0^{\xi/y} dz\, z\left[\mathcal{P}_{2, qq}^{+,[n]}\left(z,\frac{\xi}{yz}\right)+\mathcal{P}_{2, gq}^{+,[n]}\left(z,\frac{\xi}{yz}\right)\right]&=&0\,,\\
  \\
  \displaystyle \int_0^1dz\,z\left[\mathcal{P}_{1, qg}^{+,[n]}\left(z,\frac{\xi}{yz}\right)+\mathcal{P}_{1, gg}^{+,[n]}\left(z,\frac{\xi}{yz}\right)\right]+
  \int_0^{\xi/y} dz\, z\left[\mathcal{P}_{2, qg}^{+,[n]}\left(z,\frac{\xi}{yz}\right)+\mathcal{P}_{2, gg}^{+,[n]}\left(z,\frac{\xi}{yz}\right)\right]&=&0\,.
\end{array}
\label{eq:MSRoffforward}
\end{equation}
As in the case of the valence sum rule, these relations reduce to
Eq.~(\ref{eq:momsumruleP}) in the forward limit $\xi\rightarrow 0$. We
now verify that the one-loop splitting functions in
Eqs.~(\ref{eq:PqqGPD})-(\ref{eq:PggGPD}) do fulfil the equalities in
Eq.~(\ref{eq:MSRoffforward}). The explicit computation of the
integrals gives:
\begin{equation}
\begin{array}{rcl}
  \displaystyle\int_0^1dz\,z\left[\mathcal{P}^{+,[0]}_{1,
  qq}\left(z,\frac{\xi}{yz}\right)+\mathcal{P}^{+,[0]}_{1,
  gq}\left(z,\frac{\xi}{yz}\right)\right] &=&\displaystyle 
                                              -2C_F\left[\frac12\frac{\xi^2}{y^2-\xi^2}+\ln\left(\left|1-\frac{\xi^2}{y^2}\right|\right)\right]\,,\\
  \\
  \displaystyle\int_0^{\xi/y} dz\, z\left[\mathcal{P}^{+,[0]}_{2,
  qq}\left(z,\frac{\xi}{yz}\right)+\mathcal{P}^{+,[0]}_{2,
  gq}\left(z,\frac{\xi}{yz}\right)\right]&=&\displaystyle
                                             2C_F\left[\frac12\frac{\xi^2}{y^2-\xi^2}+\ln\left(\left|1-\frac{\xi^2}{y^2}\right|\right)\right]\,,
\end{array}
\end{equation}
and:
\begin{equation}
\begin{array}{rcl}
  \displaystyle\int_0^1dz\,z\left[\mathcal{P}^{+,[0]}_{1,
  qg}\left(z,\frac{\xi}{yz}\right)+\mathcal{P}^{+,[0]}_{1,
  gg}\left(z,\frac{\xi}{yz}\right)\right] &=&\displaystyle
                                              \frac{y^2\xi^2}{3(y^2-\xi^2)^2}\left[C_A\left(\frac{11\xi^2}{y^2}-4\right)+2n_fT_R\left(1-\frac{2\xi^2}{y^2}\right)\right]\\
  \\
                                          &-&\displaystyle
                                              2C_A\ln\left(\left|1-\frac{\xi^2}{y^2}\right|\right)\,,\\
  \\
  \displaystyle \int_0^{\xi/y} dz\, z\left[\mathcal{P}^{+,[0]}_{2,
  qg}\left(z,\frac{\xi}{yz}\right)+\mathcal{P}^{+,[0]}_{2,
  gg}\left(z,\frac{\xi}{yz}\right)\right]&=& \displaystyle
                                             -\frac{y^2\xi^2}{3(y^2-\xi^2)^2}\left[C_A\left(\frac{11\xi^2}{y^2}-4\right)+2n_fT_R\left(1-\frac{2\xi^2}{y^2}\right)\right]\\
  \\
                                          &+&\displaystyle 2C_A\ln\left(\left|1-\frac{\xi^2}{y^2}\right|\right)\,,
\end{array}
\end{equation}
that evidently cancel pairwise so that the equalities in
Eq.~(\ref{eq:MSRoffforward}) are satisfied. In addition, they all tend
to zero as $\xi\rightarrow 0$ as required by
Eq.~(\ref{eq:momsumruleP}).

We finally point out that the constraints in
Eqs.~(\ref{eq:valsumruleGPDkernel}) and~(\ref{eq:MSRoffforward}) can
be used to simplify the perturbative calculation of the evolution
kernels in that they allow one to determine the contribution due to
virtual diagrams by knowing the real ones. To be more specific,
virtual diagrams give rise to contributions proportional to
$\delta(1-y)$ that are naturally associated to $\mathcal{P}_1$ such
that order by order in $\alpha_s$ it can be decomposed as:
\begin{equation}
\mathcal{P}_1^{\pm,[n]}(y,\kappa)=\mathcal{P}_1^{{\rm real},\pm,[n]}(y,\kappa)-\delta(1-y) \mathcal{P}_1^{{\rm virtual},\pm,[n]}(\kappa)\,.
\end{equation}
Conversely, $\mathcal{P}_2$ only contains real-diagram contributions:
\begin{equation}
\mathcal{P}_2^{\pm,[n]}(y,\kappa)=\mathcal{P}_2^{{\rm real},\pm,[n]}(y,\kappa)\,.
\end{equation}
Taking as an example Eq.~(\ref{eq:valsumruleGPDkernel}), the
consequence of this decomposition is that:
\begin{equation}
  \int_0^1dz\,\mathcal{P}_1^{{\rm real},-,[n]}\left(z,\frac{\xi}{yz}\right)+
  \int_0^{\xi/y} dz\,
  \mathcal{P}_2^{{\rm real},-,[n]}\left(z,\frac{\xi}{yz}\right)=\mathcal{P}_1^{{\rm virtual},-,[n]}(\kappa)\,,
\end{equation}
making it unnecessary to explicitly compute the virtual-diagram
contributions. Of course, also the two equalities in
Eq.~(\ref{eq:MSRoffforward}) have to be simultaneously
fulfilled. Since by construction only $\mathcal{P}^-$ and the diagonal
terms of $\mathcal{P}^+$, \textit{i.e.} $\mathcal{P}_{qq}^+$ and
$\mathcal{P}_{gg}^+$, can get virtual corrections with the additional
constraint
$\mathcal{P}^{{\rm virtual},-,[n]}=\mathcal{P}_{qq}^{{\rm
    virtual},+,[n]}$, at each order in perturbation theory there are
two virtual contributions to be determined. On the other hand,
Eqs.~(\ref{eq:valsumruleGPDkernel}) and~(\ref{eq:MSRoffforward})
provide us with a set of three constraints. As consequence, these
equalities, not only give us access to the virtual corrections, but
also provide a strong check of the calculation of the real
contributions. However, we point out that we have explicitly computed
the one-loop virtual diagrams, verifying that the resulting
contribution agrees with the calculation obtained by means of the sum
rules. In Appendix~\ref{app:splittingfunctions}, we present this check
for the case of $\mathcal{P}_{q/q}^{[0]}$.

\subsection{Conformal moments}\label{sec:conformalmoments}

In this section, we consider the so-called conformal moments of GPDs
which in the non-singlet case are defined as~\cite{Diehl:2003ny}:
\begin{equation}
  \mathcal{C}_n^-(\xi,\mu) =
  \xi^{n}\int_{-1}^{1}dx\,C_n^{(3/2)}\left(\frac{x}{\xi}\right)F^{-}(x,\xi,\mu)\,,
\label{eq:conformalmoments}
\end{equation}
where $C_n^{(3/2)}$ are Gegenbauer polynomials of rank $3/2$ and
degree $n$ (with $n$ even). The choice of these specific moments (and
thus the underlying local conformal operators) comes from the fact
that they do not mix under renormalisation at one-loop
\cite{Ohrndorf:1981qv}. First we highlight the consequences of this
property and then sketch a way to prove it.

Multiplying Eq.~(\ref{eq:ERBLlikeeveq}) by
$\xi^{n}C_n^{(3/2)}\left(x/\xi\right)$ and integrating over $x$
between $-1$ and $1$ yields:
\begin{equation}
\frac{d \mathcal{C}_n^-(\xi, \mu)}{d\ln\mu^2} =
\frac{\alpha_s(\mu)}{4\pi}\xi^{n}\int_{-1}^1 dy
F^{-}(y,\xi,\mu)\int_{-1}^{1}\frac{dx}{|\xi|}\,C_n^{(3/2)}\left(\frac{x}{\xi}\right)
\mathbb{V}^{-,[0]}\left(\frac{x}{\xi},\frac{y}{\xi}\right) \,.
\label{eq:confmomprel}
\end{equation}
In the absence of mixing, the conformal moments of the non-singlet GPD
obey the following equality:
\begin{equation}
  \int_{-1}^1\frac{dx}{\xi}\,C_n^{(3/2)}\left(\frac{x}{\xi}\right)
  \mathbb{V}^{-,[0]}\left(\frac{x}{\xi},\frac{y}{\xi}\right)
  =\mathcal{V}_{n}^{-,[0]}(\xi)C_n^{(3/2)}\left(\frac{y}{\xi}\right)\,,
\label{eq:confmomgpd}
\end{equation}
where the anomalous dimension of the associated local conformal
operator is labelled by $\mathcal{V}_{n}^{-,[0]}$.  Looking at
Eq. \eqref{eq:confmomgpd}, one may think that
$\mathcal{V}_{n}^{-,[0]}$ generally depends on $\xi$.  However, in the
$\overline{\mbox{MS}}$ scheme, anomalous dimensions of local operators
are fixed independently of incoming or outgoing states.  Thus, as we
will see, one should expect $\mathcal{V}_{n}^{-,[0]}$ to be
$\xi$-independent. If Eq.~(\ref{eq:confmomgpd}) held true,
Eq.~(\ref{eq:confmomprel}) would then become:
\begin{equation}
  \frac{d \mathcal{C}_n^-(\xi,\mu)}{d\ln\mu^2}
  =\frac{\alpha_s(\mu)}{4\pi}\mathcal{V}_{n}^{-,[0]}(\xi)\mathcal{C}_n^-(\xi,\mu)\,,
\label{eq:confmom}
\end{equation}
making explicit the fact that GPD conformal moments evolve
multiplicatively.  An interesting indication that this is true and
also that the anomalous dimension $\mathcal{V}_{n}^{-,[0]}$ does not
depend on $\xi$ comes from considering the DGLAP ($\xi\rightarrow0$)
and the ERBL ($\xi\rightarrow1$) limits of Eq.~(\ref{eq:confmom}).

Let us start with the DGLAP limit. For $\xi\rightarrow 0$, conformal
moments coincide with Mellin moments up to a multiplicative numerical
factor. This can be seen by observing that Gegenbauer polynomials are
such that:
\begin{equation}
\lim_{\xi\rightarrow 0}
\xi^{n}C_n^{(3/2)}\left(\frac{x}{\xi}\right)=
\frac{(2n+1)!}{2^n(n!)^2}x^{n}\,.
\label{eq:gegenlimit}
\end{equation}
Therefore, the conformal moments of the non-singlet distribution in
the forward limit become:
\begin{equation}
\lim_{\xi\rightarrow 0}\mathcal{C}_n^-(\xi,\mu) = \frac{(2n+1)!}{2^n(n!)^2}[1+(-1)^n]f_{n+1}^{-}(\mu)\,,
\end{equation}
where Mellin moments of the forward distribution (PDF) are defined as:
\begin{equation}
  f_n^{-}(\mu) = \lim_{\xi\rightarrow 0} \int_0^1dx\,x^{n-1}F^{-}(x,\xi,\mu)\,,
\end{equation}
and are known to diagonalise the DGLAP equation to all orders. Using
Eq.~(\ref{eq:gegenlimit}) and the fact that:
\begin{equation}
\lim_{\xi\rightarrow 0} \frac{1}{|\xi|}\mathbb{V}^{-,[0]}\left(\frac{x}{\xi},
  \frac{y}{\xi}\right) = \left[\theta(x)\theta(y-x)-\theta(-x)\theta(x-y)\right]\frac{1}{|y|} P^{-,[0]}\left(\frac{x}{y}\right)\,,
\end{equation}
that derives from Eq.~(\ref{eq:evkernelERBLgenxi}), one finally
finds~\cite{Ellis:1996mzs}:
\begin{equation}
  \lim_{\xi\rightarrow 0} \mathcal{V}_{n}^{-,[0]}(\xi) = P_{n+1}^{-,[0]}=
  2C_F\left[\frac32+\frac{1}{(n+1)(n+2)}-2\sum_{k=1}^{n+1}\frac{1}{k}\right]\,.
\label{eq:forwardlimitconfmom}
\end{equation}

In the ERBL limit, the conformal moments yield this time~\cite{Lepage:1980fj}:
\begin{equation}
  \int_{-1}^{1}dx\,C_n^{(3/2)}\left(x\right) V_{\rm
    NS}^{-,[0]}\left(x,y\right) =2C_F\left[\frac32+\frac{1}{(n+1)(n+2)}-2\sum_{k=1}^{n+1}\frac{1}{k}\right] C_n^{(3/2)}\left(y\right)\,.
\label{eq:confmomerbl}
\end{equation}
Comparing Eq.~(\ref{eq:confmomerbl}) to
Eq.~(\ref{eq:forwardlimitconfmom}) one immediately sees that conformal
moment do not mix neither in DGLAP nor in ERBL limits and that
$\mathcal{V}_{n}$ is the same in both cases.  In order to explicitly
prove the general case, we need to compute Eq.~(\ref{eq:confmomgpd})
for a generic value of $\xi$ and for all $n$. To do so, we use the
decomposition in Eq.~(\ref{eq:evkernelERBLgenxi}) with the explicit
form of the $p_{qq}$ function given in Eq.~(\ref{eq:explicitpik}) that
yields:
\begin{equation}
\begin{array}{rcl}
  \displaystyle
  \int_{-1}^1\frac{dx}{|\xi|}\,C_n^{(3/2)}\left(\frac{x}{\xi}\right)
  \mathbb{V}^{-,[0]}\left(\frac{x}{\xi},\frac{y}{\xi}\right)&=&\displaystyle 2C_F \Bigg\{\frac{3}{2}C_n^{(3/2)}\left(\frac{y}{\xi}\right) \\
\\
&-&\displaystyle \frac12\int_{\xi}^{y}dx\left[\frac{x+\xi}{\xi(y-\xi)}C_n^{(3/2)}\left(\frac{x}{\xi}\right) -2\frac{C_n^{(3/2)}\left(x/\xi\right)-C_n^{(3/2)}\left(y/\xi\right)}{y-x}\right]\\
\\
&+&\displaystyle \frac12\int_{-\xi}^{y}dx\left[\frac{x-\xi}{\xi(y+\xi)}C_n^{(3/2)}\left(\frac{x}{\xi}\right)+2\frac{C_n^{(3/2)}\left(x/\xi\right)-C_n^{(3/2)}\left(y/\xi\right)}{y-x}\right]\Bigg\}\,.
\end{array}
\label{eq:GPDevkernelalaERBL}
\end{equation}
The explicit calculation is presented in
Appendix~\ref{app:conformalmoments} and indeed confirms that:
\begin{equation}
  \int_{-1}^1\frac{dx}{|\xi|}\,C_n^{(3/2)}\left(\frac{x}{\xi}\right)
  \mathbb{V}^{-,[0]}\left(\frac{x}{\xi},\frac{y}{\xi}\right)=2C_F\left[\frac32+\frac{1}{(n+1)(n+2)}-2\sum_{k=1}^{n+1}\frac{1}{k}\right]
  C_n^{(3/2)}\left(\frac{y}{\xi}\right)\,.
  \label{eq:ConfMomMaster}
\end{equation}
A similar calculation for the singlet sector can be achieved in a
similar fashion.

\subsection{Comparison to other calculations}

In this section, we compare our calculation to previous results for
the one-loop GPD unpolarised evolution kernels. We will show that our
calculation agrees with those already present in the literature. For
definiteness, we will concentrate on the non-singlet evolution kernel
$\mathcal{P}^{^-,[0]}$ but we have checked that agreement is found
also for the other one-loop evolution kernels.

We start with the computation by Ji presented in
Ref.~\cite{Ji:1996nm}. In that paper, the GPD evolution equations are
written in the DGLAP and in the ERBL regions separately. To find the
correspondence we use the evolution equation in the form given in
Eq.~(\ref{eq:ERBLlikeeveq}). In the DGLAP region ($x>\xi$) the
evolution kernel reduces to:
\begin{equation}
\begin{array}{rcl}
  \displaystyle \frac{1}{|\xi|}\mathbb{V}^{-,[0]}\left(\frac{x}{\xi}, \frac{1}{\xi}\right) &=& \displaystyle\theta(1-x) \left[p_{qq}\left(x, \frac{\xi}{x}\right)+p_{qq}\left(x, -\frac{\xi}{x}\right)\right]\\
  \\
                                                                                           &+&\displaystyle \delta\left(1-x\right)
                                                                                               2C_F\left[\frac32+\int_\xi^{x}\frac{dz}{z-x}+\int_{-\xi}^{x}\frac{dz}{z-x}\right]\\
  \\
                                                                                           &=&\displaystyle \theta(1-x)2C_F \frac{1+x^2-2\xi^2}{(1-x)(1-\xi^2)}\\
  \\
                                                                                           &+&\displaystyle \delta(1-x) 2C_F \left[ \frac{3}{2}+ \int_{\xi}^x\frac{dz}{z-x}+\int_{-\xi}^x\frac{dz}{z-x}\right]\,,
\end{array}
\end{equation}
Considering the shift $\xi\rightarrow\xi/2$ due to a different
definition of the external momenta and an overall factor of two due to
the fact that we are using $\alpha_s/(4\pi)$ rather than
$\alpha_s/(2\pi)$ as an expansion parameter, we exactly reproduce the
results given in Eqs.~(15)-(17) of Ref.~\cite{Ji:1996nm}. In the ERBL
region ($x<\xi$) the evolution kernel reads:
\begin{equation}
\begin{array}{rcl}
\displaystyle \frac{1}{|\xi|}\mathbb{V}^{-,[0]}\left(\frac{x}{\xi}, \frac{1}{\xi}\right) &=& \displaystyle \left[\theta(1-x) p_{qq}\left(x, \frac{\xi}{x}\right)-\theta(x-1) p_{qq}\left(x, -\frac{\xi}{x}\right)\right]\\
\\
&+&\displaystyle \delta\left(1-x\right)
    2C_F\left[\frac32+\int_\xi^{x}\frac{dz}{z-x}+\int_{-\xi}^{x}\frac{dz}{z-x}\right]\\
\\
&=&\displaystyle C_F \left[\theta(1-x) \frac{x+\xi}{\xi (1+\xi)}\left(1+\frac{2\xi}{1-x}\right)-\theta(x-1) \frac{x-\xi}{\xi (1-\xi)}\left(1-\frac{2\xi}{1-x}\right) \right]\\
\\
&+&\displaystyle \delta\left(1-x\right)
    2C_F\left[\frac32+\int_\xi^{x}\frac{dz}{z-x}+\int_{-\xi}^{x}\frac{dz}{z-x}\right]\,,
\end{array}
\end{equation}
that agrees with Eqs.~(18)-(19) of Ref.~\cite{Ji:1996nm}.

We now compare our calculation to that of Ref.~\cite{Muller:1994ses}
that is also reported in Eq.~(101) of Ref.~\cite{Diehl:2003ny}. We
again use the form of the evolution given in
Eq.~(\ref{eq:ERBLlikeeveq}) and, setting $\xi=1$ but allowing $|x|$
and $|y|$ to be larger than one, the evolution kernel becomes:
\begin{equation}
\begin{array}{rcl}
\displaystyle \mathbb{V}^{-,[0]}\left(x,y\right) &=& \displaystyle
\rho(x,y)C_F \frac{1+x}{1+y}\left(1+\frac{2}{y-x}\right)+\left(x\rightarrow
                                                      -x \atop y\rightarrow
                                                      -y\right)\\
\\
&+&\displaystyle \delta\left(y-x\right)
    C_F\left[3+2\int_1^{x}\frac{dz}{z-x}+2\int_{-1}^{x}\frac{dz}{z-x}\right]\,,
\end{array}
\end{equation}
with:
\begin{equation}
\rho(x,y) = \theta(x+1)\theta(y-x)-\theta(-x-1)\theta(x-y) =
\theta\left(\frac{y-x}{1+y}\right)
\theta\left(\frac{1+x}{1+y}\right)\mbox{sign}(1+y)\,.
\label{eq:supportfunction}
\end{equation}
It is easy to verify that:
\begin{equation}
  \int_{-\infty}^{\infty}dx\,\mathbb{V}^{-,[0]}\left(x,y\right) = 0\,.
\end{equation}
Therefore, one can rewrite:
\begin{equation}
\mathbb{V}^{-,[0]}\left(x,y\right) =
\left\{\rho(x,y)C_F \frac{1+x}{1+y}\left(1+\frac{2}{y-x}\right)+\left(x\rightarrow
                                                      -x \atop y\rightarrow
                                                      -y\right)\right\}_+\,,
                                                  \label{eq:MuellerEvKernel}
\end{equation}
where the $+$-prescription (with curly brackets) here is defined in a
yet different manner and generalises that in
Eq.~(\ref{eq:plusprescr2}) to a two-dimensional function with support
$x,y\in\mathbb{R}$ and a single pole at $x = y$:
\begin{equation}
  \{f(x,y)\}_+=f(x,y) - \delta(x-y)\int_{-\infty}^{\infty}dx\, f(x,y)\,.
  \label{eq:plusprescr3}
\end{equation}
This finally allows us to recover the results of
Refs.~\cite{Muller:1994ses, Diehl:2003ny}.

Finally, we compare our result to that of
Ref.~\cite{Belitsky:2005qn}. Adopting the notation of that reference,
one can show that:
\begin{equation}
  \vartheta_{11}^0(x_1,x_1-y_1) = \frac{2}{\xi+y}\rho\left(\frac{x}{\xi}, \frac{y}{\xi}\right)\quad\mbox{and}\quad   \vartheta_{11}^0(x_2,x_2-y_2) = \frac{2}{\xi-y}\rho\left(-\frac{x}{\xi}, -\frac{y}{\xi}\right)\,,
\end{equation}
and:
\begin{equation}
\vartheta_{111}^0(x_1,-x_2,x_1-y_1) = -\frac{\xi+x}{\xi(\xi+y)}\rho\left(\frac{x}{\xi}, \frac{y}{\xi}\right)-\frac{\xi-x}{\xi(\xi-y)}\rho\left(-\frac{x}{\xi}, -\frac{y}{\xi}\right)\,,
\end{equation}
with $\rho$ given in Eq.~(\ref{eq:supportfunction}). This allows us to
recast the evolution kernel $K_{(0)}^{qq;V}(x_1,x_2|y_1,y_2)$ given in
Eq.~(4.42) of Ref.~\cite{Belitsky:2005qn} into the following form:
\begin{equation}
  K_{(0)}^{qq;V}(x_1,x_2|y_1,y_2) = -\left\{\rho\left(\frac{x}{\xi}, \frac{y}{\xi}\right)C_F\frac{\xi+x}{\xi+y}\left(\frac{1}{\xi}+\frac{2}{y-x}\right) +\left(x\rightarrow
      -x \atop y\rightarrow
      -y\right)\right\}_+\,.
  \label{eq:RadyushkinEvKernel}
\end{equation}
Considering that a factor of two due to the different expansion
parameter ($\alpha_s/(2\pi)$ vs. $\alpha_s/(4\pi)$) is compensated by
an opposite factor that comes from the fact that in
Ref.~\cite{Belitsky:2005qn} the evolution equations are differential
w.r.t. to $\ln\mu$ rather than $\ln\mu^2$ and accounting for a minus
sign in the definition of the evolution kernels, this result coincides
with Eq.~(\ref{eq:MuellerEvKernel}) that we have already proven to
agree with our result. \footnote{We notice that, while
  Eq.~(\ref{eq:MuellerEvKernel}) applies for any $x$ and $y$ in
  $\mathbb{R}$, Eq.~(\ref{eq:RadyushkinEvKernel}) applies for
  $x,y\in[-1,1]$. However, since $x$ and $y$ in
  Eq.~(\ref{eq:RadyushkinEvKernel}) always appear in the ratios
  $x/\xi$ and $y/\xi$ with $\xi\in[0,1]$, one can rescale
  $x/\xi\rightarrow x$ and $y/\xi\rightarrow y$ in
  Eq.~(\ref{eq:RadyushkinEvKernel}) to make it coincide with
  Eq.~(\ref{eq:MuellerEvKernel}).}

\section{Numerical results}\label{sec:numerics}

In Sect.~\ref{sec:evolutionequations}, we recasted the GPD evolution
kernel in a form suitable for a straightforward implementation in the
numerical code {\tt APFEL++}~\cite{Bertone:2013vaa, Bertone:2017gds},
allowing for robust evaluations and handling of heavy-flavour
thresholds. In Sect.~\ref{sec:properties}, we detailed the theoretical
properties of this particular form. In this section, we present a
series of numerical checks aimed at establishing the validity of our
implementation to high numerical accuracy. To the best of our
knowledge, this provides the most extensive set of tests of an
implementation of GPD evolution equations ever presented in the
literature, at least with respect to publicly released codes. Although
here we are not concerned with performance and computing speed, our
implementation guarantees a fast evaluation of GPD evolution suitable
for phenomenological extractions.

\subsection{DGLAP limit and skewness dependence}\label{subsec:dgalplimit}

As discussed in Sect.~\ref{sec:DGLAPlimit}, in the forward limit,
$\xi\rightarrow 0$, our derivation of the GPD evolution equations
reproduces the one-loop DGLAP evolution. In the following, we provide
numerical evidence for this statement. In order to do so, we need to
evolve a set of distributions defined at some initial scale $\mu_0$ up
to a different scale $\mu$ using the solution of
Eq.~(\ref{eq:regdglap}). To this purpose, we use the leading-order PDF
set {\tt MMHT2014lo68cl}~\cite{Harland-Lang:2014zoa} through the
LHAPDF interface~\cite{Buckley:2014ana} with $\mu_0=1$~GeV. The
running of the strong coupling is computed at one loop using
$\alpha_s(M_Z)=0.135$, consistently with {\tt MMHT2014lo68cl}. In
addition, the evolution is performed using the variable-flavour-number
scheme, \textit{i.e.} we allow for heavy-quark-threshold crossings
during the evolution, with charm and bottom thresholds set to
$m_c=1.4$~GeV and $m_b=4.75$~GeV, respectively.

\begin{figure}[t]
  \centering
  \includegraphics[width=0.49\textwidth]{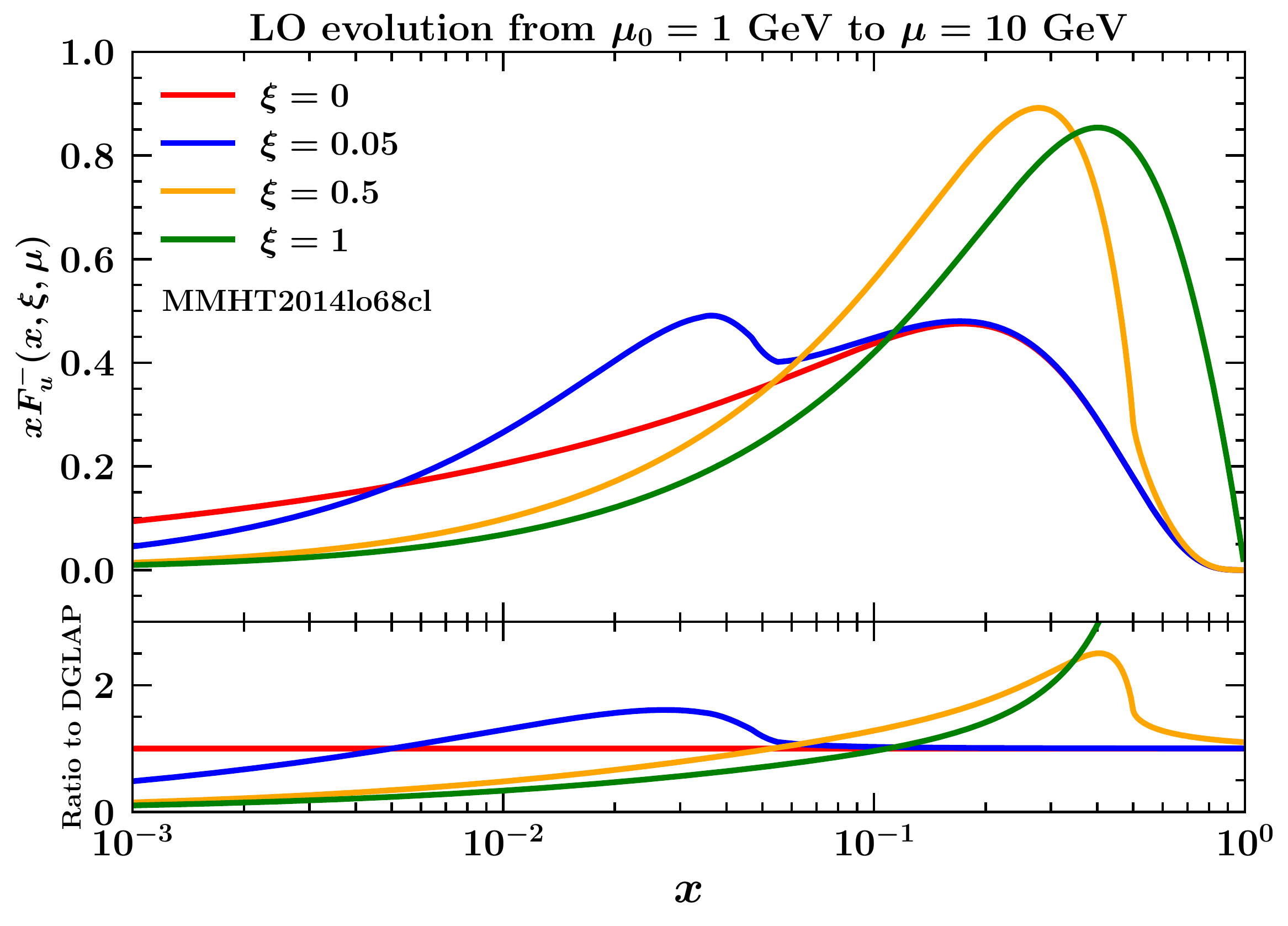}
  \includegraphics[width=0.49\textwidth]{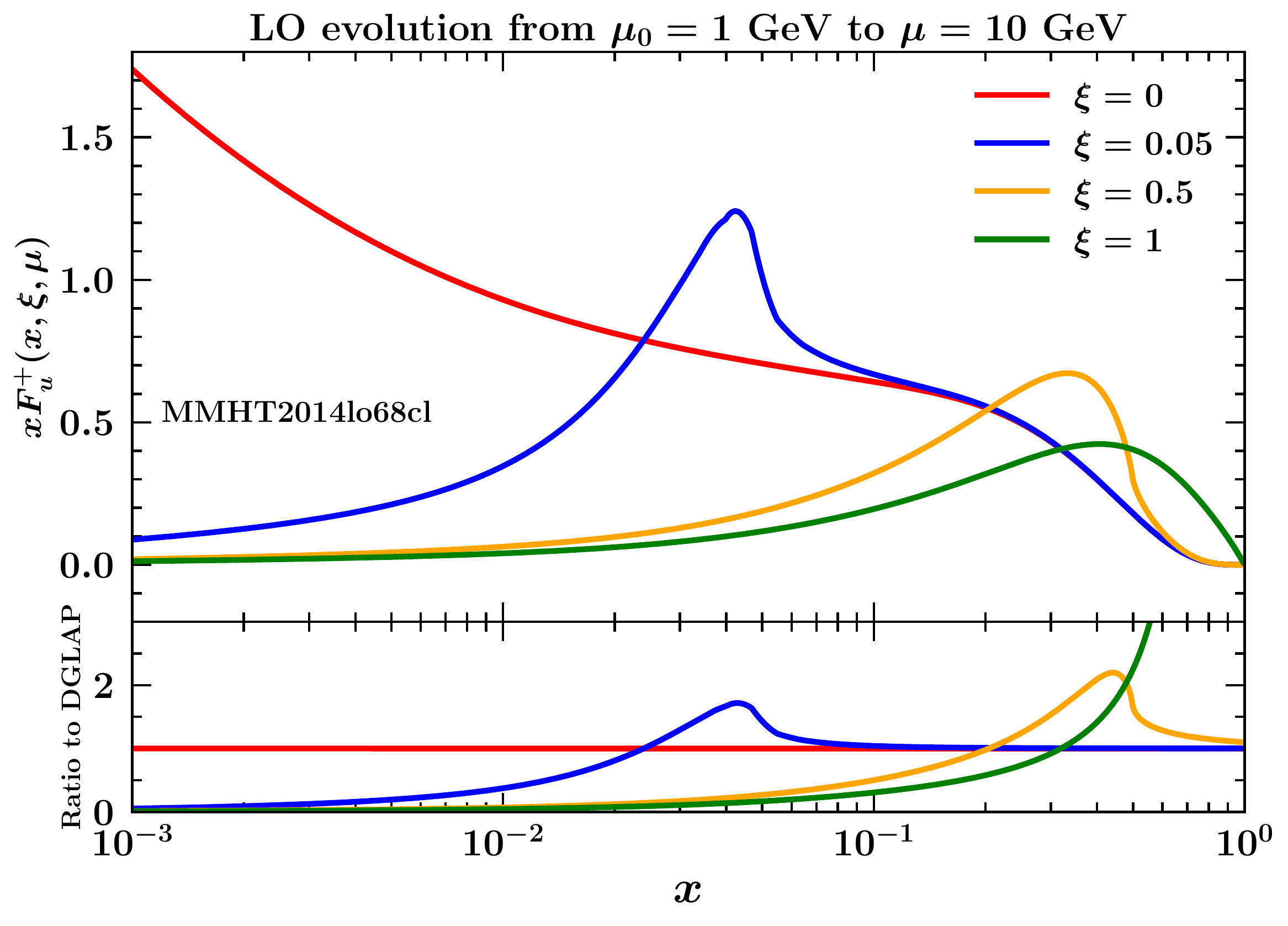}
  \includegraphics[width=0.49\textwidth]{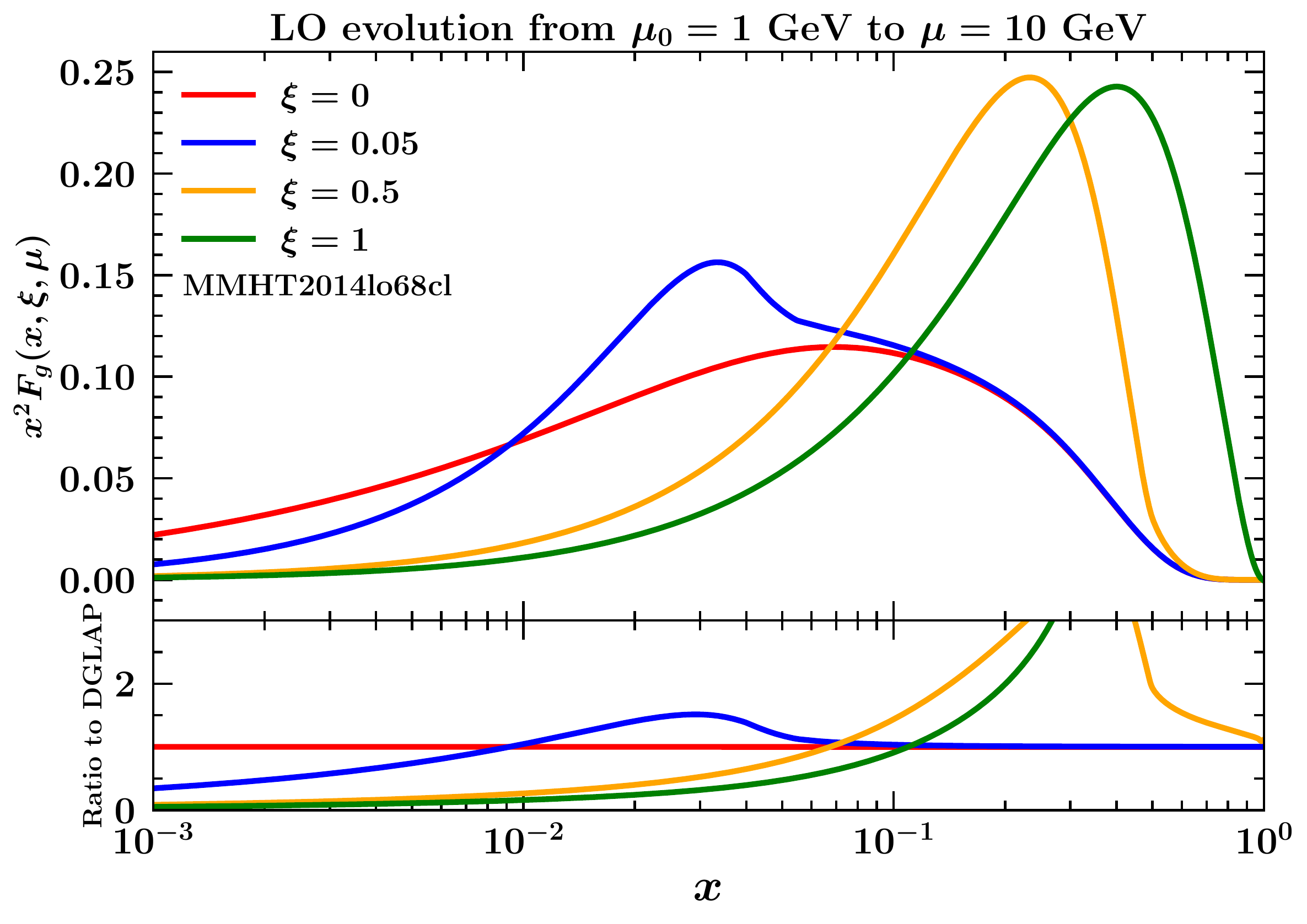}
    \vspace{15pt}
    \caption{Non-singlet up-quark (upper left) and singlet up-quark
      (upper right), and gluon (bottom) distributions evolved from
      $\mu_0=1$~GeV to $\mu=10$~GeV using the GPD evolution equations
      in Eq.~(\ref{eq:regdglap}) with different values of the skewness
      $\xi$. Initial-scale distributions are taken from the {\tt
        MMHT2014lo68cl} PDF set. To tame the fast rise of gluons at
      low-$x$, we weight the distribution with an additional power of
      $x$.  The lower inset displays the ratio to the DGLAP evolution
      as delivered by the LHAPDF grid.\label{fig:GPDEvolution}}
\end{figure}

Fig.~\ref{fig:GPDEvolution} shows the effect of evolving the {\tt
  MMHT2014lo68cl} PDF set to $\mu=10$~GeV for different values of
$\xi$, including the DGLAP limit $\xi\rightarrow0$, using the
numerical solution of Eq.~(\ref{eq:regdglap}) as implemented in {\tt
  APFEL++}. Evolution is probed for $x$ ranging between $10^{-3}$ and
1, relevant for the fixed target (Jefferson Lab, COMPASS) and collider
(EIC, EIcC) experiments, while the evolution range spans two orders of
magnitude in the hard scale $\mu^2$, from 1 to 100~GeV$^2$. The
top-left plot displays the up-quark non-singlet distribution
$F_u^-=F_u-F_{\overline{u}}$, the top-right one displays the singlet
distribution $F_u^+=F_u+F_{\overline{u}}$, while the bottom plot
displays the gluon distribution $F_g$. The upper insets show the
absolute distributions while the lower ones the ratio to the DGLAP
evolution as delivered by the LHAPDF grid~\cite{Buckley:2014ana}. The
first observation is that, as clear from the bottom insets, our GPD
evolution in the $\xi\rightarrow0$ limit reproduces the DGLAP
evolution very accurately. It is also interesting to observe how GPD
evolution modifies the shape of the distributions w.r.t. the DGLAP
when changing the skewness $\xi$. Differences are sizeable
particularly in the ERBL region, $x<\xi$, where the GPD evolution
tends to suppress the distributions w.r.t. the DGLAP one. Particularly
striking is the singlet sector in which steeply raising low-$x$
distributions are turned into decreasing distributions. In addition,
as anticipated in Sect.~\ref{subsec:continuity}, distributions are
continuous at the crossing point $x=\xi$ but develop a cusp although
the initial scale distributions are smooth.

\subsection{ERBL limit}

\begin{figure}[t]
  \centering
  \includegraphics[width=0.6\textwidth]{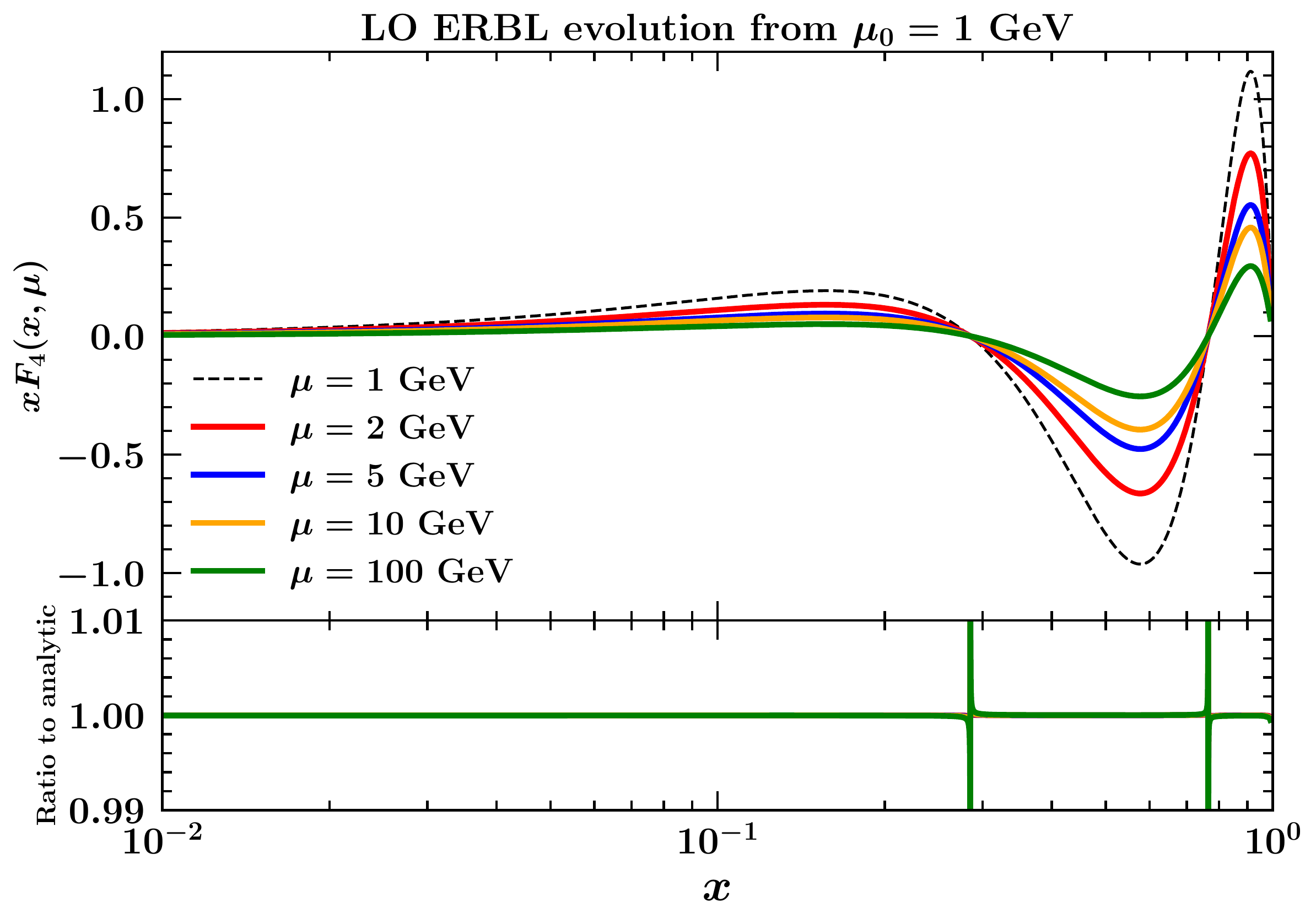}
    \vspace{15pt}
    \caption{Non-singlet leading-order ERBL evolution from
      $\mu_0=1$~GeV to different values of the final scale $\mu$ of
      the distribution in Eq.~(\ref{eq:gegenERBL}) with $n=2$. The
      upper inset displays the distributions obtained by numerically
      solving Eq.~(\ref{eq:regdglap}) with $\xi=1$, while the bottom
      inset shows the ratio to Eq.~(\ref{eq:ERBLansol}). Notice that
      the curves in the bottom inset overlap almost completely, making
      them hardly distinguishable.
      \label{fig:ERBLEvolution}}
\end{figure}

Having ascertained that using our GPD evolution equations the DGLAP
limit is recovered, we now turn to check the opposite limit,
\textit{i.e.} the ERBL limit $\xi\rightarrow 1$. To do so, we exploit
the fact that functions of this kind:
\begin{equation}
F_{2n}(x,\mu_0) = (1-x^2)C_{2n}^{(3/2)}(x)\,,
\label{eq:gegenERBL}
\end{equation}
diagonalise the (non-singlet) leading-order ERBL evolution equation
such that they evolve multiplicatively as:
\begin{equation}
F_{2n}(x,\mu) =
\exp\left[\frac{P_{2n+1}^{-,[0]}}{4\pi}\int_{\mu_0}^{\mu}d\ln\mu'^2\alpha_s(\mu')\right]F_{2n}(x,\mu_0)\,,
\label{eq:ERBLansol}
\end{equation}
with anomalous dimensions $P_{n+1}^{-,[0]}$ given in
Eq.~(\ref{eq:forwardlimitconfmom}). Fig.~\ref{fig:ERBLEvolution} shows
the non-singlet evolution of Eq.~(\ref{eq:gegenERBL}) with $n=2$ from
$\mu_0^2=1$~GeV$^2$ to a number of higher scales $\mu^2$, up to
$\mu^2=10^{4}$~GeV$^2$, using the numerical solution of
Eq.~(\ref{eq:regdglap}) with $\xi=1$. The upper panel displays the
absolute distributions including the initial-scale one, while the
lower panel displays the ratio to the analytic evolution given in
Eq.~(\ref{eq:ERBLansol}). As clear from the bottom panel, the
agreement between numerical and analytic solutions is
excellent\footnote{The spikes appearing in the lower panel of
  Fig.~\ref{fig:ERBLEvolution} correspond to the points in which the
  distributions change sign.} confirming that our implementation of
the GPD evolution gives sound results also in the ERBL limit. We could
not find other numerical tests of the recovery of the ERBL limit for
other public GPD evolution codes in the existing literature.

\subsection{Polynomiality}

GPDs enjoy the so-called polynomiality property that for quarks can be
written as:
\begin{equation}
\int_0^1 dx\,x^{2n} F_q^{-}(x,\xi,\mu) =
\sum_{k=0}^{n}A_k(\mu)\xi^{2k}\,,\quad\mbox{and}\quad \int_0^1
dx\,x^{2n+1} F_q^{+}(x,\xi,\mu) = \sum_{k=0}^{n+1}B_k(\mu)\xi^{2k}\,,
\label{Eq:polynomiality}
\end{equation}
with $F_q^\pm=F_q\pm F_{\overline{q}}$. It is important to notice that
these relations must be valid at any scale $\mu$ implying that GPD
evolution must preserve polynomiality. In this section we
quantitatively show that this is the case when using the solution of
Eq.~(\ref{eq:regdglap}).

\begin{figure}[t]
  \centering
  \includegraphics[width=0.49\textwidth]{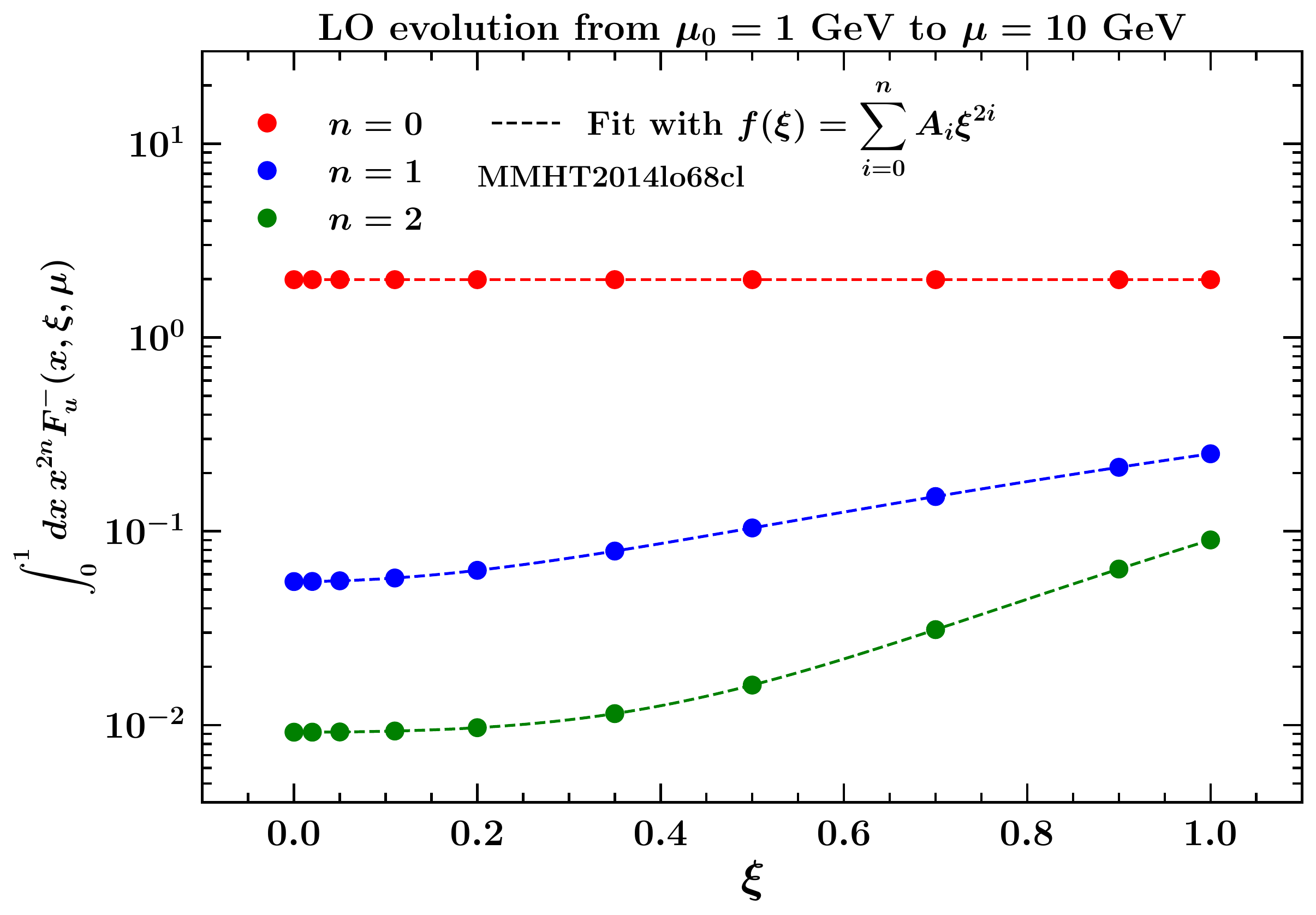}
  \includegraphics[width=0.49\textwidth]{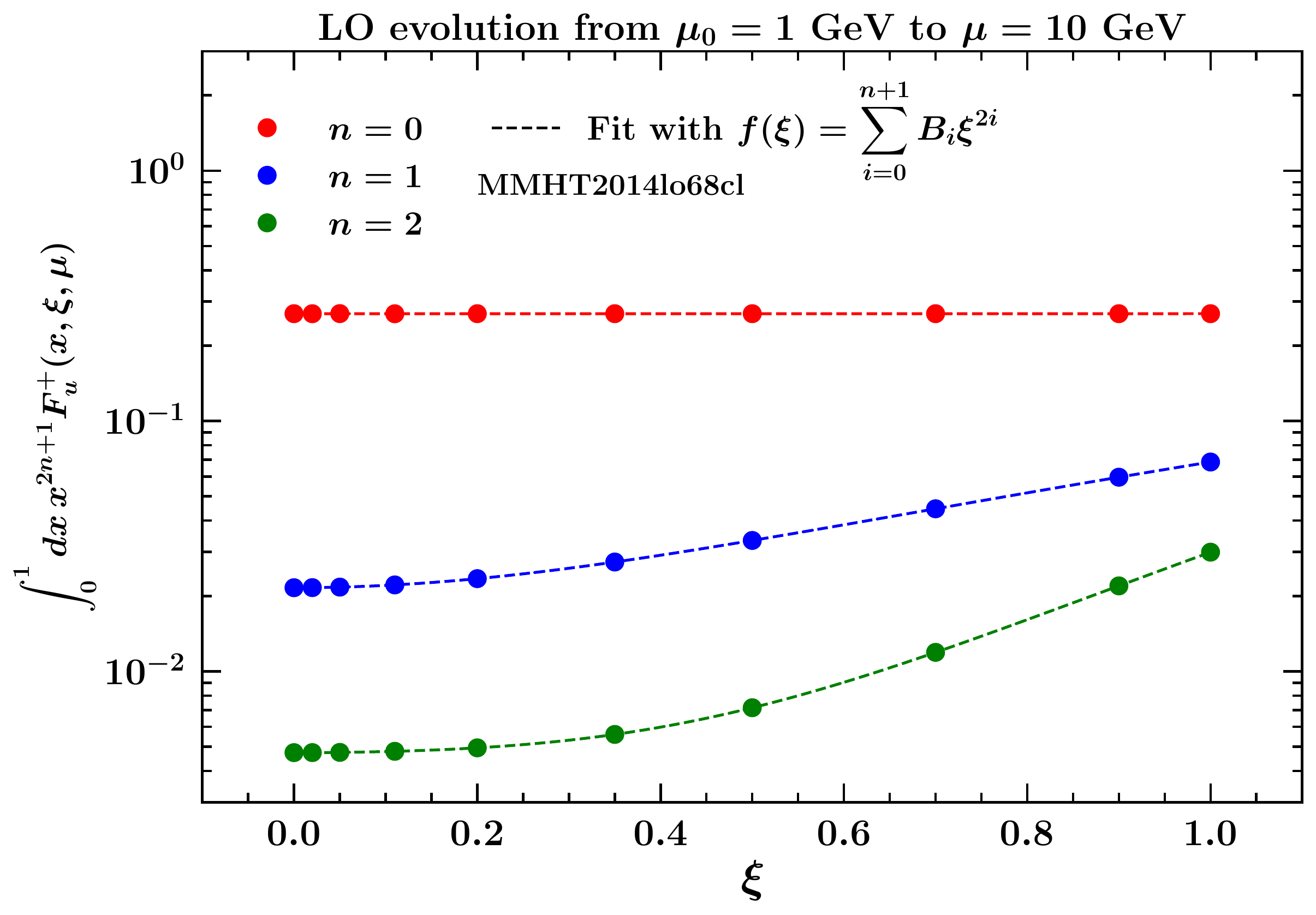}
    \vspace{15pt}
    \caption{Non-singlet (left) and singlet (right) up-quark Mellin
      moments for GPDs evolved from $\mu_0=1$~GeV to $\mu=10$~GeV
      using the GPD evolution equations in Eq.~(\ref{eq:regdglap}) as
      functions the skewness $\xi$. Initial-scale distributions are
      taken from the {\tt MMHT2014lo68cl} PDF set. Each set of points
      is fitted with the power law predicted by
      polynomiality.\label{fig:GPDMoments}}
\end{figure}

We consider the setup of Sect.~\ref{subsec:dgalplimit} in which a set
of $\xi$-independent PDFs, that thus trivially obey polynomiality, is
evolved from $\mu_0=1$~GeV to $\mu=10$~GeV. In order to check that
polynomiality is conserved, we evaluate the integrals in
Eq.~(\ref{Eq:polynomiality}) for the first three moments ($n=0,1,2$)
and for different values of $\xi$. We then fit the points thus
obtained using the expected power laws in $\xi^2$. We point out that
higher moments can be computed analogously. However, a solid check of
polynomiality requires that the number of points in $\xi$ used for the
fit be much larger than the degree of the expected polynomial in
$\xi^2$. For this reason we limit the check to the first three moments
using ten points in $\xi$: this should be enough to guarantee that the
power-law behaviours are accurately reproduced. The result is shown in
Fig.~\ref{fig:GPDMoments}. The l.h.s. plot displays the first two
moments of the up-quark non-singlet distribution $F_u^-$ as functions
of $\xi$, while the r.h.s. one shows the same for the up-quark singlet
distribution $F_u^+$. The computed values (plain dots) are
superimposed on the fitted curves proving that the expected behaviour
is obtained to a very good accuracy over the entire range in
$\xi\in[0,1]$. Some additional comments are in order. First, the first
moment ($n=0$) of the non-singlet distribution $F_u^-$ is not only
constant as expected but also equal to two which reflects the valence
sum rule for the up quark in the proton. Secondly, the first moment
($n=0$) of the singlet distribution $F_u^+$, despite being allowed to
depend on $\xi$ through a quadratic terms, is also constant. This is
due to the fact that the term proportional to $\xi^{2n+2}$ in the
second equation of Eq.~(\ref{Eq:polynomiality}) gives rise to the
so-called $D$-term~\cite{Polyakov:1999gs} that evolves
independently. Since the initial scale distributions do not include
any $D$-term, none is generated by evolution and thus only the
constant term contributes to the first moment of $F_u^+$.

\subsection{Conformal-space evolution}\label{subsec:confmomev}

In Sect~\ref{sec:conformalmoments} we have explicitly proven that the
one-loop non-singlet evolution kernel computed in this paper is such
that the evolution of the GPD conformal moments is diagonal,
\textit{i.e.} each moment evolves multiplicatively with its own
kernel. In the following, we show that our implementation of the
solution of the evolution equations numerically fulfils this property.
To do so, we consider as an initial-scale non-singlet GPD at
$\mu_0=1$~GeV the quark GPD $H^q$ given by the Radyushkin
double-distribution ansatz (RDDA) \cite{Musatov:1999xp}:
\begin{align}
  H^q(x,\xi) = \int_\Omega \textrm{d}\beta \textrm{d}\alpha \delta\left(x - \beta - \xi \alpha \right) q(|\beta|) \pi(\beta, \alpha),
\end{align}
where $\Omega$ is such that $|\alpha|+|\beta|\le 1$ and:
\begin{align}
  q(x) &= \frac{35}{32} x^{-1/2} (1-x)^3, \\
  \pi(\beta,\alpha) & = \frac{3}{4} \frac{((1-|\beta|)^2-\alpha^2)}{(1-|\beta|)^3}.
\end{align}
This simple ansatz allows us to benchmark our $x$-space evolution code
using a realistic behaviour of the non-singlet GPDs with respect to
conformal evolution.

\begin{figure}[t]
  \centering
  \includegraphics[width=0.6\textwidth]{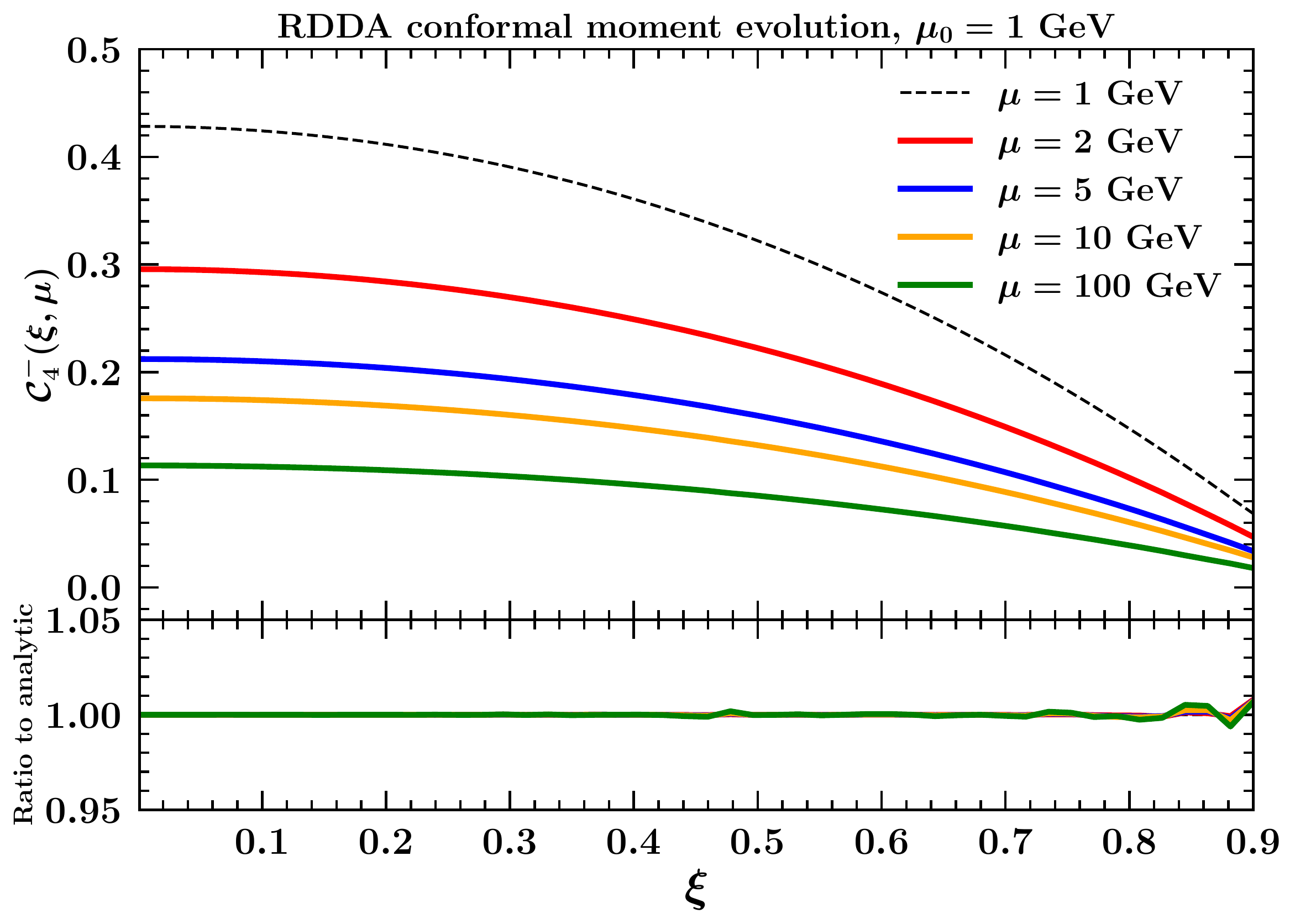}
    \vspace{15pt}
    \caption{Leading-order evolution of the second ($n=4$) conformal
      moment of the non-singlet distribution of the Radyushkin
      double-distribution ansatz (RDDA) described in the text. The
      evolution starts from $\mu_0=1$~GeV up to different values of
      the final scale $\mu$. The upper inset displays the moment as a
      function of the skewness $\xi$ obtained by numerically solving
      Eq.~(\ref{eq:regdglap}) and by computing the conformal moment of
      the final-scale distribution by means of
      Eq.~(\ref{eq:conformalmoments}) while the bottom inset shows the
      ratio to the solution of Eq.~(\ref{eq:confmom}). As in
      Fig.~\ref{fig:ERBLEvolution}, the curves in the bottom inset are
      hardly distinguishable because they all lie on top of each
      other.\label{fig:RDDA}}
\end{figure}

In Fig.~\ref{fig:RDDA}, we compare the second $(n=4)$ conformal
moment, computed by means of Eq.~(\ref{eq:conformalmoments}), of the
RDDA-based model evolved to $\mu=2,5,10,100$~GeV obtained by
numerically solving Eq.~(\ref{eq:regdglap}) to the solution of
Eq.~(\ref{eq:confmom}) with evolution kernel given in
Eq.~(\ref{eq:ConfMomMaster}). The upper panel of the plot displays the
evolved conformal moment as a function of $\xi$ for the different
values of $\mu$ computed by solving Eq.~(\ref{eq:regdglap}), while the
lower panel shows the ratio to the solution of
Eq.~(\ref{eq:confmom}). It is clear that the agreement between the two
evolution methods is excellent over the entire range in $\xi$
considered, validating our implementation in the light of
Sect~\ref{sec:conformalmoments}.

Before moving to comparing our evolution code to another
implementation, we emphasise that all the numerical tests performed in
Sects.~\ref{subsec:dgalplimit}-\ref{subsec:confmomev} turned out to be
very successful. Namely, we found that all the fundamental properties
of GPD evolution, such as DGLAP and ERBL limits, polynomiality
conservation, and equivalence with the conformal-moment approach, are
fulfilled to the sub-per-mil level or better. We regard this as a very
strong consistency check of our code.

\subsection{Comparison to Vinnikov's code}\label{subsec:vinnikov}

\begin{figure}[t]
  \centering
  \includegraphics[width=0.49\textwidth]{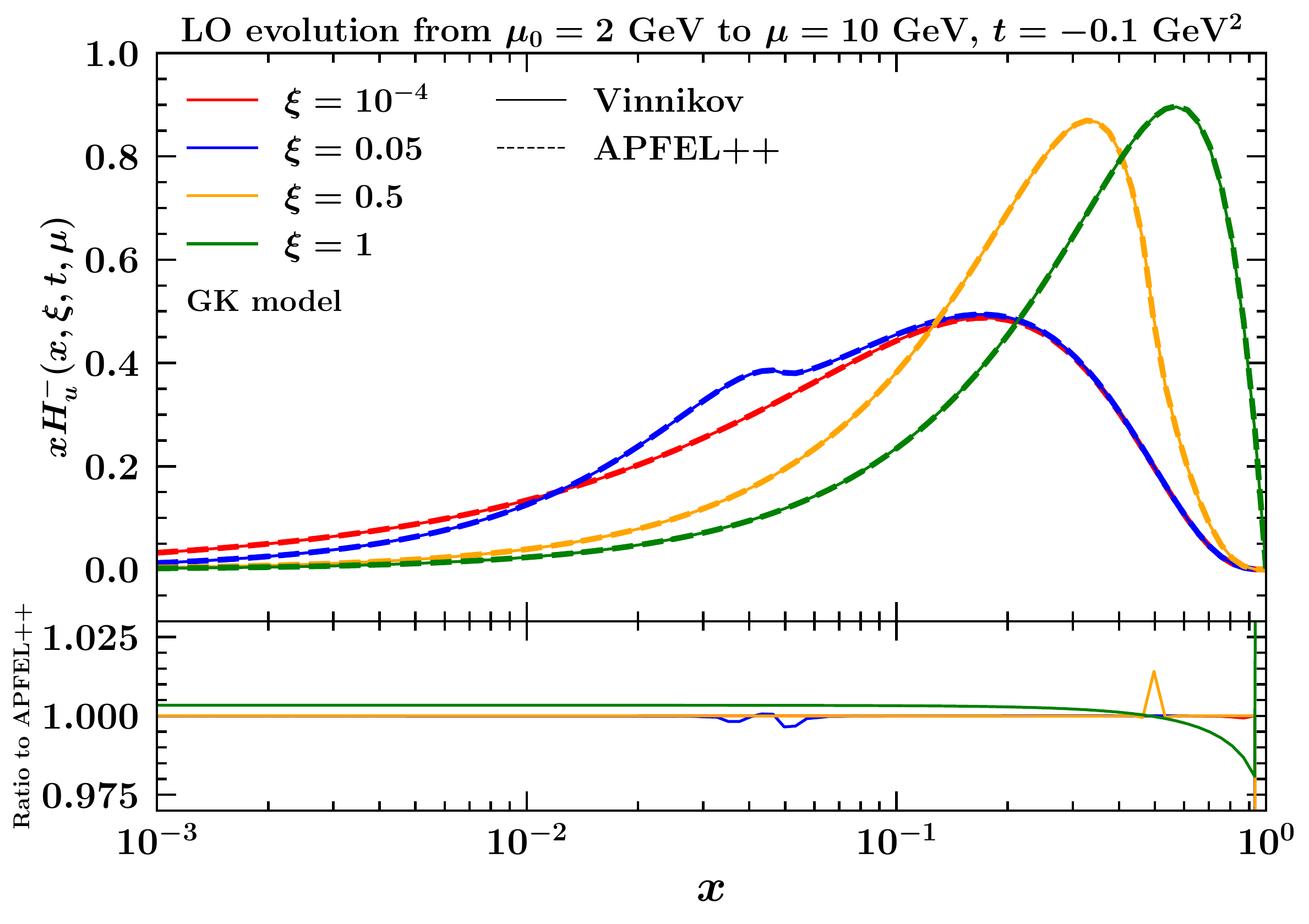}
  \includegraphics[width=0.49\textwidth]{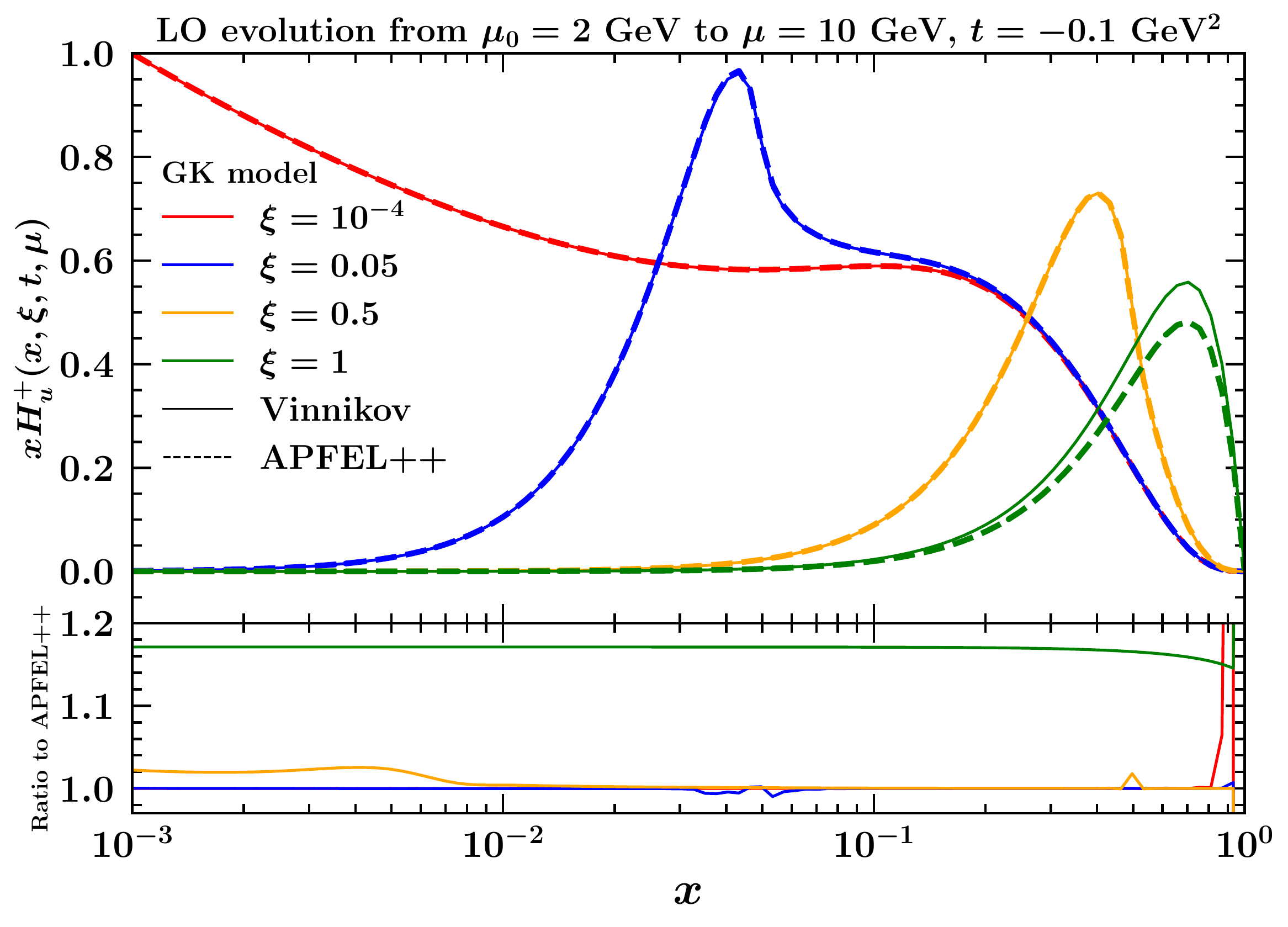}
  \includegraphics[width=0.49\textwidth]{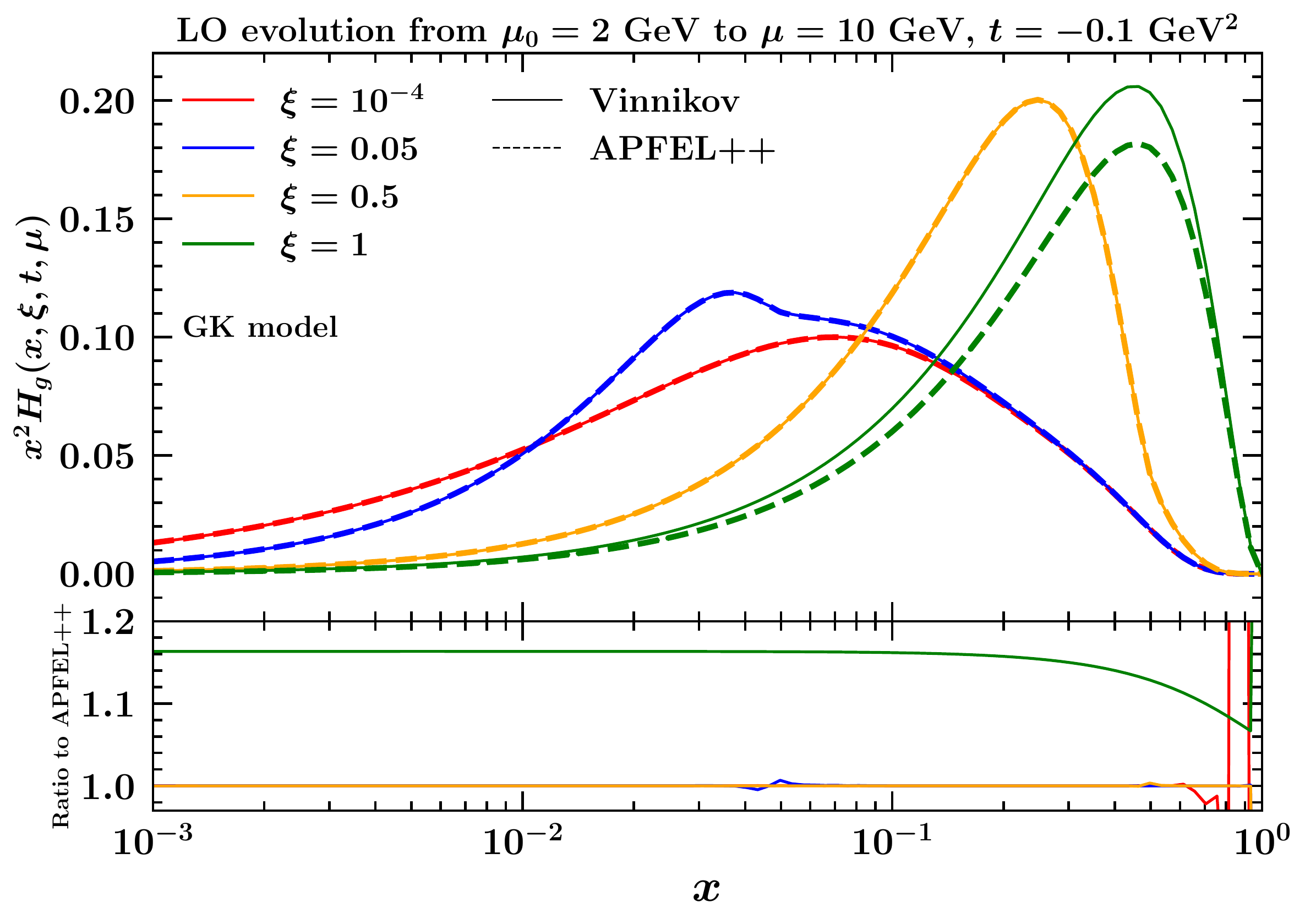}
    \vspace{15pt}
    \caption{Comparison between Vinnikov's code~\cite{Vinnikov:2006xw}
      and {\tt APFEL++}. The evolution is performed at LO in the
      $n_f=3$ scheme (no threshold crossing) between the scales
      $\mu_0=2$~GeV and $\mu = 10$~GeV. As initial-scale distributions
      we have used the model of
      Refs.~\cite{Goloskokov:2005sd,Goloskokov:2007nt,Goloskokov:2009ia}\label{fig:
        ApfelVsVinnikov} (GK model) setting $t=-0.1$~GeV$^2$ as
      momentum transfer squared.\label{fig:ApfelVsVinnikov}}
\end{figure}

In this section, we compare the evolution obtained with {\tt APFEL++}
to that presented in Ref.~\cite{Vinnikov:2006xw} that in the following
will be referred to as ``Vinnikov's code'' after its
author. Specifically, we use an implementation of Vinnikov's code
available in the {\tt PARTONS} framework~\cite{Berthou:2015oaw}. A
limitation of Vinnikov's code is that it does not implement the
variable-flavour-number scheme, \textit{i.e.} it does not allow one to
cross heavy-quark tresholds along the evolution. Therefore, for the
comparison we have used the $n_f=3$ fixed-flavour-number scheme in
which three quark flavours (up, down, and strange) are active at all
scales.  Like {\tt APFEL++}, Vinnikov's code can perform GPD evolution
only at LO. As initial-scale distributions we have used the model
presented in
Refs.~\cite{Goloskokov:2005sd,Goloskokov:2007nt,Goloskokov:2009ia}
that also depends on the momentum transfer $t$ that we set to
$t=-0.1$~GeV$^2$. In Fig.~\ref{fig:ApfelVsVinnikov} we present the
comparison for the evolution between $\mu_0=2$~GeV and $\mu=10$~GeV
for the GPDs $H_u^-$, $H_u^+$, and $H_g$. The upper panels display the
absolute distributions for four different values of the skewness
parameter $\xi=10^{-4}, 0.05,0.5,1$ with the solid lines showing the
results obtained with Vinnikov's code and the dashed lines those
obtained with {\tt APFEL++}.  The lower panels show the same curves
normalised to {\tt APFEL++}. We observe a general very good agreement
between the two codes for almost all values of $\xi$ and across the
full range in $x$ considered. The only exception is $\xi=1$ for which
a disagreement at the percent level for $H_u^-$ (non-singlet sector)
and as large as 20\% for $H_u^+$ and $H_g$ (singlet sector) is
observed. We could not identify the origin of this disagreement but,
in view of the reported numerical instabilities of Vinnikov's code in
the large-$\xi$ region~\cite{Diehl:2007zu}, we suspect that the
results of this code at $\xi=1$ might be affected by numerical
inaccuracies.  Indeed, we point out that Vinnikov's code does not
allow one neither to set $\xi =1$ nor $\xi=0$ and the smallest stable
value of $\xi$ we could find is $\xi=10^{-4}$.  Regarding the large
$\xi$ region, in the plots in Fig.~\ref{fig:ApfelVsVinnikov} we have
actually used $\xi=1-\epsilon$ with $\epsilon=10^{-6}$ for both codes.
Finally, we found severe numerical instabilities for
$0.6\lesssim \xi \lesssim 0.95$.  Therefore, we have not been able to
perform a comparison in this region.  This also justifies the need of
a new open and maintained evolution code. Moreover, the modular
architecture of {\tt APFEL++} and {\tt PARTONS} will facilitate the
integration of higher-order corrections to the evolution while this
task would probably require an almost complete rewriting of Vinnikov's
code.

\section{Conclusions}\label{sec:conclusions}

The main purpose of this paper is to provide a solid and public
implementation of GPD evolution that can be used for phenomenological
studies. To do so, we have revisited GPD evolution in view of an
efficient numerical implementation spelling out the computational
details.  We rederived the evolution equations and recomputed the
evolution kernels at one-loop accuracy in perturbative QCD. For the
calculation we adopted a Feynman-diagram approach using the operator
definition of GPDs in the light-cone gauge, which reduces the number
of diagrams to be considered, renormalised in the
$\overline{\mbox{MS}}$ scheme. Our formulation of the evolution
equations allowed us to easily study some relevant properties of the
evolution kernels. Specifically, we have shown that our calculation
correctly reproduces both the DGLAP and the ERBL limits and that it
guarantees continuity of GPDs at the cross-over point $x=\xi$. In
addition, we have worked out the consequences of the GPD sum rules on
the evolution kernels deriving equalities that need to be obeyed order
by order in perturbation theory, finally showing that our one-loop
calculation fulfils these equalities. Moreover, we have computed the
conformal moments of our non-singlet evolution kernels showing that,
as expected, they diagonalise upon Gegenbauer transform and that their
eigenvalues coincide with the well-know DGLAP and ERBL one-loop
anomalous dimensions. Finally, we have also explicitly shown that our
computation reproduces previous results present in the literature.

Our calculation has been implemented in the public code {\tt APFEL++}
that in turn has been interfaced to {\tt PARTONS}. This allowed us to
perform detailed numerical studies. We have checked that DGLAP and
ERBL evolutions are reproduced to very high accuracy in the
$\xi\rightarrow 0$ and $\xi\rightarrow 1$ limits,
respectively. Moreover, we have checked that the evolution preserves
GPD polynomiality. In addition, we have verified that our
implementation of GPD evolution agrees with the evolution computed in
conformal space. As a last check, we have compared our GPD evolution
against another existing implementation, Vinnikov's code, finding a
general good agreement.

In this paper, we limited ourselves to the one-loop (LO) evolution of
unpolarised GPDs. The next natural short-term step is the extension to
longitudinally and transversely polarised evolutions. In the longer
run, we plan to implement the two-loop (NLO) corrections to the
evolution. Facing a new era for GPD experiments at colliders, we
believe that the public release of a documented and carefully checked
implementation of GPD evolution equations meets the need of the
hadron-physics community. The code is flexible and can run with any
GPD model expressed in $x$ space. It also provides for the first time
an implementation of the variable-flavour-number scheme in a public
solver of GPD-evolution equations

We conclude by stressing once again that the implementation of GPD
evolution presented here is publicly available in the {\tt APFEL++}
code:
\begin{center}
\url{https://github.com/vbertone/apfelxx}
\end{center}
that is in turn interfaced to the {\tt PARTONS} framework:
\begin{center}
\url{https://partons.cea.fr/partons/doc/html/index.html}
\end{center}
that gives access to a large variety of GPD models some of which used
in this paper. The user can find ready-to-use example codes to evolve
any of the GPD models available in {\tt PARTONS}.

\section*{Acknowledgements}

We are grateful to M. Diehl and A. Radyushkin for stimulating
discussions and to help us interpret the results of previous
calculations.  We also thank J. Rodriguez-Quintero for useful
discussions. V.~B. is supported by the European Union’s Horizon 2020
research and innovation programme under grant agreement STRONG 2020 -
No 824093. J.~M.~M. is supported by the University of Huelva under
grant EPIT2021.

\appendix
\section{Parton-in-parton GPDs}\label{app:gpdnormlisation}

In this appendix, we introduce the unpolarised parton-in-parton GPDs
and give explicit definitions that can be used to compute them in
perturbation theory. As shown in Sect.~\ref{sec:evolutionequations},
this allows one to determine the anomalous dimensions that govern the
evolution of GPDs. We explicitly compute the tree-level contribution
to these GPDs showing that to this order they coincide with the
corresponding PDFs times a $\xi$-dependent factor thereby setting
their normalisation. In Appendix~\ref{app:splittingfunctions}, we will
use these definitions to compute the one-loop quark-in-quark anomalous
dimension.

The parton-in-parton GPDs can be easily obtained by replacing the
hadronic states in the parton-in-hadron GPDs defined in
Eq.~(\ref{eq:GPDdefinition}) with the appropriate partonic
states. Specifically, we consider on-shell massless partons moving
along the direction defined by the gauge vector $n$ with incoming
momentum $(1+\xi)p$ and outgoing momentum $(1-\xi)p$. In addition, we
also have to include an average over the colour states of the external
partons. To do so, we need to invoke for a moment the Wilson
line. Since we are working in the light-cone gauge, the Wilson line
does not contribute in the sense that it reduces to the unitary
operator in the fundamental representation of the colour group for the
quark operator and in the adjoint representation for the gluon
operator. Therefore, when averaging over the colour states of the
external partons, since the probe is assumed to be a colour singlet,
we effectively need to take the trace over the colour indices and
divide by the dimension of the colour representation. This amounts to:
\begin{equation}
\frac{1}{N_c}\mbox{Tr}_c[\dots]\,,
\end{equation}
for external quark states and to:
\begin{equation}
\frac{1}{N_c^2-1}\mbox{Tr}_c[\dots]\,,
\end{equation}
for external gluon states, where ``Tr$_c$'' indicates the trace over
the colour indices and $N_c=3$ is the number of colours. Finally, we
also need to include an average over the physical spin/helicity states
and a trace over the Dirac indices:
\begin{equation}
\frac12\sum_s \mbox{Tr}_D[\dots]\,,
\end{equation}
with $s$ running over the spin index for quark states and the helicity
index for gluon states. In the following, we will denote with ``Tr''
the trace over \textit{both} colour and Dirac indices.

In the presence of more than one massless quark flavour, one can
define seven different combinations between external partonic states
and GPD operators. We list them all below by also including the
appropriate averaging discussed above. Let us start with the gluon
operator. In this case, we can have the gluon-in-gluon GPD in which
the gluon operator acts on gluon external states:
\begin{equation}
\displaystyle \hat{F}_{g/g}(x,\xi) =\frac{n_\mu n_\nu}{2(N_c^2-1) x(n\cdot
                                       p)}\int\frac{dy}{2\pi}e^{-ix(n\cdot p)y}\sum_s{\rm Tr}\left[\left\langle
      (1-\xi)p,s\left|F_{a}^ {\mu j}\left(\frac{yn}2\right)
      F_{a}^{\nu
             j}\left(-\frac{yn}2\right)\right|(1+\xi)p,s\right\rangle_g\right]\,,
\label{eq:gluon-in-gluonGPD}
\end{equation}
where we have made the helicity index $s$ explicit in the states and
indicated with the subscript $g$ that the states refer to external
gluons. A second possibility is to bracket the gluon operator between
quark states, which defines the gluon-in-quark GPD:
\begin{equation}
\hat{F}_{g/q}(x,\xi) = \frac{n_\mu n_\nu}{2N_c x(n\cdot p)}\int\frac{dy}{2\pi}e^{-ix(n\cdot p)y}\sum_s{\rm Tr}\left[\left\langle
      (1-\xi)p,s\left|F_{a}^{\mu j}\left(\frac{yn}2\right)
      F_{a}^{\nu j}\left(-\frac{yn}2\right)\right|(1+\xi)p,s\right\rangle_q\right]\,,
\label{eq:gluon-in-quarkGPD}
\end{equation}
with the subscript $q$ denoting external quark states and the index
$s$ this time referring to the quark spin state.

\begin{figure}[t]
  \begin{centering}
    \includegraphics[width=0.9\textwidth]{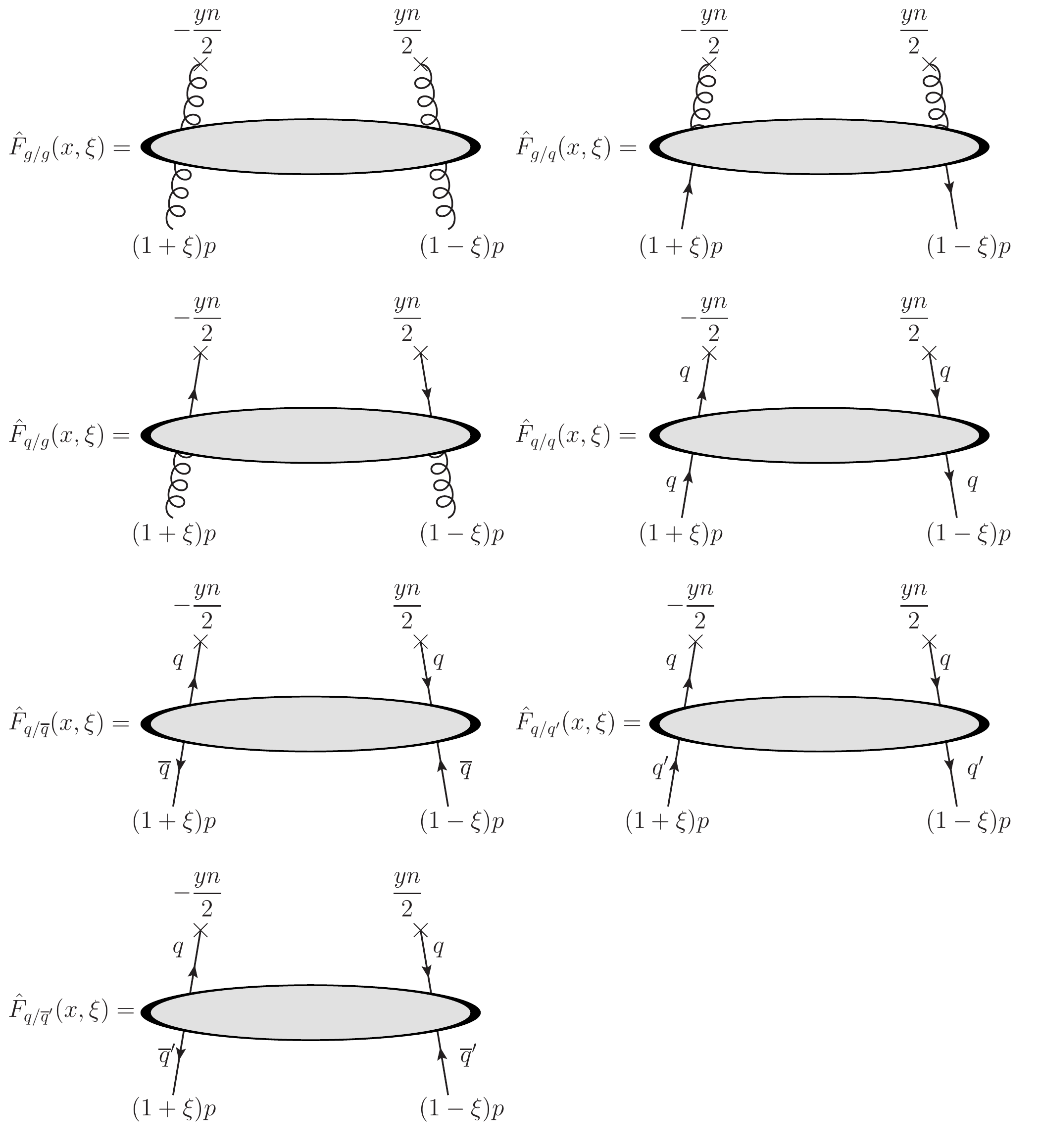}
    \vspace{15pt}
    \caption{Graphical representation of the parton-in-parton GPDs
      defined in
      Eqs.~(\ref{eq:gluon-in-gluonGPD})-(\ref{eq:quark-in-quarkS2GPD}).\label{fig:AllGPDs}}
  \end{centering}
\end{figure}

Now we move to considering the quark operator. This can be bracketed
between gluon states giving the quark-in-gluon GPD:
\begin{equation}
\hat{F}_{q/g}(x,\xi)
  =\frac{1}{2(N_c^2-1)}\int\frac{dy}{2\pi}e^{-ix(n\cdot p)y}\sum_s{\rm Tr}\left[\left\langle
      (1-\xi)p,s\left|\overline{\psi}_q\left(\frac{yn}2\right)\frac{\slashed{n}}2
      \psi_q\left(-\frac{yn}2\right)
    \right|(1+\xi)p,s\right\rangle_g\right]\,.
\label{eq:quark-in-gluonGPD}
\end{equation}
The quark operator for a specific flavour (or antiflavour) $q$ can
finally be bracketed between four different quark states:
\begin{itemize}
\item states of exactly the same flavour $q$ and charge-conjugation
  quantum number:
\begin{equation}
\hat{F}_{q/q}(x,\xi)
  =\frac{1}{2N_c}\int\frac{dy}{2\pi}e^{-ix(n\cdot p)y}\sum_s{\rm Tr}\left[\left\langle
      (1-\xi)p,s\left|\overline{\psi}_q\left(\frac{yn}2\right)\frac{\slashed{n}}2
      \psi_q\left(-\frac{yn}2\right)
    \right|(1+\xi)p,s\right\rangle_q\right]\,,
\label{eq:quark-in-quarkV1GPD}
\end{equation}
\item states of the same flavour $q$ but opposite charge-conjugation
  quantum number:
\begin{equation}
\hat{F}_{q/\overline{q}}(x,\xi)
  =\frac{1}{2N_c}\int\frac{dy}{2\pi}e^{-ix(n\cdot p)y}\sum_s{\rm Tr}\left[\left\langle
      (1-\xi)p,s\left|\overline{\psi}_q\left(\frac{yn}2\right)\frac{\slashed{n}}2
      \psi_q\left(-\frac{yn}2\right)
    \right|(1+\xi)p,s\right\rangle_{\overline{q}}\right]\,,
\label{eq:quark-in-quarkV2GPD}
\end{equation}
\item states of different flavour $q'$ but same charge-conjugation
  quantum number:
\begin{equation}
\hat{F}_{q/q'}(x,\xi)
  =\frac{1}{2N_c}\int\frac{dy}{2\pi}e^{-ix(n\cdot p)y}\sum_s{\rm Tr}\left[\left\langle
      (1-\xi)p,s\left|\overline{\psi}_q\left(\frac{yn}2\right)\frac{\slashed{n}}2
      \psi_q\left(-\frac{yn}2\right)
    \right|(1+\xi)p,s\right\rangle_{q'}\right]\,,
\label{eq:quark-in-quarkS1GPD}
\end{equation}
\item states of different flavour $q'$ and opposite charge-conjugation
  quantum number:
\begin{equation}
\hat{F}_{q/\overline{q}'}(x,\xi)
  =\frac{1}{2N_c}\int\frac{dy}{2\pi}e^{-ix(n\cdot p)y}\sum_s{\rm Tr}\left[\left\langle
      (1-\xi)p,s\left|\overline{\psi}_q\left(\frac{yn}2\right)\frac{\slashed{n}}2
      \psi_q\left(-\frac{yn}2\right)
    \right|(1+\xi)p,s\right\rangle_{\overline{q}'}\right]\,.
\label{eq:quark-in-quarkS2GPD}
\end{equation}
\end{itemize}
A graphical representation of the seven parton-in-parton GDPs listed
above is given in Fig.~\ref{fig:AllGPDs}.

It should be stressed that in general the quark and gluon fields in
Eqs.~(\ref{eq:gluon-in-gluonGPD})-(\ref{eq:quark-in-quarkS2GPD}) are
interacting fields in the sense that they can radiate and absorb
partons, possibly changing species, before interacting with the
external asymptotic states. In this way, these definitions can be used
to compute perturbative corrections in $\alpha_s$ to the anomalous
dimensions by considering diagrams with additional radiation. For any
given GPD, non-vanishing diagrams are those that have the appropriate
external free fields to annihilate the asymptotic states according to:
\begin{equation}
\psi_q^{(0)}(x) |k,s\rangle_q = e^{-ik\cdot x} u_{q,s}(k)
|0\rangle\quad\mbox{and}\quad \psi_q^{(0)}(x) |k,s\rangle_{\overline{q}} = e^{ik\cdot x} v_{q,s}(k)
|0\rangle\,,
\label{eq:asymptoticstatesq}
\end{equation}
for quarks and:
\begin{equation}
  \quad A_a^{(0),j}(x) |k,s\rangle_g =
  e^{-ik\cdot x}e_{a,s}^{j}(k)|0\rangle,
\label{eq:asymptoticstatesg}
\end{equation}
for gluons. Here $u_{q,s}(k)$ $(v_{q,s}(k))$ is the quark (antiquark)
spinor for the flavour $q$ of momentum $k$ and spin $s$, and
$e_{a,s}^{\alpha}(k)$ is the gluon polarisation vector of momentum
$k$, colour index $a$, and helicity $s$.  $\psi_q^{(0)}$ and
$A_a^{(0),\alpha}$ are respectively the quark and gluon free fields
that can be regarded as the asymptote of the original interacting
fields after radiation.

A detail worth discussing is the fact that the gluon field $A_a^{j}$
always appears through $n_\mu F_a^{\mu j}$. As discussed in
Sect.~\ref{sec:evolutionequations}, the light-cone gauge considerably
simplifies the form of this combination that reduces to
$n_\mu F_a^{\mu j}(x)=\left(n\cdot\partial\right)A_a^j(x)$. When this
operator is acting on a partonic state with plus momentum
$(1\pm\xi)p^+$, since it appears in a Fourier transform, the
derivative can be traded for a factor
$(1\pm\xi)p^+-k^+=i(x\pm \xi)p^+$. This finally allows us to write the
gluon-in-gluon and gluon-in-quark GPDs in terms of the gluon field
rather than in terms of the field strength as follows:
\begin{equation}
  \displaystyle \hat{F}_{g/g}(x,\xi) = \frac{(n\cdot p)(x^2-\xi^2)}{2(N_c^2-1)x}\int\frac{dy}{2\pi}e^{-ix(n\cdot p)y}\sum_s{\rm Tr}\left[\left\langle
      (1-\xi)p,s\left|A_{a}^ { j}\left(\frac{yn}2\right)
        A_{a}^{j}\left(-\frac{yn}2\right)\right|(1+\xi)p,s\right\rangle_g\right]\,,
  \label{eq:gluon-in-gluonGPD2}
\end{equation}
and:
\begin{equation}
  \hat{F}_{g/q}(x,\xi) = \frac{(n\cdot p) (x^2-\xi^2)}{2N_cx}\int\frac{dy}{2\pi}e^{-ix(n\cdot p)y}\sum_s{\rm Tr}\left[\left\langle
      (1-\xi)p,s\left|A_{a}^{ j}\left(\frac{yn}2\right)
        A_{a}^{j}\left(-\frac{yn}2\right)\right|(1+\xi)p,s\right\rangle_q\right]\,.
\end{equation}

\begin{figure}[t]
  \begin{centering}
    \includegraphics[width=0.45\textwidth]{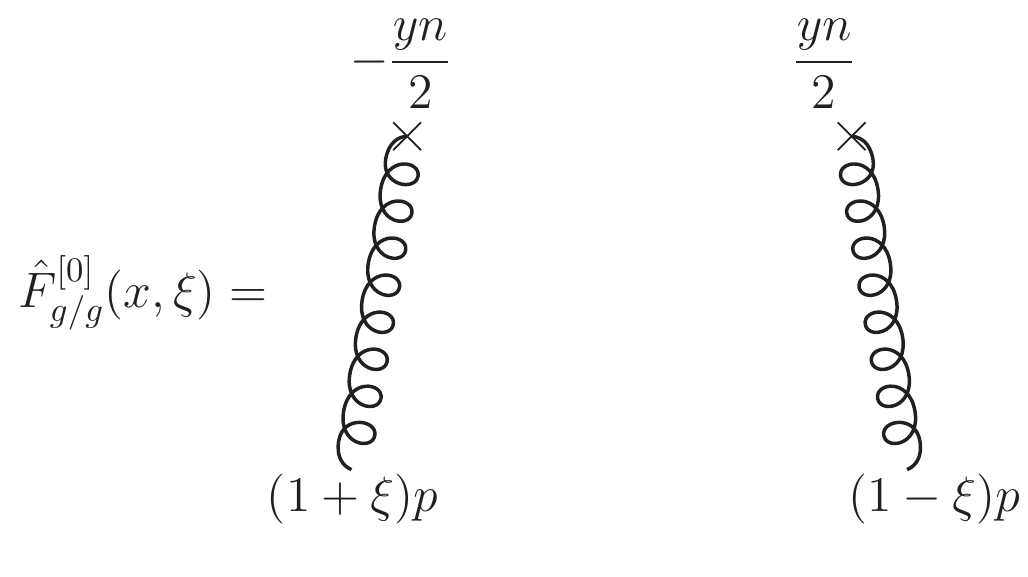}
    \includegraphics[width=0.45\textwidth]{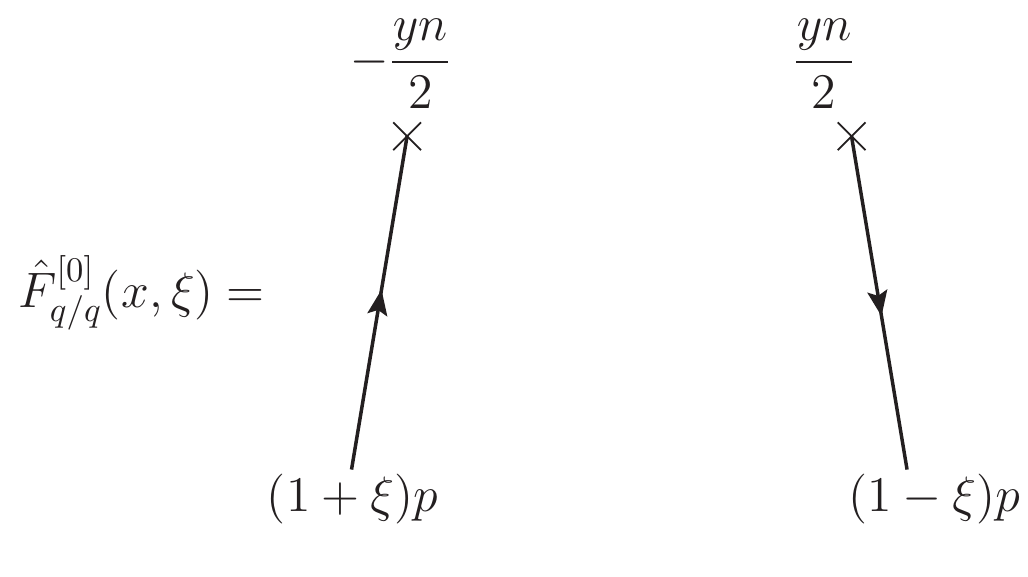}
    \vspace{15pt}
    \caption{Tree-level graphs contributing to the gluon-in-gluon GPD
      $\hat{F}_{g/g}$, Eq.~(\ref{eq:gluon-in-gluonGPD}), (left) and to
      the fully diagonal quark-in-quark GPD $\hat{F}_{q/q}$,
      Eq.~(\ref{eq:quark-in-quarkV1GPD}) (right).\label{fig:LOGPD}}
  \end{centering}
\end{figure}

Using Eqs.~(\ref{eq:asymptoticstatesq})-(\ref{eq:asymptoticstatesg})
and the orthogonality relations for quark spinors:
\begin{equation}\label{eq:spinororth}
\sum_s u_{q,s}((1+\xi)p)\overline{u}_{q,s}((1-\xi)p)=\sum_s v_{q,s}((1+\xi)p)\overline{v}_{q,s}((1-\xi)p) = \sqrt{1-\xi^2}\,\slashed{p}\,,
\end{equation}
and gluon polarisation vectors:
\begin{equation}
\sum_s e_{a,s}^j((1-\xi)p) e_{a,s}^{ j}((1+\xi)p) = -2\,,
\end{equation}
the computation of parton-in-parton GPDs reduces to integrals of this
form:
\begin{equation}
  \sqrt{1-\xi^2}\int\frac{dy}{2\pi}e^{i (1\mp x) y p\cdot n} {\rm Tr}\left[ \dots
    \mathbb{I}_c \dots\slashed{p}\right]\,,
      \label{eq:quarkrad2}
\end{equation}
for quark (minus signe) and antiquark external states (plus sign) and
to:
\begin{equation}
  -2\int\frac{dy}{2\pi}e^{i (1-x) y p\cdot n} {\rm Tr}_c\left[
    \dots\mathbb{I}_c\dots\right]\,,
  \label{eq:gluonrad2}
\end{equation}
for gluon external states. Therefore, given a specific diagram, one
just needs to replace the ellipses using standard QCD Feynman rules in
light-cone gauge. This allows parton-in-parton GPDs to have the
following perturbative expansion:
\begin{equation}
  \hat{F}_{i/j}(x,\xi) = \sum_{n=0}^{\infty}\left(\frac{\alpha_s}{4\pi}\right)^n \hat{F}_{i/j}^{[n]}(x,\xi)\,.\\
\end{equation}
At $\mathcal{O}(\alpha_s^0)$, where no additional radiation is
allowed, only the gluon-in-gluon GPD $\hat{F}_{g/g}$,
Eq.~(\ref{eq:gluon-in-gluonGPD2}), and the fully diagonal
quark-in-quark GPD $\hat{F}_{q/q}$,
Eq.~(\ref{eq:quark-in-quarkV1GPD}), are different from zero. The
corresponding Feynman diagrams are shown in Fig.~\ref{fig:LOGPD}. The
explicit computation can be done using Eqs.~(\ref{eq:quarkrad2})
and~(\ref{eq:gluonrad2}) by simply removing the ellipses and inserting
in the quark case the operator $\slashed{n}/2$. This yields:
\begin{equation}
  \begin{array}{l}
    \hat{F}_{q/q}^{[0]}(x,\xi)=\sqrt{1-\xi^2}\,\delta(1-x)\,,\\
    \\
    \hat{F}_{g/g}^{[0]}(x,\xi)=(1-\xi^2)\delta(1-x)\,,
  \end{array}
  \label{eq:LOGPDs}
\end{equation}
that compared to Eq.~(\ref{eq:normpartoninpartonGPDs}) allows us to
find that $D_q(\xi)=\sqrt{1-\xi^2}$ and $D_g(\xi)=1-\xi^2$. It should
be noted that this result, that derives from the calculation of the
disconnected diagrams in Fig.~(\ref{fig:LOGPD}), relies on imposing
the conservation of the momentum injected into the operator-insertion
vertices (see Fig.~\ref{fig:HadronGPDs}). In other words, the momentum
that flows into the vertices equals the external momentum.\footnote{We
  thank the referee for suggesting us to clarify this point.}

At $\mathcal{O}(\alpha_s)$, the interacting fields radiate one
additional parton before interacting with the external states. This
allows also $\hat{F}_{g/q}$, Eq.~(\ref{eq:gluon-in-quarkGPD}), and
$\hat{F}_{q/g}$, Eq.~(\ref{eq:quark-in-gluonGPD}) to be different from
zero while the remaining quark-in-quark GPDs
(\ref{eq:quark-in-quarkV2GPD})-(\ref{eq:quark-in-quarkS2GPD}) get
their first contribution at higher orders. Contrary to the tree-level
calculations, loop corrections to the parton-in-parton GPDs are
divergent. It is the renormalisation of these divergences that defines
the anomalous dimensions responsible for the evolution of GPDs. The
seven anomalous dimensions obtained from
Eqs.~(\ref{eq:gluon-in-gluonGPD})-(\ref{eq:quark-in-quarkS2GPD}) are
usually arranged in seven specific combinations that are convenient
for the implementation of the evolution equations. Using the same
indexing as for GPDs, there are three non-singlet anomalous
dimensions, defined as:
\begin{equation}
\begin{array}{l}
\mathcal{P}_{\pm}^{-}=\left(\mathcal{P}_{q/q}-\mathcal{P}_{q/q'}\right)\pm
  \left(\mathcal{P}_{q/\overline{q}}-\mathcal{P}_{q/\overline{q}'}\right)\,,\\
\\
\mathcal{P}_V^{-}=\mathcal{P}_{-}^-+n_f(\mathcal{P}_{q/q'}-
  \mathcal{P}_{q/\overline{q}'})\,,
\end{array}
\end{equation}
and four singlet anomalous dimensions:
\begin{equation}
\begin{array}{l}
  \mathcal{P}_{qq}^+=\mathcal{P}_{+}^-+n_f(\mathcal{P}_{q/q'}+
  \mathcal{P}_{q/\overline{q}'})\,,\\
  \\
  \mathcal{P}_{qg}^+=2n_f\mathcal{P}_{q/g}\,,\\
  \\
  \mathcal{P}_{gq}^+=\mathcal{P}_{g/q}\,,\\
  \\
  \mathcal{P}_{gg}^+=\mathcal{P}_{g/g}\,.
\end{array}
\end{equation}
As mentioned above, at one loop one finds
$\mathcal{P}_{q/\overline{q}}=\mathcal{P}_{q/q'}=\mathcal{P}_{q/\overline{q}'}=0$
that in turn implies
$\mathcal{P}_{+}^-=\mathcal{P}_{-}^-=\mathcal{P}_V^-=\mathcal{P}_{qq}^+=\mathcal{P}_{q/q}$.

\section{One-loop quark-in-quark anomalous
  dimension}\label{app:splittingfunctions}

In this appendix, we present the details of the calculation of the
one-loop anomalous dimensions in the $\overline{\mbox{MS}}$
renormalisation scheme using the light-cone gauge. As discussed in
Sect.~\ref{sec:evolutionequations}, the anomalous dimensions can be
determined by extracting the pole part of appropriately defined
parton-in-parton GPDs that can be computed in perturbation theory. In
App.~\ref{app:gpdnormlisation} we have introduced the parton-in-parton
GPDs and carried out the tree-level computation. For illustrative
purposes, here we consider the one-loop correction to the
quark-in-quark GPD $\hat{F}_{q/q}^{[1]}$ and using
Eq.~(\ref{eq:splittingfromGPD}) we immediately obtain the one-loop
quark-in-quark anomalous dimension $\mathcal{P}_{q/q}^{[0]}$. The
remaining one-loop anomalous dimensions can be extracted in an
analogous way by simply considering the appropriare parton-in-parton
GPDs.

\begin{figure}[h]
  \begin{centering}
    \includegraphics[width=0.5\textwidth]{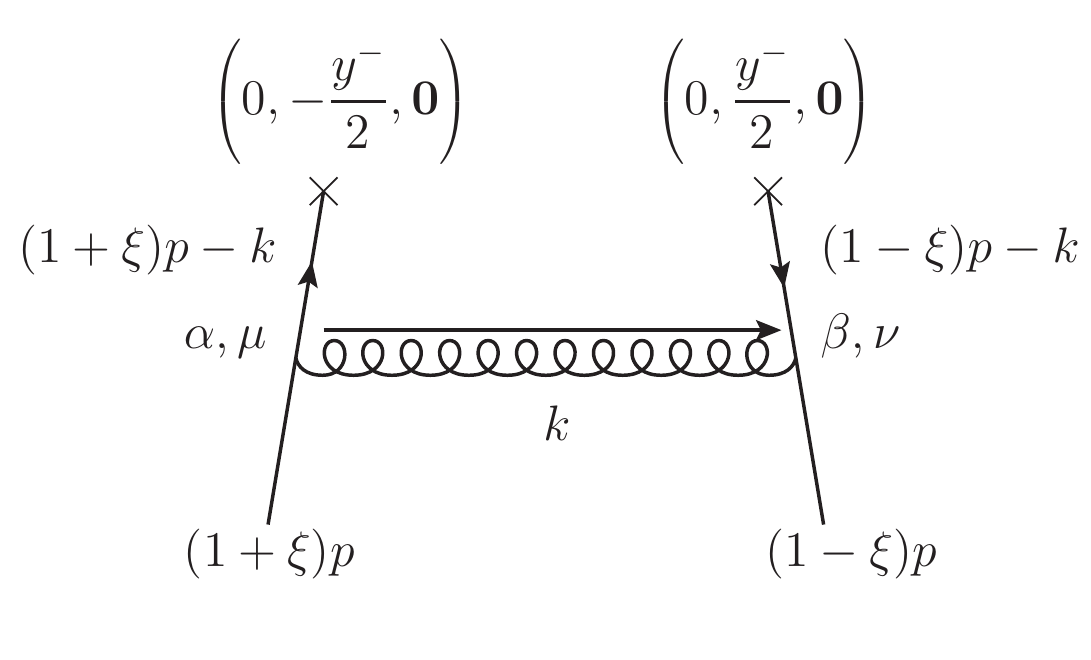}
    \vspace{15pt}
    \caption{Real graph contributing to the quark-in-quark GPD at one loop.\label{fig:NLOGPDc}}
  \end{centering}
\end{figure}

The advantage of using the light-cone gauge is a reduction of the
number of diagrams to be considered. Specifically,
$\hat{F}_{q/q}^{[1]}$ results from the computation of one single
diagram displayed in Fig.~\ref{fig:NLOGPDc}.
For the calculation, we will use a gauge vector that in light-cone
coordinates\footnote{Given a four-vector $v^\mu=(t,x,y,z)$, its
  light-cone-coordinate representation is
  $v^\mu=(v^+,v^-,\mathbf{v}_T)$ with $v^{\pm}=(t\pm z)/\sqrt{2}$ and
  $\mathbf{v}_T=(x,y)$.} reads $n^\mu=(0,1,\mathbf{0}_T)$, such that
the scalar product of any vector $v$ with $n$ gives $n\cdot
v=v^+$. Using the definition in Eq.~(\ref{eq:quark-in-quarkV1GPD}) and
the manipulation in Eq.~(\ref{eq:quarkrad2}), we obtain:
\begin{equation}
\frac{g^2}{16\pi^2}\hat{F}_{q/q}^{[1]}(x,\xi) =
\frac{\sqrt{1-\xi^2}}{2N_c}\int\frac{dy}{2\pi}
e^{i(1-x)yp^+}\mbox{Tr}\left[\mathbb{I}_cR^{(1)}(y,\xi)\slashed{p}\right]\,,
\label{eq:FqqR}
\end{equation}
with:
\begin{equation}
\begin{array}{rcl}
  R^{(1)}(y,\xi) &=&\displaystyle
                     \int\frac{d^{4-2\epsilon}k}{(2\pi)^{4-2\epsilon}}e^{-ik^+y}\delta_{\alpha\beta}i\mathcal{D}_{\mu\nu}(k)\\
  \\
                 &\times&\displaystyle (-ig\mu^\epsilon\gamma^\mu
                          t_\alpha)\frac{i((1+\xi)\slashed{p}-\slashed{k})}{((1+\xi)p-k)^2+i\varepsilon}\frac{\gamma^+}{2} \frac{-i((1-\xi)\slashed{p}-\slashed{k})}{((1-\xi)p-k)^2+i\varepsilon}(ig\mu^\epsilon\gamma^\nu t_\beta) \,,
\end{array}
\end{equation}
and where $\mathcal{D}_{\mu\nu}$ is defined in
Eq.~(\ref{eq:lightconepropagator}) and $t_\alpha$ are the SU(3)
generators. Expressing the integration measure in light-cone
coordinates:
\begin{equation}
  d^{4-2\epsilon}k = dk^+dk^-d^{2-2\epsilon}\mathbf{k}_T\,,
\end{equation}
the integral reduces to:
\begin{equation}
\begin{array}{rcl}
  R^{(1)}(y,\xi) &=& \displaystyle \frac{g^2}{16\pi^2}2i  t_\alpha
                     t_\alpha\mu^{2\epsilon}\int\frac{d^{2-2\epsilon}\mathbf{k}_T}{(2\pi)^{2-2\epsilon}}dk^+dk^-e^{-ik^+y}\mathcal{D}_{\mu\nu}(k)\\
  \\
                 &\times&\displaystyle \frac{\gamma^\mu[(1+\xi)\slashed{p}-\slashed{k}]\slashed{n} [(1-\xi)\slashed{p}-\slashed{k}]\gamma^\nu}{[((1+\xi)p-k)^2+i\varepsilon][((1-\xi)p-k)^2+i\varepsilon]}\,.
\end{array}
\label{eq:lightconecoordinatesintegral}
\end{equation}
The gluon propagator $\mathcal{D}_{\mu\nu}$ has two components (see
Eq.~(\ref{eq:lightconepropagator})): one proportional to the metric
tensor $g_{\mu\nu}$ and one to the gauge vector $n_\mu$. It is thus
convenient to split the integral into two components that we
respectively denote with the superscripts $(g)$ and $(n)$:
\begin{equation}
\hat{F}_{q/q}^{[1]}(x,\xi) = \hat{F}_{q/q}^{[1],(g)}(x,\xi) +
\hat{F}_{q/q}^{[1],(n)}(x,\xi)\,.
\label{eq:Fsplit}
\end{equation}
Integrating over $k^+$, using the identity
$\mbox{Tr}_c[t_\alpha t_\alpha]=N_cC_F$, plus some additional
manipulations, we obtain:
\begin{equation}
\begin{array}{rcl}
  \hat{F}_{q/q}^{[1],(g)}(x,\xi) &=&\displaystyle 
                                     \frac{iC_F \sqrt{1-\xi^2}}{(p^+)^2(1-x)(x^2-\xi^2)}\mu^{2\epsilon}\int\frac{d^{2-2\epsilon}\mathbf{k}_T}{(2\pi)^{2-2\epsilon}}\mathbf{k}_T^2\,I(\mathbf{k}_T^2)\,,\\
  \\
  \hat{F}_{q/q}^{[1],(n)}(x,\xi) &=&\displaystyle \frac{2x}{1-x} \hat{F}_{q/q}^{[1],(g)}(x,\xi) + \frac{4i  C_F \sqrt{1-\xi^2}}{p^+
                                     (1-x)^2}\mu^{2\epsilon}\int\frac{d^{2-2\epsilon}\mathbf{k}_T}{(2\pi)^{2-2\epsilon}}\,J(\mathbf{k}_T^2)\,,
\end{array}
\label{eq:Fsingleterms}
\end{equation}
where:
\begin{equation}
\begin{array}{rcl}
I(\mathbf{k}_T^2)&=&\displaystyle \int_{-\infty}^{+\infty}\frac{dk^-}{(k^-- k_1^-)
                     (k^--k_2^-)(k^--k_3^-)}\,,\\
  \\
  J(\mathbf{k}_T^2)&=&\displaystyle \int_{-\infty}^{+\infty}\frac{k^-dk^-}{(k^-- k_1^-)
  (k^--k_2^-)(k^--k_3^-)}\,,
\end{array}
\label{eq:GPDintegralsIJ}
\end{equation}
with:
\begin{equation}
  k_1^- = \frac{\mathbf{k}_T^2}{2 (1-x)p^+ }-i(1-x)\varepsilon\,,\quad k_2^- =
  - \frac{\mathbf{k}_T^2}{2(x +\xi) p^+}+i (x+\xi)\varepsilon\,,\quad k_3^- =
  - \frac{\mathbf{k}_T^2}{2(x -\xi) p^+}+i(x-\xi)\varepsilon\,.
\label{eq:poleskminus}
\end{equation}
In defining $k_2^-$ and $k_3^-$, we have multiplied the term
$i\varepsilon$ coming from the quark propagators (see \textit{e.g.}
Eq.~(\ref{eq:lightconecoordinatesintegral})) by $(x\pm\xi)$ to account
for the correct sign of these terms. This derives from the fact that,
precisely like the finite term $\mathbf{k}_T^2$, also the
infinitesimal contribution $i\varepsilon$ gets a factor
$1/[2(x\pm\xi)p^+]$. Since we are only interested in the position of
the pole w.r.t. the integration path, \textit{i.e.} the real axis, we
only need to know the sign of the factor
$1/[2(x\pm\xi)p^+]$. Considering that $p^+$ is positive, in the limit
$\epsilon\rightarrow 0^+$, this is equivalent to multiplying
$i\varepsilon$ by $(x\pm\xi)$, hence the definitions in
Eq.~(\ref{eq:poleskminus}). This is crucial to determine the pole
configuration of the integrand in $k^-$ as a function of the relative
position of $x$ and $\xi$. In addition, for the same reason, we have
multiplied the $i\varepsilon$ term of $k_1^-$ by $(1-x)$.

In order to compute these integrals, we need to consider different
configurations depending on the position of the poles relative to the
real axis. We close the integration path upwards in such a way that
the it runs anticlockwise and all the residues get a factor $+2\pi
i$. We start by assuming $-\xi<x<1$. In this configuration the
relevant cases are:
\begin{itemize}
\item $x>\xi$: In this case the position of the poles is shown in the
  left plot of Fig.~\ref{fig:kminus}. The contour picks up the poles
  in $k_2^-$ and $k_3^-$, producing:
  \begin{equation}
    \begin{array}{l}
      \displaystyle I (\mathbf{k}_T^2) \mathop{=}^{x>\xi} \frac{2\pi
      i}{k_2^--k_3^-}\times\left[\frac{1}{k_2^--k_1^-}-\frac{1}{k_3^--k_1^-}\right]\,,\\
      \\
      \displaystyle J (\mathbf{k}_T^2) \mathop{=}^{x>\xi} \frac{2\pi
      i}{k_2^--k_3^-}\times\left[\frac{k_2^-}{k_2^--k_1^-}-\frac{k_3^-}{k_3^--k_1^-}\right]\,.
    \end{array}
  \end{equation}
\item $x<\xi$: The poles are placed as shown in the right plot of
  Fig.~\ref{fig:kminus}. The poles in $k_1^-$ and $k_3^-$ are now
  external and only the pole in $k_2^-$ contributes, giving:
  \begin{equation}
    \begin{array}{l}
      \displaystyle
    I (\mathbf{k}_T^2) \mathop{=}^{x<\xi} \frac{2\pi
      i}{k_2^--k_3^-}\times\frac{1}{k_2^--k_1^-}\,,\\
\\
      \displaystyle J (\mathbf{k}_T^2) \mathop{=}^{x<\xi} \frac{2\pi i}{k_2^--k_3^-}\times\frac{k_2^-}{k_2^--k_1^-}\,.
    \end{array}
  \end{equation}
  \begin{figure}[h]
  \begin{centering}
    \includegraphics[width=0.49\textwidth]{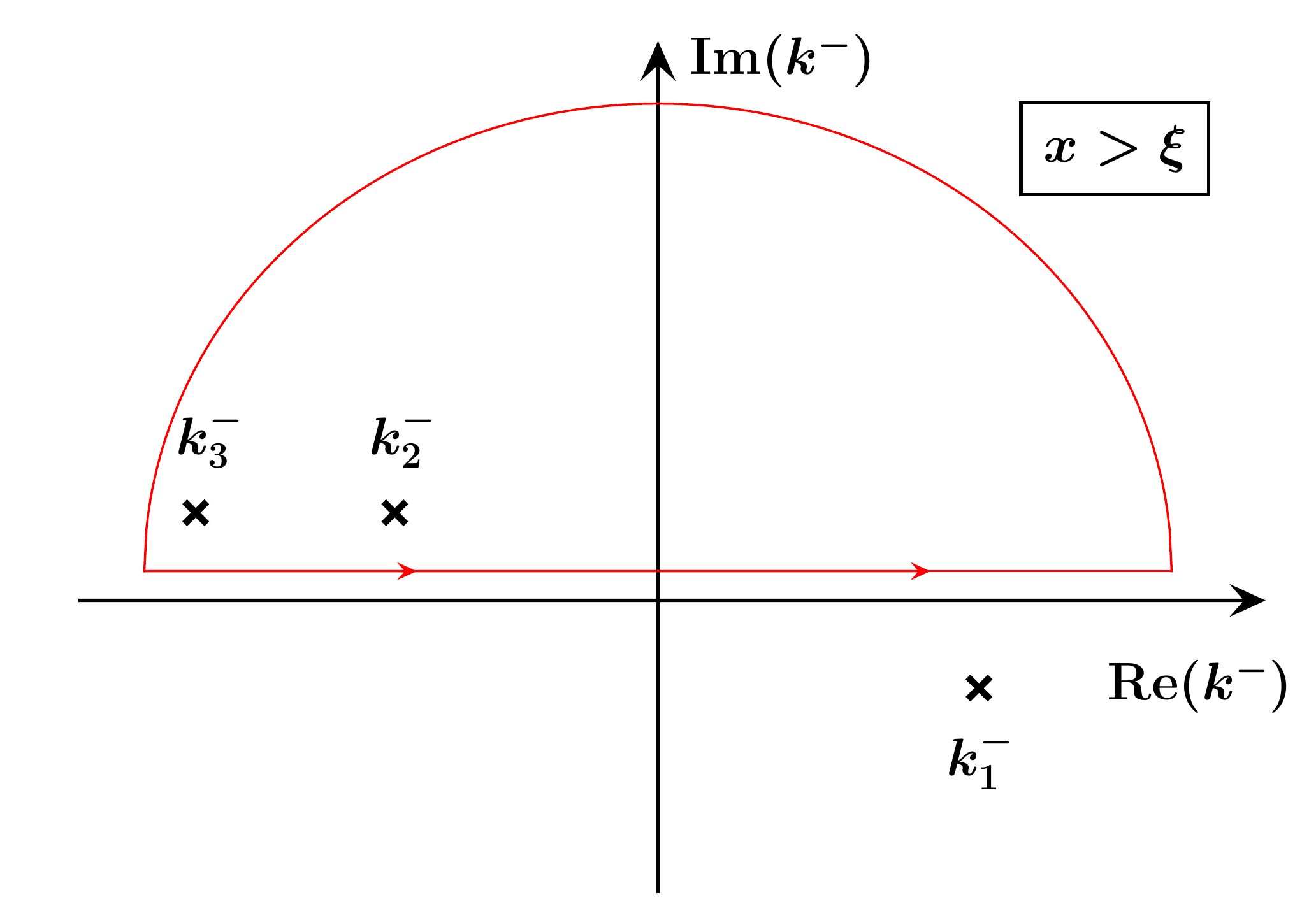}
    \includegraphics[width=0.49\textwidth]{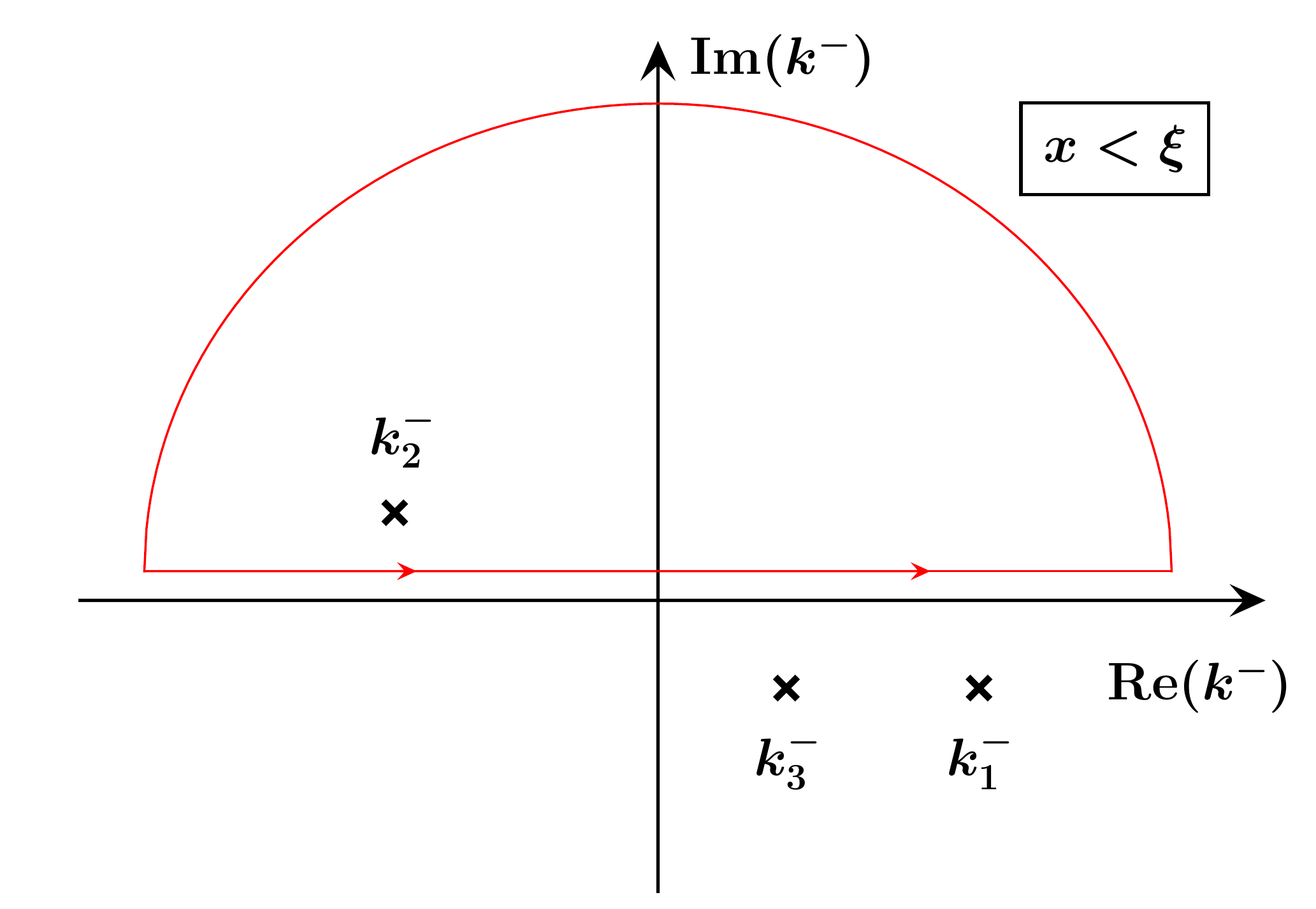}
    \vspace{15pt}
    \caption{Position of the poles in the complex plane defined by
      $k^-$ in the integrals in Eq.~(\ref{eq:GPDintegralsIJ}). The red
      arrows indicate the integration path.\label{fig:kminus}}
  \end{centering}
\end{figure}
\end{itemize}
The net result is that the integrals $I$ and $J$ can generally be
written as:
\begin{equation}
  \begin{array}{rcl}
    I (\mathbf{k}_T^2)&=&\displaystyle  \frac{2\pi
                          i}{k_2^--k_3^-}\left[\frac{1}{k_2^--k_1^-}-\theta(x-\xi)\frac{1}{k_3^--k_1^-}\right]\\
    \\
                      &=&\displaystyle -\frac{4\pi
                          i
                          (p^+)^2(1-x)(x^2-\xi^2)}{\xi\mathbf{k}_T^4}\left[\frac{x+\xi}{1+\xi}-\theta(x-\xi)
                          \frac{x-\xi}{1-\xi}\right]\,,\\
    \\
J (\mathbf{k}_T^2)&=&\displaystyle \frac{2\pi i}{k_2^--k_3^-}\left
                      [\frac{ k_2^- }{k_2^--k_1^-}-\theta(x-\xi)
                      \frac{k_3^-}{k_3^--k_1^-}\right]\\
    \\
    &=&\displaystyle\frac{2\pi i p^+(1-x)(x^2-\xi^2)}{\xi\mathbf{k}_T^2}\left[\frac{1}{1+\xi}-\theta(x-\xi) \frac{1}{1-\xi}\right]\,.
  \end{array}
  \label{eq:solutionIJ}
\end{equation}
These results have been obtained under the assumption $-\xi<x<1$. If
$x>1$, the pole in $k_1^-$ moves into the upper half of the complex
plane in $k^-$. In this configuration, the case $x>\xi$ in the
l.h.s. of Fig.~\ref{fig:kminus} produces a vanishing result because
all poles lie above the integration path that can then be closed
downwards were there are no poles. In addition, (assuming $\xi<1$) the
conditions $x>1$ and $x<\xi$ cannot be simultaneously fulfilled.
Therefore, the r.h.s. configuration of Fig.~\ref{fig:kminus} is ruled
out. In conclusion, the results in Eq.~(\ref{eq:solutionIJ})
effectively multiply $\theta(1-x)$. If $x<-\xi$, the pole in $k_2^-$
moves into the lower half of the complex plane. In this way, for
$x<\xi$, all poles are below the integration path thus yielding again
a vanishing result, while the configuration $x>\xi$ is ruled
out. Therefore, the results above also multiply a factor
$\theta(x+\xi)$. This allows us to recast Eq.~(\ref{eq:solutionIJ}) as
follows:
\begin{equation}
  \begin{array}{rcl}
  I (\mathbf{k}_T^2)&=&\displaystyle  -\frac{4\pi
                          i
                          (p^+)^2(1-x)(x^2-\xi^2)}{\xi\mathbf{k}_T^4}\theta(1-x)\left[\theta(x+\xi)\frac{x+\xi}{1+\xi}-\theta(x-\xi)
                        \frac{x-\xi}{1-\xi}\right]\,,\\
    \\
  J (\mathbf{k}_T^2) &=&\displaystyle\frac{2\pi i
    p^+(1-x)(x^2-\xi^2)}{\xi\mathbf{k}_T^2}\theta(1-x)\left[\theta(x+\xi)\frac{1}{1+\xi}-\theta(x-\xi)
    \frac{1}{1-\xi}\right]\,.
    \end{array}
\end{equation}
We can finally put everything together using Eq.~(\ref{eq:Fsplit}) and
Eq.~(\ref{eq:Fsingleterms}) to obtain the one-loop real correction to
the bare quark-in-quark GPD:
\begin{equation}
\begin{array}{rcl}
  \hat{F}_{q/q}^{[1]}(x,\xi) &=&
                                     \hat{F}_{q/q}^{[1],  (g)}(x,\xi)+\hat{F}_{q/q}^{[1],  (n)}(x,\xi)\\
  \\
                                 &=&\displaystyle 
                                     \frac{C_F
                                     \sqrt{1-\xi^2}\theta(1-x)}{\xi(1-x)}\\
  \\
  &\times&\displaystyle \left[\theta(x+\xi)\frac{(x+\xi)
                                     (1-x+2\xi)}{1+\xi}-\theta(x-\xi) \frac{(x-\xi)
                                     (1-x-2\xi)}{1-\xi}\right]\mu^{2\epsilon}S_{\epsilon}\int\frac{dk_T^2}{k_T^{2+2\epsilon}}\,,
     \label{eq:FullDiagramGPD}
\end{array}
\end{equation}
where for the $(2-2\epsilon)$-dimensional integral in $\mathbf{k}_T$,
we have used the identity:
\begin{equation}
\int\frac{d^{2-2\epsilon}\mathbf{k}_T}{(2\pi)^{2-2\epsilon}}\frac1{\mathbf{k}_T^2}
=\frac{S_\epsilon}{4\pi}\int_0^\infty
\frac{dk_T^2}{k_T^{2+2\epsilon}}\,,
\end{equation}
with $S_\epsilon$ given in Eq.~(\ref{eq:Sepsilon}). It turns out that
the integral over $k_T$ in Eq.~(\ref{eq:FullDiagramGPD})
vanishes~\cite{Collins:1984xc}. This result can be regarded as the
consequence of the cancellation of two divergences due to the use of
dimensional regularisation to regularise both the UV divergence when
$k_T\rightarrow \infty$ and the IR divergence when $k_T\rightarrow
0$. Therefore, this integral can be interpreted as:
\begin{equation}
  \int_0^\infty
  \frac{dk_T^2}{k_T^{2+2\epsilon}}\sim\frac{1}{\epsilon_{\rm UV}}-\frac{1}{\epsilon_{\rm IR}}\,.
\end{equation}
This structure could be more clearly highlighted by regularising UV
and IR divergences differently as for example done in
Refs.~\cite{Stewart:2010qs, Echevarria:2011epo}. For infrared-safe
observables, the IR divergence cancels against an opposite divergence
produced in the calculation of the partonic cross section. Therefore,
what we are concerned with is the UV divergence that needs to be
cancelled by a renormalisation constant. To this purpose, we discard
the IR divergence and write the result of the calculation above as:
\begin{equation}
  \hat{F}_{q/q}^{[1]}(x,\xi) =
  \frac{C_F \sqrt{1-\xi^2} \theta(1-x)}{\xi(1-x)}\left[\theta(x+\xi)\frac{(x+\xi)
      (1-x+2\xi)}{1+\xi}-\theta(x-\xi) \frac{(x-\xi)
      (1-x-2\xi)}{1-\xi}\right]\frac{\mu^{2\epsilon} S_{\epsilon}}{\epsilon_{\rm
      UV}}\,.
\end{equation}

The calculation of the one-loop quark-in-quark GPD is still incomplete
because so far we have only considered the ``real'' contribution in
Fig.~\ref{fig:NLOGPDc}. We still need to include the ``virtual''
contribution. As discussed in Sect.~\ref{subsec:sumrules}, an explicit
calculation is unnecessary in that this contribution can be obtained
from the knowledge of the real one. As a matter of fact, writing:
\begin{equation}
  \hat{F}_{q/q}^{[1]}(x,\xi) \rightarrow \hat{F}_{q/q}^{[1]}(x,\xi) + A(\xi)\delta(1-x)\,,
\end{equation}
and imposing the valence sum rule gives:\footnote{Here we are also
  using the fact that $\hat{F}_{\overline{q}/q}^{[1]}(x,\xi)$ is zero
  to identify the non-singlet GPD with $\hat{F}_{q/q}^{[1]}(x,\xi)$.}
\begin{equation}
 A(\xi)=2C_F \sqrt{1-\xi^2}\left[\frac32-2\int_0^1\frac{dz}{1-z}-\ln(|1-\xi^2|)\right]\frac{\mu^{2\epsilon} S_{\epsilon}}{\epsilon_{\rm
     UV}}\,.
 \label{eq:virtualsumrule}
\end{equation}
Once the full one-loop quark-in-quark GPD has been computed, we can
use Eq.~(\ref{eq:splittingfromGPD}), with $D_q(\xi)=\sqrt{1-\xi^2}$,
to extract the anomalous dimension finally obtaining
$\mathcal{P}_{q/q}^{[0]}$ as in Eq.~(\ref{eq:polepartGPDFs}).

\begin{figure}[h]
  \begin{centering}
    \includegraphics[width=0.8\textwidth]{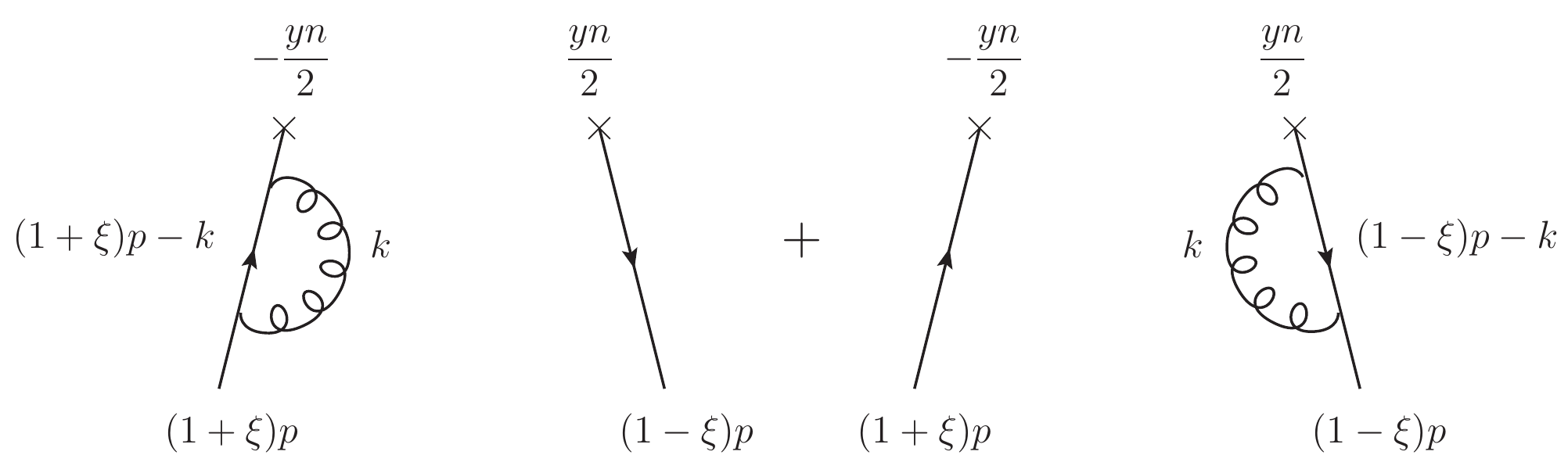}
    \vspace{15pt}
    \caption{Virtual graphs contributing to the quark-in-quark GPD at one loop.\label{fig:NLOGPDvirt}}
  \end{centering}
\end{figure}

It is however instructive to perform the calculation of the virtual
contribution to $\mathcal{P}_{q/q}^{[0]}$ to explicitly verify that,
at least at one-loop accuracy, the constraints discussed in
Sect.~\ref{subsec:sumrules} are actually
fulfilled. Fig.~\ref{fig:NLOGPDvirt} displays the relevant diagrams.

These diagrams correspond to self-energy corrections to the external
legs. As a consequence, they can be included by means of the
Lehmann-Symanzik-Zimmermann (LSZ) reduction
formula~\cite{Lehmann:1954rq}. Specifically, virtual corrections to
the one-loop quark-in-quark GPD are included by computing:
\begin{equation}
  \frac{Z_F(1+\xi)+Z_F(1-\xi)}{2}\hat{F}_{q/q}(x,\xi)\,,
\label{eq:includevirt}
\end{equation}
where $\hat{F}_{q/q}(x,\xi)$ is computed with amputated external legs
and $Z_F$ is the residue of the quark propagator. We have included a
correction for each external leg along with a factor of $1/2$ as a
consequence of the LSZ reduction formula. As we will show below, the
explicit dependence of $Z_F$ on the longitudinal momentum fractions
$1\pm\xi$ emerges from the regularisation of the $1/(nk)$ divergence
caused by the light-cone gluon propagator.

\begin{figure}[h]
  \vspace{-10pt}
  \begin{centering}
    \includegraphics[width=0.4\textwidth]{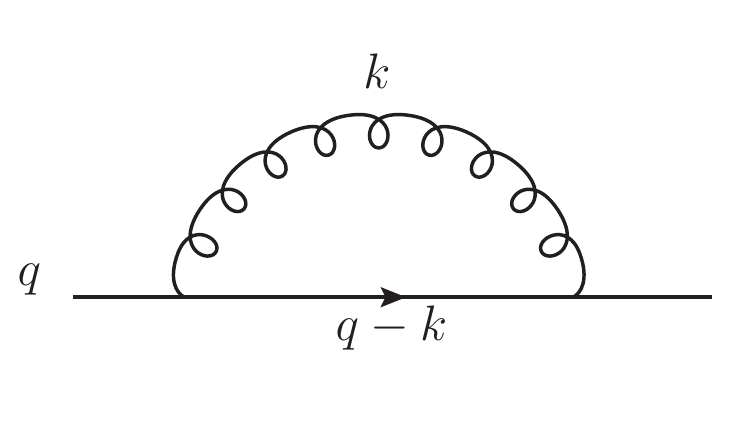}
    \caption{One-loop quark self energy.\label{fig:QuarkSelfEnergy}}
  \end{centering}
\end{figure}

In order to identify the residue $Z_F$ we follow
Ref.~\cite{Curci:1980uw}. The quark propagator in momentum space in
the vicinity of the pole behaves as follows:\footnote{Notice that
  $D_F$ is a matrix in both Dirac and colour space. However, since it
  is diagonal in colour space, we omit the corresponding indices
  implying that it multiplies the identity matrix
  $\mathbb{I}_{N_c \times N_c}$.}
\begin{equation}
D_F(q) \mathop{=}_{q^2\sim 0} \frac{i
  Z_F}{\slashed{q}}+\mbox{finite corrections}\,,
\label{eq:proppole}
\end{equation}
which effectively defines the residue $Z_F$ and where the finite
corrections are related to the quark spectral function in the
continuum region~\cite{Zwicky:2016lka}. As is well known, the 1PI
contribution to the self-energy, $\Sigma$, can be resummed to all
orders producing:
\begin{equation}
  D_F(q) = \frac{i}{\slashed{q}} +\frac{i}{\slashed{q}}(-i\Sigma(q))
  \frac{i}{\slashed{q}}+  \dots = \frac{i}{\slashed{q}-\Sigma(q)}\,.
\label{eq:propresummed}
\end{equation}
We now show how $\Sigma$ is related to the residue $Z_F$. First of
all, we observe that in light-cone gauge $\Sigma$ must have this
structure:
\begin{equation}
  \Sigma(q) = A\slashed{q}+B\slashed{n}\frac{q^2}{2(nq)}\,,
  \label{eq:selfenergydecomposition}
\end{equation}
where $A$ and $B$ are scalar coefficients. Plugging this equation into
Eq.~(\ref{eq:propresummed}), one finds:
\begin{equation}
D_F(q) =
\frac{i}{(1-A)\slashed{q}-B\slashed{n}\frac{q^2}{2(nq)}}=\frac{1}{1-A-B}\left[\frac{i}{\slashed{q}}-\frac{B}{1-A}\frac{i\slashed{n}}{2(nq)}\right]\,.
\end{equation}
By comparison with Eq.~(\ref{eq:proppole}), one immediately sees that:
\begin{equation}
Z_F = \frac{1}{1-A-B}\,.
\end{equation}
Since $\Sigma$ starts at $\mathcal{O}(\alpha_s)$:
\begin{equation}
  \Sigma(q) = \sum_{n=1}^{\infty}a_s^n \Sigma^{[n]}(q)\,,
\end{equation}
so do $A$ and $B$:
\begin{equation}
  A = \sum_{n=1}^{\infty}a_s^n A^{[n]}\,,\quad\mbox{and}\quad B = \sum_{n=1}^{\infty}a_s^n B^{[n]}\,.
\end{equation}
This implies that the perturbative expansion of the residue reads:
\begin{equation}
  Z_F = 1+a_s(A^{[1]}+B^{[1]})+\mathcal{O}(\alpha_s^2)\,.
\end{equation}
Using this equality in Eq.~(\ref{eq:includevirt}), one finds that the
contribution to the one-loop correction of the quark-in-quark GPD due
to the virtual diagrams in Fig.~\ref{fig:NLOGPDvirt} amounts to:
\begin{equation}
\hat{F}_{q/q}^{\rm virt}(x,\xi) =
\frac{[A^{[1]}(1+\xi)+B^{[1]}(1+\xi)]+[A^{[1]}(1-\xi)+B^{[1]}(1-\xi)]}{2}\sqrt{1-\xi^2}\delta(1-x)\,,
\label{eq:oneloopvirt}
\end{equation}
where we have used Eq.~(\ref{eq:LOGPDs}) for
$\hat{F}_{q/q}^{[0]}$. The values of $A^{[1]}$ and $B^{[1]}$ can be
extracted by computing the diagram in Fig.~\ref{fig:QuarkSelfEnergy}.

\noindent Using light-cone-gauge Feynman rules, this diagram evaluates
to:
\begin{equation}
\begin{array}{rcl}
\displaystyle \frac{g^2}{16\pi^2}(-i\Sigma^{[1]}(q)) &=&\displaystyle
                                                                \int\frac{d^{4-2\epsilon}k}{(2\pi)^{4-2\epsilon}} (-ig\mu^{\epsilon}\gamma^\nu
                 t_\beta)i\delta_{\beta \alpha}\mathcal{D}_{\mu\nu}(k)\frac{i(\slashed{q}-\slashed{k})}{(q-k)^2} (-ig \mu^{\epsilon}\gamma^\mu
                 t_\alpha)\\
  \\
             &\equiv&\displaystyle  \frac{g^2}{16\pi^2}C_F \frac{(4\pi^2\mu^2)^{\epsilon}}{\pi^{2}}\left[\Sigma_{F}+\Sigma_{A}\right]\,,
\end{array}
\end{equation}
where we have defined:
\begin{equation}\label{eq:SigmaF}
  \Sigma_F=
  2(1-\epsilon)\left[\slashed{q}\int \frac{d^{4-2\epsilon}k}{k^2(q-k)^2}-\gamma_\mu\int \frac{d^{4-2\epsilon}k\,k^\mu}{k^2(q-k)^2}\right]\,,
\end{equation}
and:
\begin{equation}
  \Sigma_A=
  \int d^{4-2\epsilon}k \frac{\slashed{k} (\slashed{q}-\slashed{k})\slashed{n}+\slashed{k} (\slashed{q}-\slashed{k})\slashed{n}}{(nk)k^2(q-k)^2}\,.
\end{equation}
The numerator of the integrand of $\Sigma_A$ can be rearranged as
follows:
\begin{equation}
\slashed{k} (\slashed{q}-\slashed{k})\slashed{n}+\slashed{k}
(\slashed{q}-\slashed{k})\slashed{n} = \slashed{k} \slashed{q}\slashed{n}+\slashed{k}
\slashed{q}\slashed{n} -2k^2\slashed{n} = 2(q^2-(q-k)^2)\slashed{n}-\slashed{q}\slashed{k} \slashed{n}-\slashed{q}\slashed{k}\slashed{n} \,.
\end{equation}
The term proportional to $(q-k)^2$ can be discarded because it cancels
one of the poles in the denominator. This leaves a single pole in
$k^-$ (or $k^+$) that produces a vanishing result because the
integration path can be closed in a way that it contains no
poles. Finally, we have:
\begin{equation}\label{eq:SigmaA}
  \Sigma_A=
  2 q^2\slashed{n}\int \frac{d^{4-2\epsilon}k}{(nk)k^2(q-k)^2}-(\slashed{q}\gamma_\mu \slashed{n}+\slashed{n}\gamma_\mu \slashed{q})\int \frac{d^{4-2\epsilon}k\,k^\mu}{(nk)k^2(q-k)^2}\,.
\end{equation}
In view of the use of the Feynman-parameter method to solve the
integrals in Eqs.~(\ref{eq:SigmaF}) and~(\ref{eq:SigmaA}), we have
omitted the $i\epsilon$ terms from the propagators. Denoting:
\begin{equation}
  \begin{array}{ll}
    J_F =\displaystyle  \int \frac{d^{4-2\epsilon}k}{k^2(q-k)^2}\,,
    &\quad  J_F^\mu =\displaystyle  \int \frac{d^{4-2\epsilon}k\,k^\mu}{k^2(q-k)^2}\,,\\
    \\
    J_A =\displaystyle  \int
    \frac{d^{4-2\epsilon}k}{(nk)k^2(q-k)^2}\,,&\quad 
    J_A^\mu =\displaystyle  \int \frac{d^{4-2\epsilon}k\,k^\mu}{(nk)k^2(q-k)^2}\,,
  \end{array}
\end{equation}
and using the Feynman-parameter identity:
\begin{equation}
\frac{1}{QR} = \int_0^1\frac{dx}{[xQ+(1-x)R]^2}\,,
\end{equation}
with $Q=k^2$ and $R=(q-k)^2$, allows us to recast these integrals
follows:
\begin{equation}
J_{F,A}^{(\mu)} = \int_0^1dx I_{F,A}^{(\mu)}(x)\,,
\end{equation}
with:
\begin{equation}
  \begin{array}{rcl}
    I_F(x) &=&\displaystyle  \int \frac{d^{4-2\epsilon}k}{[k^2-2(1-x)(p k)+(1-x) q^2]^2}\,,\\
    \\
    I_F^\mu(x) &=&\displaystyle  \int \frac{d^{4-2\epsilon}k\,k^\mu}{[k^2-2(1-x)(p k)+(1-x) q^2]^2}\,,\\
    \\
    I_A(x) &=&\displaystyle  \int \frac{d^{4-2\epsilon}k}{(nk) [k^2-2(1-x)(p k)+(1-x) q^2]^2}\,,\\
    \\
    I_A^\mu(x) &=&\displaystyle  \int \frac{d^{4-2\epsilon}k\,k^\mu}{(nk) [k^2-2(1-x)(p k)+(1-x) q^2]^2}\,.
  \end{array}
\end{equation}
These integrals can finally be computed using, for example,
Eqs.~(A.1), (A.2), (A.6) and (A.7) of
Ref.~\cite{Pritchard:1978ts}. The result is:
\begin{equation}
  \begin{array}{rcl}
    I_F(x) &=&\displaystyle  i\pi^{2-\epsilon}e^{\epsilon i\pi}\frac{\Gamma(\epsilon)}{[x(1-x) q^2]^\epsilon}\,,\\
    \\
    I_F^\mu(x) &=&\displaystyle i\pi^{2-\epsilon}e^{\epsilon i\pi}\frac{\Gamma(\epsilon)}{[x(1-x) q^2]^\epsilon}(1-x) q^\mu\,,\\
    \\
    I_A(x) &=&\displaystyle  i\pi^{2-\epsilon}e^{\epsilon i\pi}\frac{\Gamma(\epsilon)}{[x(1-x) q^2]^\epsilon(nq)}\frac{1}{1-x}\,,\\
    \\
    I_A^\mu(x) &=&\displaystyle  i\pi^{2-\epsilon}e^{\epsilon i\pi}\frac{\Gamma(\epsilon)}{[x(1-x) q^2]^\epsilon(nq)}\left[q^\mu+\frac{x}{1-x}\frac{q^2}{2(nq)}n^\mu\right]\,.
  \end{array}
\end{equation}
Gathering all pieces, we obtain:
\begin{equation}
  \Sigma_F= 2 i\pi^{2}\Gamma(\epsilon)(1-\epsilon)\int_0^1
  \frac{dx}{[\pi x(1-x) q^2]^\epsilon}(1-x) \slashed{q}\,,
\end{equation}
and:
\begin{equation}
  \Sigma_A=
  2i\pi^{2}e^{\epsilon i\pi}\Gamma(\epsilon)\int_0^1
  \frac{dx}{[\pi x(1-x) q^2]^\epsilon} \frac{2x}{1-x}\frac{q^2\slashed{n}}{2(nq)}\,,
\end{equation}
so that:
\begin{equation}
  \begin{array}{rcl}
    \Sigma^{[1]}(q)&=&\displaystyle  -2C_F e^{\epsilon i\pi} \Gamma(\epsilon) \int_0^1
                       dx \left[\frac{4\pi\mu^2}{x(1-x) q^2}\right]^\epsilon
                       \left[ (1-\epsilon) (1-x) \slashed{q}
                       +\frac{2x}{1-x}\slashed{n}\frac{q^2}{2(n
                       q)}\right]\\
    \\
                   &=&\displaystyle  -2C_F \frac{\mu^{2\epsilon}
                       S_\epsilon}{\epsilon_{\rm UV}} \int_0^1
                       dx \left[ (1-x) \slashed{q}
                       +\frac{2x}{1-x}\slashed{n}\frac{q^2}{2(n
                       q)}\right] +\mathcal{O}(\epsilon^0) \,.
  \end{array}
\end{equation}
The integral in $x$ is clearly divergent because of the singularity at
$x=1$ caused by the $1/(nk)$ term in the gluon propagator. Therefore,
before identifying the coefficients $A$ and $B$ using
Eq.~(\ref{eq:selfenergydecomposition}), it is first necessary to make
this integral convergent. To do so, we first note that any
regularisation needs somehow to rely on the light-cone projection of
the incoming/outgoing parton, \textit{i.e.} $(np)$, that defines the
direction along which the so-called ``rapidity'' divergences take
place~\cite{Collins:2011}.  Therefore, we introduce a \textit{generic}
regularisation that we denote by the subscript ``Reg$(np)$'':
\begin{equation}
  \begin{array}{rcl}
    \Sigma^{[1]}(q) &=&\displaystyle  -2C_F \frac{\mu^{2\epsilon}
                        S_\epsilon}{\epsilon_{\rm UV}} \int_0^1
                        dx \left[ (1-x) \slashed{q}
                        +\frac{2x}{1-x}\slashed{n}\frac{q^2}{2(n
                        q)}\right]_{{\rm Reg}(np)}\,.
  \end{array}
\end{equation}
With this at hand, we can extract from the regularisation sign
anything that is not affected by the regularisation itself. In
addition, we take $(nq)=y(np)$, with $y=1\pm\xi$, as required by
Eq.~(\ref{eq:includevirt}). This leads to:
\begin{equation}
  \begin{array}{rcl}
    \Sigma^{[1]}(q) &=&\displaystyle  -2C_F \frac{\mu^{2\epsilon}
                        S_\epsilon}{\epsilon_{\rm UV}} \left\{\frac12\slashed{q}
                        + \left[-2+2 \int_0^1dx\,y \left[\frac{1}{y(1-x)}\right]_{{\rm Reg}(np)}\right]\slashed{n}\frac{q^2}{2(nq)}\right\}\,.
  \end{array}
\end{equation}
It is important to notice that the factor $1/y$ that comes from
$1/(nq)$ must remain \textit{inside} the regularisation sign. We are
therefore forced to multiply and divide by $y$ \textit{outside} the
regularisation sign to reconstruct $1/(nq)$ leaving a leftover factor
of $y$.  This finally allows us to identify the one-loop coefficients
$A^{[1]}$ and $B^{[1]}$ by inspection of
Eq.~(\ref{eq:selfenergydecomposition}), whose sum relevant to
Eq.~(\ref{eq:oneloopvirt}) is:
\begin{equation}
A^{[1]}(y)+B^{[1]}(y) = 2C_F \frac{\mu^{2\epsilon}
                        S_\epsilon}{\epsilon_{\rm UV}} \left[\frac32-2 \int_0^1dx\,y \left[\frac{1}{y(1-x)}\right]_{{\rm Reg}(np)}\right]\,.
  \label{eq:afterregularisation}
\end{equation}
We now focus on the integral in $x$ in
Eq.~(\ref{eq:afterregularisation}) and manipulate it as follows:
\begin{equation}
  \int_0^1
  dx\, y\left[\frac{1}{y(1-x)}\right]_{{\rm
      Reg}(np)} = \int_0^{y}
  dt \left(\frac{1}{t}\right)_{{\rm
      Reg}(np)} =\int_0^{1}\frac{dz}{1-z}+\ln y\,.
\end{equation}
In the first equality we have made the change of variable
$y(1-x) = t$. In the second equality we have extended the integral in
$t$ to the interval $[0,1]$ and subtracted the residual that can now
be integrated giving $\ln y$. We have then made another change of
variable, $t=1-z$, and removed the regularisation sign to match the
notation in Eq.~(\ref{eq:virtualsumrule}). This finally gives:
\begin{equation}
  A^{[1]}(y)+B^{[1]}(y) =2C_F \frac{\mu^{2\epsilon}
    S_\epsilon}{\epsilon_{\rm UV}} \left[\frac32-2 \int_0^{1}\frac{dz}{1-z}-2\ln y\right]\,.
\end{equation}
Plugging this identity into Eq.~(\ref{eq:oneloopvirt}) finally yields:
\begin{equation}
  \hat{F}_{q/q}^{\rm virt}(x,\xi) = 2C_F\sqrt{1-\xi^2} \left[\frac{3}{2}-2\int_0^{1}\frac{dz}{1-z}-\ln(1-\xi^2)\right]\frac{\mu^{2\epsilon}
  S_\epsilon}{\epsilon_{\rm UV}}\delta(1-x)\,,
\end{equation}
that agrees with Eq.~(\ref{eq:virtualsumrule}).

\section{Diagonalisation of the conformal moments}\label{app:conformalmoments}

In this appendix we provide a general proof of
Eq.~(\ref{eq:ConfMomMaster}). To do so, we define $z=y/\xi$ and, using
the change of variable $v=x/\xi$ in the integrals, rewrite the
r.h.s. of Eq.~(\ref{eq:GPDevkernelalaERBL}) without the factor $2C_F$
as follows:
\begin{equation}
\begin{array}{rcl}
  \displaystyle
I &=&\displaystyle \frac{3}{2}C_n^{(3/2)}\left(z\right) - \frac12\int_{1}^{z}dv\left[\frac{v+1}{z-1}C_n^{(3/2)}\left(v\right) -2\frac{C_n^{(3/2)}\left(v\right)-C_n^{(3/2)}\left(z\right)}{z-v}\right]\\
\\
&+&\displaystyle \frac12\int_{-1}^{z}dv\left[\frac{v-1}{z+1}C_n^{(3/2)}\left(v\right)+2\frac{C_n^{(3/2)}\left(v\right)-C_n^{(3/2)}\left(z\right)}{z-v}\right]\,.
\end{array}
\label{eq:integralmoms}
\end{equation}
Now, we use the fact that $C_n^{(3/2)}$ is indeed a polynomial of
degree $n$ whose expansion reads:
\begin{equation}
  C_n^{(3/2)}(x) = \sum_{k=0}^{\lfloor n/2 \rfloor} a_k^{(n)}x^{\ell}\,\quad\mbox{with}\quad  a_k^{(n)} =
(-1)^k2^{\ell}\frac{\Gamma(\ell+k+3/2)}{\Gamma(3/2)k!\ell!}\,,
\end{equation}
with $\ell = n-2k$. This allows us to write:
\begin{equation}
  \begin{array}{rcl}
    I&=&\displaystyle \frac{3}{2}C_n^{(3/2)}\left(z\right)\\
    \\
    &-& \displaystyle \frac12\sum_{k=0}^{\lfloor n/2 \rfloor}
        a_k^{(n)}\left[\int_{1}^{z}dv\frac{v^{\ell+1}+v^{\ell}}{z-1}-\int_{-1}^{z}dv\frac{v^{\ell+1}-v^{\ell}}{z+1}
        +2 \int_{1}^{z}dv\frac{z^{\ell}-v^{\ell}}{z-v}+2
        \int_{-1}^{z}dv\frac{z^{\ell}-v^{\ell}}{z-v}\right]\,.
\end{array}
\end{equation}
Let us now solve all the integrals in the r.h.s. of this equation. The
first gives:
\begin{equation}
\int_{1}^{z}dv\frac{v^{\ell+1}+v^{\ell}}{z-1} = \frac1{\ell+2}\frac{1-z^{\ell+2}}{1-z}+\frac1{\ell+1}\frac{1-z^{\ell+1}}{1-z}=\frac{1}{\ell+2}\sum_{j=0}^{\ell+1}z^{j}+\frac{1}{\ell+1}\sum_{j=0}^{\ell}z^{j}\,,
\end{equation}
and similarly, the second:
\begin{equation}
\int_{-1}^{z}dv\frac{v^{\ell+1}-v^{\ell}}{z+1}=-\frac{(-1)^{\ell+2}}{\ell+2}\sum_{j=0}^{\ell+1}(-z)^{j}+\frac{(-1)^{\ell+1}}{\ell+1}\sum_{j=0}^{\ell}(-z)^{j}\,,
\end{equation}
where we have used the geometric series:
\begin{equation}
\sum_{j=0}^{n} v^{j}=\frac{1-v^{n+1}}{1-v}\,.
\end{equation}
Their combination evaluates to:
\begin{equation}
  \int_{1}^{z}dv\frac{v^{\ell+1}+v^{\ell}}{z-1}-\int_{-1}^{z}dv\frac{v^{\ell+1}-v^{\ell}}{z+1}
  = \left[\frac{1}{\ell+2}+\frac{1}{\ell+1} \right]\sum_{j=0}^{\ell}\left[1+(-1)^{\ell-j}\right]z^{j}\,.
\end{equation}
Notice that the $z^{\ell+1}$ term vanishes because the projector is
null for $j=\ell+1$. Now we turn to the third and fourth integrals in
Eq.~(\ref{eq:integralmoms}):
\begin{equation}
  \begin{array}{rcl}
\displaystyle \int_{1}^{z}dv\frac{z^{\ell}-v^{\ell}}{z-v} &=&\displaystyle
z^{\ell-1}\int_{1}^{z}dv\frac{1-(v/z)^{\ell}}{1-v/z}=
\sum_{j=0}^{\ell-1}z^{\ell-j-1}\int_{1}^{z}dv\,v^j =
                                                \sum_{j=0}^{\ell-1}\frac{z^{\ell}-z^{\ell-j-1}}{j+1} \\
    \\
                                            &=&\displaystyle -\sum_{j=0}^{\ell-1}\frac{z^{j}}{\ell-j}+z^{\ell}\sum_{j=1}^{\ell}\frac{1}{j}\,,
  \end{array}
\end{equation}
and:
\begin{equation}
\int_{-1}^{z}dv\frac{z^{\ell}-v^{\ell}}{z-v} =-\sum_{j=0}^{\ell-1}\frac{(-1)^{\ell-j}z^{j}}{\ell-j}+z^{\ell}\sum_{j=1}^{\ell}\frac{1}{j}\,,
\end{equation}
so that their combination gives:
\begin{equation}
2\int_{1}^{z}dv\frac{z^{\ell}-v^{\ell}}{z-v}+ 2\int_{-1}^{z}dv\frac{z^{\ell}-v^{\ell}}{z-v}=-2\sum_{j=0}^{\ell-1}\frac{1+(-1)^{\ell-j}}{\ell-j} z^{j}+z^{\ell}\sum_{j=1}^{\ell}\frac{4}{j}\,.
\end{equation}
Gathering all pieces, one finds:
\begin{equation}
\begin{array}{rcl}
  \displaystyle
  I &=&\displaystyle \frac{3}{2}C_n^{(3/2)}\left(z\right)\\
  \\
  &-&\displaystyle  \sum_{k=0}^{\lfloor n/2 \rfloor} a_k^{(n)}\left[\sum_{j=0}^{\ell-1}\left(\frac{1}{\ell+2}+\frac{1}{\ell+1} -\frac{2}{\ell-j}\right)\frac{1+(-1)^{\ell-j}}{2}z^{j} +\left(\frac{1}{\ell+2}+\frac{1}{\ell+1} +\sum_{j=0}^{\ell-1}\frac{2}{j+1}\right)z^{\ell}\right]\,.
\end{array}
\end{equation}
Now we exchange the sums over $k$ and the first sum over $j$ in the
second line of the equation above by using the following equality:
\begin{equation}
  \sum_{k=0}^{\lfloor n/2 \rfloor}
  \sum_{j=0}^{\ell-1}\dots=\sum_{k=0}^{\lfloor n/2 \rfloor}
  \sum_{j=0}^{n-2k-1}\dots = \sum_{j=0}^{n-1}\sum_{k=0}^{\lfloor \frac{n-j-1}2 \rfloor}\dots\,,
\end{equation}
finding:
\begin{equation}
\begin{array}{rcl}
  \displaystyle
  I &=&\displaystyle \frac{3}{2}C_n^{(3/2)}\left(z\right)\\
  \\
  &-&\displaystyle \sum_{j=0}^{n-1}z^{j} \frac{1+(-1)^{n-j}}{2}\sum_{k=0}^{\lfloor \frac{n-j-1}2 \rfloor} a_k^{(n)}\left(\frac{1}{n-2k+2}+\frac{1}{n-2k+1}
      -\frac{2}{n-2k-j}\right) \\
\\
&-&\displaystyle \sum_{k=0}^{\lfloor n/2 \rfloor} z^{\ell} a_k^{(n)}\left(\frac{1}{\ell+2}+\frac{1}{\ell+1} +\sum_{j=0}^{\ell-1}\frac{2}{j+1}\right)\,.
\end{array}
\end{equation}
where in the first line we have made explicit $\ell=n-2k$ and used the
equality:
\begin{equation}
\frac{1+(-1)^{\ell-j}}{2} = \frac{1+(-1)^{n-2k-j}}{2}=\frac{1+(-1)^{n-j}}{2}\,.
\end{equation}
This projector nullifies all the terms in the first series over $j$
for which $n-j$ is odd selecting only the even ones. Therefore we can
identify the combination $n-j$ with an even index, \textit{i.e.}
$n-j=2h$, and remove the projector. Replacing the summation index $k$
with $h$ and making explicit the index $\ell$ also in the second line
gives:
\begin{equation}
\begin{array}{rcl}
  \displaystyle
  I &=&\displaystyle \frac{3}{2}C_n^{(3/2)}\left(z\right)\\
  \\
  &-&\displaystyle \sum_{h=0}^{\lfloor n/2 \rfloor}a_h^{(n)}z^{n-2h}
      \Bigg[\frac{1}{n-2h+2}+\frac{1}{n-2h+1}+
      2\sum_{j=1}^{n-2h}\frac{1}{j}\\
\\
&+&\displaystyle \sum_{j=1}^{h} \frac{a_{h-j}^{(n)}}{a_h^{(n)}}\left(\frac{1}{n-2h+2j+2}+\frac{1}{n-2h+2j+1}-\frac1j\right)\Bigg]\,.
\end{array}
\label{eq:almostthere}
\end{equation}
It turns out that the term in the square brackets is independent of
the summation index $h$ (this statement can be easily verified
numerically). Therefore, without loss of generality, we can set $h=0$
inside the square brackets and pull it out from the summation symbol,
obtaining:
\begin{equation}
  I = \left[\frac{3}{2}-\frac{1}{n+2}-\frac{1}{n+1}- 2\sum_{j=1}^{n}\frac{1}{j}\right]C_n^{(3/2)}\left(z\right) = \left[\frac{3}{2}+\frac{1}{(n+1)(n+2)}- 2\sum_{j=1}^{n+1}\frac{1}{j}\right]C_n^{(3/2)}\left(z\right)\,,
\end{equation}
which finally proves the identity in Eq.~(\ref{eq:ConfMomMaster}).

\bibliographystyle{ieeetr}
\bibliography{Bibliography}

\end{document}